\documentclass[11pt]{article}
\usepackage[margin=0.8in]{geometry} 
\usepackage[toc,page]{appendix}

\usepackage{graphicx}				
\graphicspath{{Figures/}{../figures/}}

\usepackage{subfiles}				

\usepackage{amsmath}
\usepackage{amsfonts}
\usepackage{graphicx}
\usepackage{caption}
\usepackage{subcaption}
\usepackage{color}
\usepackage{esint}
\usepackage{xcolor}

\usepackage[T1]{fontenc}

\newcommand{\bsub}{\begin{subequations}}
\newcommand{\esub}{\end{subequations}$\!$}
\renewcommand{\theequation}{\arabic{section}.\arabic{equation}}

\definecolor{plot_blue}{HTML}{1F77B4}
\definecolor{plot_orange}{HTML}{FF7F0E}
\definecolor{plot_green}{HTML}{2CA02C}

\newtheorem{theorem}{Proposition}[section]

\begin{document}

\title{The Linear Stability of Symmetric Spike Patterns for a
  Bulk-Membrane Coupled Gierer-Meinhardt Model} \author{Daniel
  Gomez\thanks{Dept. of Mathematics, UBC, Vancouver,
    Canada. (corresponding author {\tt dagubc@math.ubc.ca})}\enspace,
  Michael J. Ward\thanks{Dept. of Mathematics, UBC, Vancouver,
    Canada. {\tt ward@math.ubc.ca}}\enspace, Juncheng
  Wei\thanks{Dept. of Mathematics, UBC, Vancouver, Canada. {\tt
      jcwei@math.ubc.ca}}} \baselineskip=16pt

\baselineskip=16pt
	
\maketitle

\begin{abstract}
We analyze a coupled bulk-membrane PDE model in which a scalar linear 2-D bulk diffusion process is coupled through a linear Robin boundary condition to a two-component 1-D reaction-diffusion (RD) system with Gierer-Meinhardt (nonlinear) reaction kinetics defined on the domain boundary. For this coupled model, in the singularly perturbed limit of a long-range inhibition and short-range activation for the membrane-bound species, asymptotic methods are used to analyze the existence of localized steady-state multi-spike membrane-bound patterns, and to derive a nonlocal eigenvalue problem (NLEP) characterizing ${\mathcal O}(1)$ time-scale instabilities of these patterns. A central, and novel, feature of this NLEP is that it involves a membrane Green's function that is coupled nonlocally to a bulk Green's function. When the domain is a disk, or in the well-mixed shadow-system limit corresponding to an infinite bulk diffusivity, this Green's function problem is analytically tractable, and as a result we will use a hybrid analytical-numerical approach to determine unstable spectra of this NLEP. This analysis characterizes how the 2-D bulk diffusion process and the bulk-membrane coupling modifies the well-known linear stability properties of steady-state spike patterns for the 1-D Gierer-Meinhardt model in the absence of coupling. In particular, phase diagrams in parameter space for our coupled model characterizing either oscillatory instabilities due to Hopf bifurcations, or competition instabilities due to zero-eigenvalue crossings are constructed. Finally, linear stability predictions from the NLEP analysis are confirmed with full numerical finite-element simulations of the coupled PDE system.
\end{abstract}

\let\thefootnote\relax\footnote{M.\@ Ward and J.\@ Wei acknowledge the support of the NSERC Discovery Grant Program. D.\@ Gomez was supported by an NSERC Doctoral Fellowship.}



\setcounter{equation}{0}
\setcounter{section}{0}
\section{Introduction}\label{sec:intro}

Pattern formation is readily observed in a variety of physical
and biological phenomena. It is widely believed that, for
systems modeled by reaction diffusion (RD) equations, the driving mechanism
behind pattern formation is a diffusion driven (or Turing)
instability. First described in 1952 by Alan\@ M.\@ Turing
\cite{turing_1952}, this mechanism relies on a difference in the
diffusivities of two interacting and diffusing species in order to
drive the system away from a spatially homogeneous, and kinetically
stable, equilibrium solution to one exhibiting spatial patterns. One
of the key insights of Turing is the notion that diffusion, an
intuitively smoothing and stabilizing process, can in fact lead to
spatial instabilities. Following Turing's original work, a substantial
body of literature detailing diffusion-driven instabilities in the
context of a variety of models has been developed. Most pertinent to
our present study is the activator-inhibitor model of Gierer and
Meinhardt \cite{gierer_1972}. 

While Turing instability analysis has been successful in predicting
the onset of spatially periodic instabilities, it does not provide a
full account of pattern formation phenomena. Indeed, a complete
picture requires a characterization of the spatially
periodic patterns that emerge from a Turing instability. To
  do so, one approach has been to use techniques of weakly-nonlinear
analysis where the asymptotically small parameter describes some
distance in parameter space from the Turing instability
bifurcation point. A significant hurdle in such an analysis
occurs when the ratio of activator to inhibitor diffusivities is
small, owing to the fact that the standard Turing-type
  analysis reveals a large band of unstable modes with approximately
equal growth rates. Our focus will instead be on the alternative
theoretical framework that assumes an asymptotically small ratio of
activator to inhibitor diffusivities. In this context, strongly
localized spatial patterns emerge, which are characterized by an
activator that is concentrated in regions of small spatial
  extent.  This strongly localized character of the activator
solution greatly facilitates the asymptotic construction of
steady-state patterns by reducing the problem to that of finding the
spike locations and their heights. Furthermore, similar techniques can
be used to study the linearized stability of strongly localized
patterns (cf.~\cite{iron_2001}, \cite{doelgm}, \cite{ward_n_2003},
\cite{vpd}). These asymptotic reductions provide a framework for a
rigorous existence and linear stability theory of spike patterns
(cf.~\cite{wei_1999}, \cite{wei-book}).

Motivated by various specific biological cell signalling problems with
surface receptor binding (cf.~\cite{levine_2005},
\cite{madzvamuse_2015}, \cite{ratz_2012}, \cite{ratz_2013},
\cite{ratz_2014}, \cite{gomez2007}, \cite{madz2018}, \cite{elliott}),
a more recent focus for research has been to analyze pattern formation
aspects associated with coupled bulk-surface RD systems. Given some
bounded domain, these models consist of an RD system posed in the
interior that is coupled to an additional system posed on the domain
boundary. The coupling for the interior, or bulk, problem is directed
through the boundary conditions, whereas on the boundary, or membrane,
it takes the form of source or ``feed'' terms. It is worth noting that
these coupled systems are to be understood as a leading order
approximation in the limit of a small, but nonzero, membrane
width. One key motivation for studying these models is that in
specific applications the difference in the diffusivities of
two species may not be substantial enough to lead to a Turing
instability. On the other hand the bulk, or cytosolic,
diffusivities are typically substantially larger than their membrane
counterparts. It is proposed, therefore, that it is this large
difference between the bulk and membrane diffusivities that can lead
to a Turing instability and ultimately pattern formation
(\cite{ratz_2012}, \cite{ratz_2014}, \cite{madzvamuse_2015},
\cite{madzvamuse_2016}).

The primary goal of this paper is to initiate detailed asymptotic
studies of strongly localized patterns in coupled bulk-surface RD
systems.  To this end, we introduce such a PDE model in which a scalar
linear 2-D bulk diffusion process is coupled through a linear Robin
boundary condition to a two-component 1-D RD system with
Gierer-Meinhardt (nonlinear) reaction kinetics defined on the domain
boundary or ``membrane''. Similar, but more complicated, coupled
bulk-surface models, some with nonlinear bulk reaction kinetics and in
higher space dimensions, have previously been formulated and studied
through either full PDE simulations or from a Turing instability
analysis around some patternless steady-state (cf.~\cite{ratz_2012},
\cite{ratz_2013}, \cite{ratz_2014}, \cite{madzvamuse_2015},
\cite{madzvamuse_2015}, \cite{ratz_2014}, \cite{macdonald2013}).  Our
coupled model, formulated below, provides the first analytically
tractable PDE system with which to investigate how the bulk diffusion
process and the bulk-membrane coupling influences the existence and
linear stability of localized ``far-from-equilibrium''
(cf.~\cite{nishiura}) steady-state spike patterns on the membrane. In
the limit where the bulk and membrane are uncoupled, our PDE system
reduces to the well-studied 1-D Gierer-Meinhardt RD system on the
membrane with periodic boundary conditions. The existence and linear
stability of steady-state spike patterns for this limiting uncoupled
problem is well understood (cf.~\cite{wei_1999}, \cite{iron_2001},
\cite{doelgm}, \cite{vpd}, \cite{ward_n_2003}).

Our model is formulated as follows: Given some 2-D bounded domain
$\Omega$ we pose on its boundary an RD system with Gierer-Meinhardt kinetics
\begin{subequations}\label{eq:bsrde2d}
\begin{gather}
  \partial_t u = \varepsilon^2 \partial_\sigma^2 u - u + u^p/v^q\,,
  \qquad 0 < \sigma < L\,, \quad t > 0\,, \label{eq:bsrde2d_u}\\
 \tau_s \partial_t v = D_v \partial_\sigma^2 v - (1+K)v + KV +
 \varepsilon^{-1}u^m/v^s\,, \qquad 0<\sigma<L\,,\quad t > 0\,,
 \label{eq:bsrde2d_v}
\end{gather}
where $\sigma$ denotes arclength along the boundary of length $L$, and
where both $u$ and $v$ are $L$-periodic. In $\Omega$ we consider the
linear 2-D bulk diffusion process
\begin{equation}
  \tau_b \partial_t V = D_b\Delta V - V\,,\qquad x\in\Omega\,,\qquad
  D_b\partial_n V + K V = K v\,,\qquad x\in\partial\Omega\,,
  \label{eq:bsrde2d_V}
\end{equation}
\end{subequations}
where the coupling to the membrane is through a Robin condition. The
Gierer-Meinhardt exponent set
$(p,q,m,s)$ is assumed to satisfy the usual conditions
(cf.~\cite{wei_1999,iron_2001})
\begin{equation}\label{eq:gm_expon}
  p>1\,, \qquad q>0\,, \qquad m>0\,,  \qquad s\ge 0\,, \qquad 0 <
 \frac{p-1}{q} <\frac{m}{s+1} \,.
\end{equation}
In this model $\tau_b$ and $\tau_s$ are time constants associated with
the bulk and membrane diffusion process, $D_b$ and $D_v$ are the
diffusivities of the bulk and membrane inhibitor fields, and $K>0$ is
the bulk-membrane coupling parameter.

The paper is organized as follows. In \S\ref{sec:general} we
use the method of matched asymptotic expansions to derive a
nonlinear algebraic system for the spike locations and heights of a
multi-spike steady-state pattern for the membrane-bound species. A
singular perturbation analysis is then used to derive an NLEP
characterizing the linear stability of these localized steady-states
to ${\mathcal O}(1)$ time-scale instabilities.  A more explicit
analysis of both the nonlinear algebraic system and the NLEP
requires the calculation of a novel 1-D membrane Green's function
that is coupled nonlocally to a 2-D bulk Green's function. Although
intractable analytically in general domains, this Green's function
problem is explicitly studied in two special cases: the
\mbox{well-mixed limit}, $D_b\gg 1$, for the bulk diffusion field in
an arbitrary bounded 2-D domain, and when $\Omega$ is a disk of
radius $R$ with finite $D_b$.

\begin{figure}[b!]
	\centering
	\includegraphics[height=2.25in]{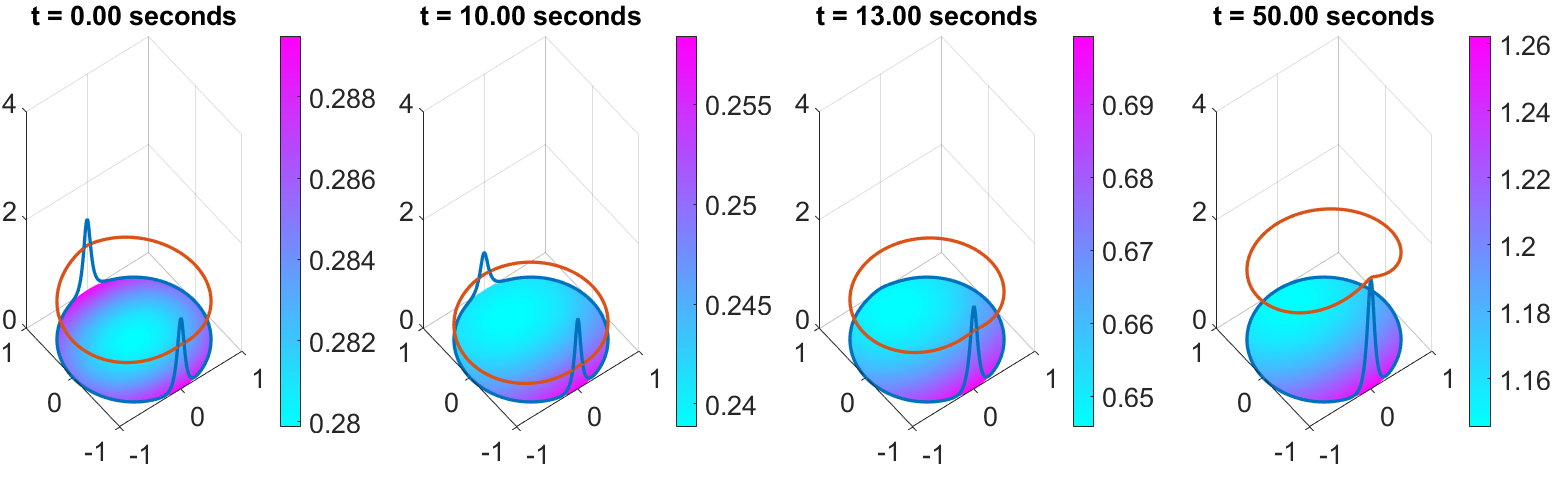}
	\caption{Snapshots of the numerically computed solution of \eqref{eq:bsrde2d} starting from a 2-spike equilibrium for the unit disk with $D_b=10$, $\tau_s=0.6$, $\tau_b=0.1$, $K = 2$, and $D_v = 10$ (this corresponds to point $2$ in the left panel of Figure \ref{fig:disk_legend_and_snapshots}). The bulk inhibitor is shown as the colourmap, whereas the lines along the boundary indicate the activator ({\color{plot_blue}blue}) and inhibitor ({\color{plot_orange}orange}) membrane concentrations. The results show a competition instability, leading to the annihilation of a spike.}\label{fig:disk_snapshots_p_2_n_2}
\end{figure}

In \S\ref{sec:symm} we restrict our steady-state and NLEP
analysis to these two special cases, and consider only symmetric
$N$-spike patterns characterized by equally-spaced spikes on the 1-D
membrane, for which the nonlinear algebraic system is readily
solved. In this restricted scenario, by using a hybrid
analytical-numerical method on the NLEP we are then able to provide
linear stability thresholds for either synchronous or asynchronous
perturbations of the steady-state spike amplitudes. More
specifically, we provide phase diagrams in parameter space
characterizing either oscillatory instabilities of the spike
amplitudes, due to Hopf bifurcations, or asynchronous (competition)
instabilities, due to zero-eigenvalue crossings, that trigger spike
annihilation events. These linear stability phase diagrams show that
the bulk-membrane coupling can have a diverse effect on the linear
stability of symmetric $N$-spike patterns. In each case we find that
stability thresholds are typically increased (making the system more
stable) when the bulk-membrane coupling parameter $K$ is relatively
small, whereas the stability thresholds are decreased as $K$
continues to increase. This nontrivial effect is further complicated
when studying synchronous instabilities, for which there appears to
be a complex interplay between the membrane and bulk timescales,
$\tau_s$ and $\tau_b$, as well as with the coupling $K$. At various
specific points in these phase diagrams for both the well-mixed case
(with $D_b$ infinite) and the case of the disk (with $D_b$ finite),
our linear stability predictions are confirmed with full numerical
finite-element simulations of the coupled PDE system
(\ref{eq:bsrde2d}).
 
\begin{figure}[t]
	\centering
	\includegraphics[height=2.25in]{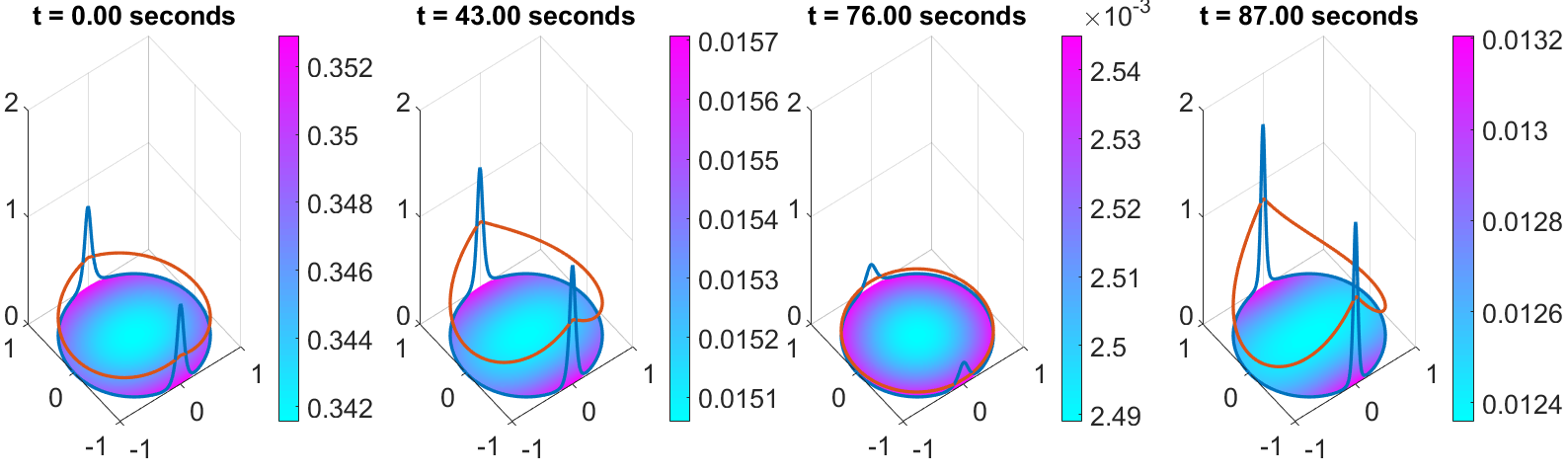}
	\caption{Snapshots of the numerically computed solution of \eqref{eq:bsrde2d} starting from a 2-spike equilibrium for the unit disk with $D_b=10$, $\tau_s=0.6$, $\tau_b=0.1$, $K = 0.025$, and $D_v = 1.8$ (this corresponds to point $5$ in the left panel of Figure \ref{fig:disk_legend_and_snapshots}). The bulk inhibitor is shown as the colourmap, whereas the lines along the boundary indicate the activator ({\color{plot_blue}blue}) and inhibitor ({\color{plot_orange}orange}) membrane concentrations. The results show a synchronous oscillatory instability of the spike amplitudes.}\label{fig:disk_snapshots_p_5_n_2}
\end{figure}
 
As an illustration of spike dynamics resulting from full
PDE simulations, in Figures \ref{fig:disk_snapshots_p_2_n_2} and
\ref{fig:disk_snapshots_p_5_n_2} we show results
computed for the unit disk with $D_b=10$, showing competition and oscillatory instabilities for a two-spike solution, respectively. The parameter values are given in the figure captions and correspond to specific points in the linear
stability phase diagram given in the left panel of
Figure \ref{fig:disk_legend_and_snapshots}.

In \S \ref{sec:boundary} we use a regular perturbation analysis to
show the effect on the asynchronous instability thresholds of
introducing a small smooth perturbation of the boundary of the unit
disk. This analysis, which requires a detailed calculation of the
perturbed 1-D membrane Green's function, shows that a two-spike
pattern can be stabilized by a small outward peanut-shaped
deformation of a circular disk.  Finally, in \S\ref{sec:discussion}
we briefly summarize our results and highlight some open problems and
directions for future research.

\setcounter{equation}{0}
\setcounter{section}{1}
\section{Spike Equilibrium and its Linear Stability: General Asymptotic Theory}\label{sec:general}

\subsection{Asymptotic Construction of $N$-Spike Equilibria}\label{sec:eq_theory}

In this section we provide an asymptotic construction of an
$N$-spike steady-state solution to \eqref{eq:bsrde2d}. Specifically,
we consider the steady-state problem for the membrane species
\begin{subequations}\label{eq:bsrde2d_equilibrium}
\begin{gather}
  \varepsilon^2 \partial_\sigma^2 u_e  - u_e + u_e^p/v_e^q = 0\,, \qquad
         0 < \sigma < L \,,  \quad u\,\, \mbox{is $L$-periodic}\,, \\
  D_v \partial_\sigma^2 v_e - (1+K)v_e + KV_e +\varepsilon^{-1}u_e^m/v_e^s =
     0\,,  \qquad  0<\sigma<L\,,\quad v\,\, \mbox{is $L$-periodic}\,,
   \label{eq:bsrde2d_equilibrium_vet}
\end{gather}
which is coupled to the steady-state bulk-diffusion process by
\begin{equation}
  D_b\Delta V_e - V_e = 0\,,\quad x\in\Omega\,; \qquad
  D_b\partial_n V_e + K V_e = K v_e\,,\quad x\in\partial\Omega \,.\label{eq:b_v}
\end{equation}
\end{subequations}

From (\ref{eq:b_v}), the bulk-inhibitor evaluated on the
membrane is readily expressed in terms of a Green's function as
\begin{equation}\label{eq:V_e}
  V_e(\sigma) = K \int_0^L G_\Omega(\sigma,\tilde{\sigma})
  v_e(\tilde{\sigma})\,d\tilde{\sigma}\,,
\end{equation}
where we have used arc-length to parameterize the boundary. Here,
$G_\Omega(\sigma,\tilde{\sigma})$ is the Green's function satisfying
\begin{equation}
  D_b\Delta_x G_\Omega(x,\tilde{\sigma}) - G_\Omega(x,\tilde{\sigma}) = 0\,,
  \quad x\in\Omega\,,\qquad D_b\partial_n G_\Omega(\sigma,\tilde{\sigma}) +
  K G_\Omega(\sigma,\tilde{\sigma}) = \delta(\sigma-\tilde{\sigma})\,,\quad
  0<\sigma<L\,.
\end{equation}
We remark that the values of the bulk-inhibitor field within the bulk can
likewise be obtained with a Green's function whose source
is in the interior. However, for our purposes it is only
the restriction to the boundary that is important.

At this stage the steady-state membrane problem takes the form
\begin{equation}    \label{eq:bsrde2d_equilibrium_ve}
\begin{split}
 \varepsilon^2 \partial_\sigma^2 u_e  - u_e + & u_e^p/v_e^q = 0\,, \qquad
                  0 < \sigma < L\,, \\
 D_v \partial_\sigma^2 v_e - (1+K)v_e + & K^2\int_0^L
    G_\Omega(\sigma,\tilde{\sigma}) v_e(\tilde{\sigma})\,d\tilde{\sigma}
    + \varepsilon^{-1}u_e^m/v_e^s = 0\,, \qquad 0<\sigma<L\,,
  \end{split}
\end{equation}
which differs from the problem studied in \cite{iron_2001}
for the uncoupled ($K=0$) case only by the addition of the non-local
term. This additional term leads to difficulties in the construction
of spike patterns. In particular, it complicates the concept of a
"symmetric" pattern since, in general, the non-local term will not be
translation invariant. Moreover, in the well-mixed
and disk case, the construction of \textit{asymmetric} patterns is
more intricate as a result of the non-local term.

We now construct an $N$-spike steady-state pattern for
(\ref{eq:bsrde2d_equilibrium_ve}) characterized by an activator
concentration that is localized at $N$ distinct spike
locations $0\leq \sigma_1<...<\sigma_N<L$ to be determined.
We assume that the spikes are well-separated in the sense that
$|\sigma_{\lbrace{(i+1)\mod N\rbrace}} - \sigma_i \mod L| \gg
\varepsilon$ for $i=1,\ldots,N$. Upon introducing
stretched coordinates $y = \varepsilon^{-1}(\sigma-\sigma_j)$, we
deduce that the inhibitor field is asymptotically constant near
each spike, i.~e.
\begin{equation}
v_e\sim v_{ej} \equiv v_e(\sigma_j)\,.
\end{equation}
In addition, the activator concentration is determined in
terms of the unique solution $w(y)$ to the core problem
\begin{equation}\label{eq:core}
  w^{\prime\prime} - w + w^p = 0\,,\quad y\in\mathbb{R}\,,\qquad w^{\prime}(0)=0\,,
  \quad w(0)>0 \,, \quad w(y)\rightarrow 0 \quad \mbox{as} \quad
  |y|\rightarrow\infty\,.
\end{equation}
Since the solution to the core problem decays exponentially as
$y\rightarrow \pm\infty$ we deduce that
\begin{equation}
  u_e(\sigma) \sim \sum_{j=1}^{N}v_{ej}^\gamma
  w\bigl(\varepsilon^{-1}[\sigma-\sigma_j]\bigr)\,,\qquad
  \mbox{as} \quad \varepsilon\rightarrow 0 \,,
\end{equation}
where $\gamma \equiv q/(p-1)$. The solution to (\ref{eq:core})
is given explicitly as
\begin{equation}\label{eq:w}
  w(y) = \biggl(\frac{p+1}{2}\biggr)^{\frac{1}{p-1}}
  \biggl[\text{sech}\biggl(\frac{p-1}{2}y\biggr)\biggr]^{\frac{2}{p-1}}.
\end{equation}

Next, since $u_e$ is localized, we have in the sense of
distributions that
\begin{equation*}
  \varepsilon^{-1}u_e^m/v_e^s\longrightarrow\omega_m\sum_{j=1}^{N}
  [v_e(\sigma_j)]^{\gamma m - s}\delta(\sigma-\sigma_j)\qquad
   \mbox{as} \quad \varepsilon\rightarrow 0 \,,
\end{equation*}
where we have defined
\begin{equation}
\omega_m \equiv \int_{-\infty}^{\infty}[w(y)]^m \, dy \,.
\end{equation}
In this way, for $\varepsilon\rightarrow 0$, we
obtain from (\ref{eq:bsrde2d_equilibrium_ve}) the following
integro-differential equation for the inhibitor field:
\begin{equation}\label{eq:bsrde2d_ve_asy_eq}
  D_v \partial_\sigma^2 v_e  - (1+K)v_e  + K^2\int_0^L
  G_\Omega(\sigma,\tilde{\sigma})v_e(\tilde{\sigma})\, d\tilde{\sigma} =
  -\omega_m\sum_{j=1}^{N}v_{ej}^{\gamma m - s}\delta(\sigma-\sigma_j) \,.
\end{equation}
To conveniently represent the solution to this equation we
introduce the Green's function $G_{\partial\Omega}(\sigma,\zeta)$
satisfying
\begin{equation}
  D_v\partial_\sigma^2 G_{\partial\Omega}(\sigma,\zeta) -
  (1+K) G_{\partial\Omega}(\sigma,\zeta) + K^2\int_0^L
  G_\Omega(\sigma,\tilde{\sigma})G_{\partial\Omega}(\tilde{\sigma},\zeta)
  \, d\tilde{\sigma} = - \delta(\sigma-\zeta)\,,\qquad 0<\sigma,\zeta< L \,.
\end{equation}
In terms of this Green's function, the membrane inhibitor
field is given by
\begin{equation}\label{eqs:ve}
  v_e(\sigma) = \omega_m \sum_{j=1}^{N} v_{ej}^{\gamma m - s}
  G_{\partial\Omega}(\sigma,\sigma_j)\,.
\end{equation}
Substituting $\sigma = \sigma_i$, and recalling the
definition $v_{ei} \equiv v_e(\sigma_i)$, (\ref{eqs:ve}) yields the
$N$ self-consistency conditions
\begin{equation}
  v_{ei} - \omega_m \sum_{j=1}^{N}v_{ej}^{\gamma m - s}
  G_{\partial\Omega}(\sigma_i,\sigma_j) = 0\,,\qquad i=1,\ldots,N\,.
\end{equation}
These conditions provide the first $N$ algebraic equations for
 our overall system in $2N$ unknowns to be completed below. The
remaining $N$ equations arise from solvability conditions
when performing a higher-order matched asymptotic expansion analysis of
the steady-state solution.

To this end, we again introduce stretched coordinates
$y = \varepsilon^{-1}\bigl(\sigma-\sigma_j)$, but we now
introduce a two-term inner expansion for the surface bound species
for $\varepsilon\to 0$ as
\begin{equation}
  u_e(y) \sim v_{ej}^\gamma w(y) + \varepsilon u_1(y) +
  {\mathcal O}(\varepsilon^2)\,,
  \qquad v_e(y) \sim v_{ej} + \varepsilon v_1(y) + {\mathcal O}(\varepsilon^2)\,,
  \qquad V_e \sim {\mathcal O}(1)\,.
\end{equation}
Upon substituting this expansion into
\eqref{eq:bsrde2d_equilibrium}, and collecting the
${\mathcal O}(\varepsilon)$ terms, we get
\begin{equation}
  \mathcal{L}_0 u_1 \equiv  u_1^{\prime\prime} - u_1 + pw^{p-1}u_1 =
  qv_{ej}^{\gamma - 1} w^p v_1\,, \qquad
      D_v v_1^{\prime\prime} + v_{ej}^{\gamma m - s} w^m = 0\,.
                         \label{eq:higher_order_v_1}
\end{equation}
Since $\mathcal{L}_0 w^{\prime} = 0$, the solvability
condition for the first equation yields that
\begin{equation*}
  q v_{ej}^{\gamma - 1} \int_{-\infty}^{\infty} w^p w^{\prime} v_1 \, dy = 0\qquad
  \Longleftrightarrow \qquad \int_{-\infty}^{\infty} (w^{p+1})^{\prime} v_1 \,
  dy = 0\,.
\end{equation*}
Then, we integrate by parts twice, use the exponential decay of $w(y)$
as $|y|\rightarrow\infty$, and substitute \eqref{eq:higher_order_v_1}
for $v_1^{\prime\prime}$. This yields that
\begin{equation*}
  I_p(y)v_1^{\prime}(y)\biggr|_{-\infty}^{\infty} + \frac{v_{ej}^{\gamma m - s}}{D_v}
  \int_{-\infty}^{\infty} I_p(y)[w(y)]^m \, dy = 0\,,
\end{equation*}
where we have defined $I_p(y) \equiv \int_0^y [w(z)]^{p+1}dz$. Since
$w$ is even, while $I_p$ is odd, the integral above vanishes, and we get
\begin{equation*}
v_1^{\prime}(+\infty) + v_1^{\prime}(-\infty) = 0\,.
\end{equation*}
In this way, a higher order matching process between the
inner and outer solutions yields the \textit{balance conditions},
\begin{equation*}
  \partial_\sigma v_e(\sigma_i + 0) + \partial_\sigma v_e(\sigma_i - 0) = 0\,,
  \qquad i=1,\ldots,N\,.
\end{equation*}
By using (\ref{eqs:ve}) for $v_e$, we can write these
balance equations in terms of the Green's function
$G_{\partial\Omega}$ as
\begin{equation}
v_{ei}^{\gamma m - s} \bigl[ \partial_{\sigma} G_{\partial\Omega}(\sigma_i + 0,\sigma_i)
+ \partial_\sigma G_{\partial\Omega}(\sigma_i - 0,\sigma_i)\bigr] +
2 \sum_{j\neq i} v_{ej}^{\gamma m - s} \partial_\sigma
G_{\partial\Omega}(\sigma_i,\sigma_j) = 0\,,\qquad i=1,\ldots,N\,.
\end{equation}
We summarize the results of this formal asymptotic construction
in the following proposition:

\begin{theorem} As $\varepsilon\rightarrow 0$ an $N$-spike
  steady-state solution to
  \eqref{eq:bsrde2d_equilibrium} with spikes centred at
  $\sigma_1,...,\sigma_N$ is asymptotically given by
\begin{subequations}\label{eq:asymptotic_equilibrium}
\begin{align}
  & u_e(\sigma) \sim \sum_{j=1}^{N} v_{ej}^\gamma
    w\bigl(\varepsilon^{-1}[\sigma-\sigma_j])\,, \qquad
   v_e(\sigma) \sim \omega_m \sum_{j=1}^{N}v_{ej}^{\gamma m - s}
    G_{\partial\Omega}(\sigma,\sigma_j)\,, \\
  & V_e(\sigma) \sim \omega_m K \sum_{j=1}^{N}v_{ej}^{\gamma m - s}\int_{0}^{L}
    G_{\Omega}(\sigma,\tilde{\sigma})G_{\partial\Omega}(\tilde{\sigma},\sigma_j)\,
    d\tilde{\sigma}\,,
\end{align}
\end{subequations}
where $\omega_m \equiv \int_{-\infty}^{\infty}[w(y)]^m \, dy$ and
$\gamma \equiv {q/(p-1)}$. Here the steady-state spike
  locations $\sigma_1,...,\sigma_N$ and $v_{e1},...,v_{eN}$, which
  determine the heights of the spikes, are to be found from
  the following non-linear algebraic system:
\begin{subequations}\label{eq:equi_alg_system}
\begin{gather}
   v_{ei} - \omega_m \sum_{j=1}^{N}v_{ej}^{\gamma m - s}
    G_{\partial\Omega}(\sigma_i,\sigma_j) = 0\,, \qquad  i=1,\ldots,N\,,
                \label{eq:equi_alg_system_consistency}\\
    v_{ei}^{\gamma m - s} \bigl[ \partial_{\sigma} G_{\partial\Omega}
    (\sigma_i + 0,\sigma_i) + \partial_\sigma G_{\partial\Omega}
    (\sigma_i - 0,\sigma_i)\bigr] + 2 \sum_{j\neq i} v_{ej}^{\gamma m - s}
    \partial_\sigma G_{\partial\Omega}(\sigma_i,\sigma_j) = 0\,, \qquad
    i=1,\ldots,N\,.  \label{eq:equi_alg_system_balance} 
\end{gather}
\end{subequations}
\end{theorem}

\subsection{Linear Stability of $N$-Spike Equilibria}

\paragraph{}In our linear stability analysis, given below,
of $N$-spike equilibria we make two simplifying assumptions. First,
we focus exclusively on the case $s=0$. Second, we consider only
instabilities that arise on an ${\mathcal O}(1)$
timescale. Therefore, we do not consider very weak instabilities,
occurring on asymptotically long time-scales in $\varepsilon$, that
are due to any unstable small eigenvalue that tends to zero as
$\varepsilon\to 0$.

Let $u_e(\sigma)$, $v_e(\sigma)$, and $V_e(x)$ denote the
the steady-state constructed in \S \ref{sec:eq_theory}. For
$\lambda\in\mathbb{C}$, we consider a perturbation of the form
\begin{equation*}
  u(\sigma) = u_e(\sigma) + e^{\lambda t}\phi(\sigma)\,,\qquad v(\sigma) =
  v_e(\sigma) + e^{\lambda t} \psi(\sigma)\,,\qquad V(x) = V_e(x) +
  e^{\lambda t}\eta(x) \,,
\end{equation*}
where $\phi$, $\psi$, and $\eta$ are small. Upon substituting into
\eqref{eq:bsrde2d} and linearizing, we obtain the eigenvalue problem
\begin{subequations}
\begin{gather}
   \varepsilon^2 \partial_\sigma^2 \phi - \phi + p u_e^{p-1}v_e^{-q}\phi -
   qu_e^pv_e^{-(q+1)}\psi = \lambda\phi\,, \qquad 0<\sigma< L\,,
   \label{eq:stability_PDE_phi}\\
    D_v\partial_\sigma^2\psi - \mu_{s\lambda}^2\psi + K\eta = -
    m\varepsilon^{-1}u_e^{m-1}\phi\,, \qquad 0<\sigma< L\,,
    \label{eq:stability_PDE_psi}\\
   D_b\Delta\eta - \mu_{b\lambda}^2\eta = 0\,, \qquad x \in \Omega\,, \\
   D_b\partial_n \eta + K\eta = K\psi\,,  \qquad x \in \partial\Omega\,,
\end{gather}
\end{subequations}
where we have defined $\mu_{s\lambda}$ and $\mu_{b\lambda}$ by
\begin{equation}
  \mu_{s\lambda} = \sqrt{1+K+\tau_s\lambda}\,,\qquad \mu_{b\lambda} =
  \sqrt{1+\tau_b\lambda}\,.
\end{equation}
The bulk inhibitor field evaluated on the boundary is represented
as
\begin{equation*}
  \eta(\sigma) = K\int_0^L G^\lambda_\Omega(\sigma,\tilde{\sigma})
  \psi(\tilde{\sigma})\, d\tilde{\sigma} \,,
\end{equation*}
where $G^\lambda_\Omega$ is the $\lambda$-dependent bulk Green's function satisfying
\begin{equation}\label{eq:g_lam_surf}
  D_b\Delta_x G_\Omega^\lambda(x,\tilde{\sigma}) - \mu_{b\lambda}^2
  G_\Omega^\lambda(x,\tilde{\sigma}) = 0\,,\quad x\in\Omega\,,\qquad
  D_b \partial_n G_\Omega^\lambda(\sigma,\tilde{\sigma}) + K
  G_\Omega^\lambda(\sigma,\tilde{\sigma}) = \delta(\sigma-\tilde{\sigma})\,,
  \quad 0<\sigma<L\,.
\end{equation}
Next, we seek a localized activator perturbation of the form
\begin{equation}
  \phi(\sigma) \sim \sum_{j=1}^{N}\phi_j\bigl(\varepsilon^{-1}
  [\sigma-\sigma_j]\bigr)\,,
\end{equation}
where we impose that $\phi_j(y)\rightarrow 0$ as
$|y|\rightarrow \infty$. With this form, we evaluate in the
  sense of distributions that
\begin{equation*}
  \varepsilon^{-1}mu_e^{m-1}\phi\longrightarrow m\sum_{j=1}^N v_{ej}^{\gamma(m-1)}
  \left( \int_{-\infty}^{\infty}[w(y)]^{m-1}\phi_j(y)\, dy\right)\,
  \delta(\sigma-\sigma_j)
  \qquad \mbox{as} \quad \varepsilon\rightarrow 0 \,.
\end{equation*}
By using this limiting result in (\ref{eq:stability_PDE_psi}), the
problem for $\psi$ becomes
\begin{equation*}
  D_v\partial_\sigma^2 \psi - \mu_{s\lambda}^2\psi + K^2 \int_0^L G_\Omega^\lambda
  (\sigma,\tilde{\sigma})\psi(\tilde{\sigma})\, d\tilde{\sigma} = -
  m\sum_{j=1}^{N}v_{ej}^{\gamma(m - 1)}
  \left(\int_{-\infty}^{\infty} [w(y)]^{m-1}\phi_j(y)\,dy\right)\, 
  \delta(\sigma-\sigma_j)\,.
\end{equation*}
The solution to this problem is represented as
\begin{equation}\label{lin:psi}
  \psi(\sigma) = m\sum_{j=1}^{N}v_{ej}^{\gamma(m-1)} G_{\partial\Omega}^\lambda
  (\sigma,\sigma_j)\int_{-\infty}^{\infty} [w(y)]^{m-1}\phi_j(y)\, dy\,,
\end{equation}
where $G_{\partial\Omega}^\lambda$ is the $\lambda$-dependent
membrane Green's function satisfying
\begin{equation}\label{eq:g_lam_memb}
  D_v\partial_\sigma^2 G_{\partial\Omega}^\lambda(\sigma,\zeta) -
  \mu_{s\lambda}^2 G_{\partial\Omega}^\lambda(\sigma,\zeta) +
  K^2\int_0^L G_\Omega^\lambda(\sigma,\tilde{\sigma})G_{\partial\Omega}^\lambda
  (\tilde{\sigma},\zeta)\, d\tilde{\sigma} = - \delta(\sigma-\zeta)\,,
  \qquad 0<\sigma,\zeta< L \,.
\end{equation}

Next, it is convenient to re-scale $v_e$ as
\begin{equation}
  v_e(\sigma) =\omega_m^{\frac{1}{1-\gamma m}}\hat{v}_e(\sigma)\,,\qquad
  v_{ej} = \omega_m^{\frac{1}{1-\gamma m}}\hat{v}_{ej}\,.
\end{equation}
In the stretched coordinates $y = \varepsilon^{-1}(\sigma-\sigma_j)$,
we use \eqref{lin:psi} to obtain that
\eqref{eq:stability_PDE_phi} becomes
\begin{equation*}
  \phi_i^{\prime\prime} - \phi_i + pw^{p-1}\phi_i -
  mqw^p\sum_{j=1}^{N}\hat{v}_{ei}^{\gamma - 1}
  G_{\partial\Omega}^\lambda(\sigma_i,\sigma_j)\hat{v}_{ej}^{\gamma(m-1)}
  \frac{\int_{-\infty}^\infty w^{m-1}\phi_j\,dy}{\int_{-\infty}^\infty w^m\,dy} =
  \lambda \phi_i\,.
\end{equation*}
To recast this spectral problem in vector form we define
\begin{equation}\label{lin:mat_def}
  \pmb{\phi} \equiv \begin{pmatrix} \phi_1 \\ \vdots \\ \phi_N\end{pmatrix}\,,
  \quad
  \hat{\mathcal{V}}_e \equiv
  \begin{pmatrix} \hat{v}_{e1} &  & 0 \\ & \ddots & \\ 0 & & \hat{v}_{eN}
  \end{pmatrix}\,,\quad
  \mathcal{G}_{\partial\Omega}^\lambda \equiv
  \begin{pmatrix} G_{\partial\Omega}^\lambda(\sigma_1,\sigma_1) & \cdots
    & G_{\partial\Omega}^\lambda(\sigma_1,\sigma_N) \\
    \cdots & \ddots & \vdots \\
    G_{\partial\Omega}^\lambda(\sigma_N,\sigma_1) & \cdots
    & G_{\partial\Omega}^\lambda(\sigma_N,\sigma_N)
  \end{pmatrix}\,,
\end{equation}
and we introduce the matrix $\mathcal{E}$ by
\begin{equation}\label{eq:mate}
  \mathcal{E} \equiv \hat{\mathcal{V}}_e^{\gamma - 1}
  \mathcal{G}_{\partial\Omega}^\lambda \hat{\mathcal{V}}_e^{\gamma (m-1)} \,.
\end{equation}
In this way, we deduce that $\pmb{\phi}$ must solve the vector
nonlocal eigenvalue problem (NLEP) given by
\begin{equation}\label{lin:vec}
  \pmb{\phi}^{\prime\prime}(y) - \pmb{\phi}(y) + pw^{p-1}\pmb{\phi}(y) -
  mqw^p\frac{\int_{-\infty}^{\infty}[w(y)]^{m-1}\mathcal{E}\pmb{\phi}(y)\,dy}
  {\int_{-\infty}^{\infty} [w(y)]^m\,dy} = \lambda \pmb{\phi}(y)\,.
\end{equation}
We can reduce this vector NLEP to a collection of scalar NLEPs by
diagonalizing it. Specifically, we seek perturbations of the form
$\pmb{\phi} = \phi\pmb{c}$ where $\pmb{c}$ is an eigenvector of
$\mathcal{E}$, that is
\begin{equation}\label{lin:mate_eig}
\mathcal{E}\pmb{c} = \chi(\lambda)\pmb{c} \,.
\end{equation}
Then, it readily follows that the vector NLEP (\ref{lin:vec})
can be recast as the scalar NLEP
\begin{equation}\label{eq:NLEP_general_scalar}
  \mathcal{L}_0\phi - mq \chi(\lambda) w^p\frac{\int_{-\infty}^{\infty}[w(y)]^{m-1}
    \phi(y)\,dy}{\int_{-\infty}^{\infty} [w(y)]^m\,dy} = \lambda \phi\,,
\end{equation}
where $\chi(\lambda)$ is any eigenvalue of $\mathcal{E}$. In
(\ref{eq:NLEP_general_scalar}), the operator $\mathcal{L}_0$, referred to
as the local operator, is defined by
\begin{equation}
\mathcal{L}_0 \phi \equiv\phi^{\prime\prime}(y) - \phi(y) + pw^{p-1}\phi(y)\,.
\end{equation}
Notice that we obtain a (possibly) different NLEP for each eigenvalue
$\chi(\lambda)$ of $\mathcal{E}$. Therefore, the spectrum of
the matrix $\mathcal{E}$ will be central in the analysis below for
classifying the various types of instabilities that can occur.

\subsection{Reduction of NLEP to an Algebraic Equation and an Explicitly Solvable Case}
Next, we show how to reduce the determination of the spectrum of
the NLEP (\ref{eq:NLEP_general_scalar}) to a root-finding problem. To this
end, we define $c_m$ by
\begin{equation}\label{red:cm}
c_m \equiv mq \chi(\lambda) \frac{\int_{-\infty}^{\infty}[w(y)]^{m-1}\phi(y)\,dy}
{\int_{-\infty}^{\infty} [w(y)]^m\,dy}\,,
\end{equation}
and write the NLEP as
  $(\mathcal{L}_0 - \lambda) \phi = c_m w^p$, so that
  $\phi = c_m (\mathcal{L}_0 - \lambda)^{-1} w^p$. Upon multiplying
  both sides of this expression by $w^{m-1}$, we integrate over the
  real line and substitute the resulting expression back into
  (\ref{red:cm}). For eigenfunctions for which
  $\int_{-\infty}^{\infty} w^{m-1}\phi \, dy\neq 0$, we readily obtain
  that $\lambda$ must be a root of $\mathcal{A}(\lambda)=0$, where
\begin{equation}\label{eq:NLEP_algebraic}
  \mathcal{A}(\lambda) \equiv \mathcal{C}(\lambda) - \mathcal{F}(\lambda)
  \qquad \mathcal{C} \equiv \frac{1}{\chi(\lambda)}\,, \qquad
  \mathcal{F}(\lambda)\equiv mq\frac{\int_{-\infty}^{\infty}[w(y)]^{m-1}
(\mathcal{L}_0-\lambda)^{-1}[w(y)]^p \, dy}{\int_{-\infty}^{\infty}[w(y)]^m\,dy}\,.
\end{equation}
Since, it is readily shown that there are no unstable eigenvalues of
the NLEP (\ref{eq:NLEP_general_scalar}) for eigenfunctions for which
$\int_{-\infty}^{\infty} w^{m-1}\phi \, dy=0$, the roots of
$\mathcal{A}(\lambda)=0$ will provide all the unstable eigenvalues of
the NLEP (\ref{eq:NLEP_general_scalar}).

For general Gierer-Meinhardt exponents, the spectral theory of the
operator $\mathcal{L}_0$ leads to some detailed properties of the term
$\mathcal{F}(\lambda)$ for various exponent sets
(cf.~\cite{ward_n_2003}).  In addition, to make further progress on
the root-finding problem \eqref{eq:NLEP_algebraic}, we need some
explicit results for the multiplier $\chi(\lambda)$.

For special sets of Gierer-Meinhardt exponents, known as the
``explicitly solvable cases'' (cf.~\cite{nec}), the term
$\mathcal{F}(\lambda)$ can be evaluated explicitly. We focus
specifically on one such set $(p,q,m,0) = (3,1,3,0)$ for which 
the key identity $\mathcal{L}_0 w ^2 = 3 w^2$ holds, where
$w=\sqrt{2}\,\mbox{sech}{y}$ from (\ref{eq:w}). Thus, after
integrating by parts we obtain
$$
\int_{-\infty}^{\infty} w^2 (\mathcal{L}_0 - \lambda)^{-1}w^3 dy =
\frac{\int_{-\infty}^{\infty}(\mathcal{L}_0 - \lambda)w^2
  (\mathcal{L}_0 - \lambda)^{-1}w^3 dy}{3-\lambda} =
\frac{\int_{-\infty}^{\infty}w^2 (\mathcal{L}_0 -
  \lambda)(\mathcal{L}_0 - \lambda)^{-1}w^3 dy}{3-\lambda} =
\frac{\int_{-\infty}^{\infty}w^5 dy}{3-\lambda}.
$$
By making use of the identities
$$
\int_{-\infty}^{\infty}w^5 dy = \dfrac{3\pi}{\sqrt{2}},\qquad
\int_{-\infty}^\infty w^3 dy = \sqrt{2}\pi,
$$
we obtain that $\mathcal{F}(\lambda)={9/\left[2(3-\lambda)\right]}$,
so that the root-finding problem (\ref{eq:NLEP_algebraic})
reduces to determining $\lambda$ such that
\begin{equation}\label{eq:g_explicitly_solvable}
  \mathcal{A}(\lambda) \equiv \frac{1}{\chi(\lambda)} -
  \frac{9/2}{3-\lambda} = 0\,.
\end{equation}

In addition to the explicitly solvable case $(p,q,m,s)=(3,1,3,0)$, the
root-finding problem \eqref{eq:NLEP_algebraic} simplifies
considerably for a general Gierer-Meinhardt exponent set, when we
focus on determining parameter thresholds for zero-eigenvalue
crossings (corresponding to asynchronous instabilities). Since
$\mathcal{L}_0 w = w^{\prime\prime} - w + p w^p = (p-1)w^p$, it follows that
$\mathcal{L}_0^{-1} w^p = \tfrac{1}{p-1} w$, from which we calculate
$$
  \mathcal{F}(0) =
  mq\frac{\int_{-\infty}^{\infty}w^{m-1}\mathcal{L}_0^{-1}w^{p}\,dy}
  {\int_{-\infty}^{\infty}  w^m \, dy} = \frac{mq}{p-1}\,.
$$
Therefore, a zero-eigenvalue crossing for a general Gierer-Meinhardt
exponent set occurs when
\begin{equation}\label{eq:a_zero_eigenvalue}
	\mathcal{A}(0) = \frac{1}{\chi(0)} - \frac{mq}{p-1} = 0\,.
\end{equation}

\setcounter{equation}{0}
\setcounter{section}{2}
\section{Symmetric $N$-Spike Patterns: Equilibria and Stability}\label{sec:symm}

\paragraph{}
For the remainder of this paper we will focus exclusively on
\textit{symmetric} $N$-spike steady-states that are characterized by
equidistant (in arc-length) spikes of equal heights. Due to the
bulk-membrane coupling it is unclear whether such symmetric patterns
will exist for a general domain. Indeed it may be that a spike pattern
with spikes of equal heights may require the equidistant requirement
to be dropped. These more general considerations can perhaps be better
approached by requiring that a certain Green's matrix admit the
eigenvector $\mathbf{e} = (1,...,1)^T$.

Avoiding these additional complications, we focus instead on two
distinct cases for which symmetric spike patterns, as we have defined
them, can be constructed. The first case is the disk of
  radius $R$, denoted by $\Omega = B_R(0)$, and the second case
  corresponds to the \textit{well-mixed} limit for which
  $D_b\rightarrow\infty$ in an arbitrary bounded domain. In both
cases the Green's function is invariant under translations, satisfying
$$
G_{\partial\Omega}(\sigma+\vartheta\mod L,\zeta+\vartheta\mod L) =
G_{\partial\Omega}(\sigma,\zeta)\,,\qquad \forall\,\, \sigma\,,
\zeta\in[0,L)\,,\qquad \vartheta\in\mathbb{R}\,.
$$
By using this key property in
\eqref{eq:equi_alg_system_consistency}, we calculate the common
spike height as
\begin{equation}
  v_{ej} = v_{e0} = \biggl[\omega_m\sum_{k=0}^{N-1}G_{\partial\Omega}
  \biggl(\frac{k L}{N},0\biggr) \biggr]^{\frac{1}{1-\gamma m + s}}\,.
\end{equation}
With a common spike height, the balance equations
\eqref{eq:equi_alg_system_balance} then reduce to
\begin{equation}
  \bigl[ \partial_{\sigma} G_{\partial\Omega}(0^+,0) + \partial_\sigma
  G_{\partial\Omega}(0^-,0)\bigr] + 2 \sum_{k=1}^{N-1} \partial_\sigma
  G_{\partial\Omega}\biggl(\frac{kL}{N},0\biggr) = 0\,,
\end{equation}
which can be verified either explicitly or by using the symmetry of the
Green's function.

For a symmetric $N$-spike steady-state the NLEP
\eqref{eq:NLEP_general_scalar} can be simplified significantly. First the
matrix $\mathcal{E}$, defined in (\ref{eq:mate}), simplifies to
\begin{equation*}
\mathcal{E} = \hat{v}_{e0}^{\gamma m - 1} \mathcal{G}_{\partial\Omega}^{\lambda}\,.
\end{equation*}
Therefore, from (\ref{lin:mate_eig}) it follows that
$\chi(\lambda) = \hat{v}_{e0}^{\gamma m - 1}\mu(\lambda)$, where
$\mu(\lambda)$ is an eigenvalue of the Green's matrix
$\mathcal{G}_{\partial\Omega}^\lambda$ defined in
(\ref{lin:mat_def}). Furthermore, by using the bi-translation
invariance and symmetry of $G_{\partial\Omega}^\lambda$, we can define
\begin{equation}
  H_{|j-i|}^\lambda \equiv G_{\partial\Omega}^\lambda(|\sigma_i-\sigma_j|,0) =
  G_{\partial\Omega}^\lambda(|i-j|L/N,0)\,,
\end{equation}
which allows us to write the Green's matrix as
\begin{equation*}
  \mathcal{G}_{\partial\Omega}^{\lambda} =
  \begin{pmatrix}
    H_0^\lambda & H_1^\lambda & H_2^\lambda & \cdots &  H_{N-1}^\lambda \\
    H_{N-1}^\lambda & H_0^\lambda & H_1^\lambda & \cdots & H_{N-2}^\lambda \\
    \vdots & \vdots & \vdots & \ddots & \vdots \\
    H_1^\lambda & H_2^\lambda & H_3^\lambda & \cdots & H_0^\lambda
 \end{pmatrix}\,,
\end{equation*}
which we recognize as a \textit{circulant matrix}. As a result,
  the matrix spectrum of $\mathcal{G}_{\partial\Omega}^{\lambda}$ is readily
  available as
\begin{equation}\label{eq:greens_eigenpair}
  \mu_k(\lambda) = \sum_{j=0}^{N-1} H_{j}^\lambda e^{i\frac{2\pi j k}{N}}\,,
  \qquad \pmb{c}_k(\lambda) = \biggl( 1 , \cdots, e^{i\frac{2\pi(N-1)k}{N}}
  \biggr)^T \,, \qquad  k=0,\ldots,N-1\,.
\end{equation}

For each value of $k=0,\ldots,N-1$ we obtain a corresponding
NLEP problem from (\ref{eq:NLEP_general_scalar}). Since
$\pmb{c}_0 = (1,\ldots,1)^T$ we can interpret this ``mode'' as a
\textit{synchronous} perturbation. In contrast, the values
$k=1,\ldots,N-1$ for $N\geq 2$ correspond to \textit{asynchronous}
perturbations, since the corresponding eigenvectors
$\pmb{c}_k(\lambda)$ are all orthogonal to $(1,\ldots,1)^T$. Any
unstable asynchronous ``mode'' of this type is referred to as a
\textit{competition} instability, in the sense that the linear
stability theory predicts that the heights of individual spikes may
grow or decay, but that the overall sum of all the spike heights
remains fixed.  For each value of $k$, the NLEP
\eqref{eq:NLEP_general_scalar} becomes
\begin{equation}\label{eq:NLEP_symmetric}
  \mathcal{L}_0 \phi - mq\chi_k(\lambda)w^p\dfrac{\int_{-\infty}^{\infty}
    [w(y)]^{m-1}\phi(y)\,dy}{\int_{-\infty}^{\infty}[w(y)]^m\,dy} =
  \lambda\phi\,,\qquad \mbox{where} \quad \chi_k(\lambda) \equiv
  \frac{\mu_k(\lambda)}{\sum_{j=1}^{N-1}G_{\partial\Omega}(jL/N,0)} =
  \frac{\mu_k(\lambda)}{\mu_0(0)}\,.
\end{equation}
Further analysis requires details of the Green's function
$G_{\partial\Omega}^\lambda$, which are available in our two special
cases.

\subsection{NLEP Multipliers for the Well-Mixed Limit}
\paragraph{}
In the well-mixed limit, $D_b\to \infty$, the membrane Green's
function, satisfying (\ref{eq:g_lam_memb}), is given by (see
(\ref{eq:SL_G_Membrane_expansion}) of Appendix
\ref{app:Greens_Functions})
\begin{equation}\label{eq:SL_G_Membrane_Asy_0}
  G_{\partial\Omega}^\lambda(\sigma,\zeta) = \Gamma^\lambda(|\sigma-\zeta|) +
  \frac{K^2}{\mu_{s\lambda}^2 A} \frac{1}{\mu_{s\lambda}^2(\mu_{b\lambda}^2+\beta)-
    K\beta} = \Gamma^\lambda(|\sigma-\zeta|) +
  \frac{\gamma_\lambda}{\mu_{s\lambda}^2}\,,\quad \gamma_\lambda \equiv
  \frac{K^2/A}{\mu_{s\lambda}^2(\mu_{b\lambda}^2+\beta)-K\beta}\,,
\end{equation}
where $\beta \equiv {KL/A}$. Here $\Gamma^\lambda$ is the periodic
Green's function for the uncoupled ($K=0$) problem, which is given
explicitly by (\ref{eq:Uncoupled_G_Membrane}) of Appendix
\ref{app:Greens_Functions} as
\begin{equation*}
  \Gamma^\lambda(x) = \frac{1}{2\sqrt{D_v}\mu_{s\lambda}}
  \coth\biggl(\frac{\mu_{s\lambda} L}{2\sqrt{D_v}}\biggr)
  \cosh\biggl(\frac{\mu_{s\lambda}}{\sqrt{D_v}}|x|\biggr) -
  \frac{1}{2\sqrt{D_v}\mu_{s\lambda}}\sinh\biggl(\frac{\mu_{s\lambda}}{\sqrt{D_v}}
  |x|\biggr)\,.
\end{equation*}
After some algebra we use (\ref{eq:greens_eigenpair}) to
calculate the eigenvalues $\mu_k(\lambda)$ of the Green's matrix as
\begin{equation*}
  \mu_k(\lambda) = \sum_{j=0}^{N-1} \Gamma^\lambda(jL/n) e^{i\frac{2\pi j k}{N}} +
  \delta_{k0}\frac{N\gamma_\lambda}{\mu_{s\lambda}^2} = \frac{1}{2\sqrt{D_v}
    \mu_{s\lambda}}\frac{\cosh\bigl(\frac{\mu_{s\lambda} L}{2N\sqrt{D_v}}\bigr)
    \sinh\bigl(\frac{\mu_{s\lambda}L}{2N\sqrt{D_v}}\bigr)}
  {\sinh\bigl(\frac{\mu_{s\lambda}L}{2N\sqrt{D_v}} + \frac{i\pi k}{N}\bigr)
    \sinh\bigl(\frac{\mu_{s\lambda}L}{2N\sqrt{D_v}} - \frac{i\pi k}{N}\bigr)} +
  \delta_{k0}\frac{N\gamma_\lambda}{\mu_{s\lambda}^2} \,,
\end{equation*}
where $\delta_{k0}$ is the Kronecker symbol. In this way,
we obtain from (\ref{eq:NLEP_symmetric}) that the NLEP multipliers are
given by
\begin{equation}\label{eq:SL_symmetric_chi_k}
  \chi_0(\lambda) = \frac{\frac{1}{2\sqrt{D_v}\mu_{s\lambda}}
    \coth\biggl(\frac{\mu_{s\lambda}L}{2N\sqrt{D_v}}\biggr) +
    \frac{N\gamma_\lambda}{\mu_{s\lambda}^2}}{\frac{1}{2\sqrt{D_v}\mu_{s0}}
    \coth\biggl(\frac{\mu_{s0}L}{2N\sqrt{D_v}}\biggr) +
    \frac{N\gamma_0}{\mu_{s0}^2}}\,,\qquad \chi_k(\lambda) =
  \frac{\frac{1}{2\sqrt{D_v}\mu_{s\lambda}}\frac{\cosh\bigl(\frac{\mu_{s\lambda}L}
      {2N\sqrt{D_v}}\bigr)\sinh\bigl(\frac{\mu_{s\lambda}L}{2N\sqrt{D_v}}\bigr)}
    {\sinh\bigl(\frac{\mu_{s\lambda}L}{2N\sqrt{D_v}} + \frac{i\pi k}{N}\bigr)
      \sinh\bigl(\frac{\mu_{s\lambda}L}{2N\sqrt{D_v}} - \frac{i\pi k}{N}\bigr)}}
  {\frac{1}{2\sqrt{D_v}\mu_{s0}}\coth\biggl(\frac{\mu_{s0}L}{2N\sqrt{D_v}}\biggr)
    + \frac{N\gamma_0}{\mu_{s0}^2}}\,,
\end{equation}
for $k=1,...,N-1$. We observe from the $\chi_0(\lambda)$
term in (\ref{eq:SL_symmetric_chi_k}), that any synchronous
instability will depend on the membrane diffusivity $D_v$ only in the
form $N^2 D_v$. This shows that a synchronous instability parameter
threshold will be fully determined by the one-spike case upon rescaling by
$1/N^2$. We remark here that the numerator for $\chi_k(\lambda)$ can be simplified by using the identity $\sinh(z+ia)\sinh(z-ia) =\tfrac{1}{2}[\cosh(2z) - \cos(2a)]$ so that $\chi_k(\lambda)$ is real valued whenever $\text{Im}\lambda =0$.

\subsection{NLEP Multipliers for the Disk}
\paragraph{}
In the disk we can calculate the membrane Green's function as a
Fourier series (see (\ref{eq:Disk_G_Membrane_Series}) of Appendix
\ref{app:Greens_Functions})
\begin{equation}
  G_{\partial\Omega}^\lambda(\sigma,\zeta) = \frac{1}{2\pi R}
  \sum_{n=-\infty}^{\infty} g_n^\lambda e^{i\tfrac{n}{R}(\sigma-\zeta)}\,,
\end{equation}
where $g_n^\lambda$ is given explicitly by
\begin{equation}\label{eq:gnlambda}
  g_n^\lambda = \frac{1}{D_v\bigl(\frac{n}{R}\bigr)^2 +\mu_{s\lambda}^2 -
    K^2 a_n^\lambda}\,,\quad a_n^\lambda = \frac{1}{D_b P_n^{\prime}(R) + K}\,,
  \quad P_n(r) \equiv
  \frac{I_{|n|}(\omega_{b\lambda}r)}{I_{|n|}(\omega_{b\lambda} R)}\,, \quad
  \omega_{b\lambda} \equiv \frac{\mu_{b\lambda}}{\sqrt{D_b}} \,.
\end{equation}
Here $I_n(z)$ is the $n^\text{th}$ modified Bessel function of the
first kind. From \eqref{eq:greens_eigenpair} the eigenvalues
of the Green's matrix become
\begin{equation*}
  \mu_k(\lambda) = \frac{1}{2\pi R}\sum_{n=1}^{\infty}
  g_n^\lambda\sum_{j=0}^{N-1}e^{i\frac{2\pi(k+n)j}{N}} \,.
\end{equation*}
By using the identities
\begin{equation*}
  \sum_{j=0}^{N-1}e^{i\frac{2\pi (k+n) j}{N}} = \begin{cases} N & n\in N\mathbb{Z}-k\,,
    \\ 0 & \text{otherwise}\end{cases}\,,\qquad\text{and}\qquad
  g_{-n}^\lambda = g_{n}^\lambda\,,
\end{equation*}
the eigenvalues are given explicitly by
\begin{equation*}
  \mu_k(\lambda) = \frac{N}{2\pi R}g_k^\lambda + \frac{N}{2\pi R}
  \sum_{n=1}^{\infty}\bigl( g_{nN +k}^\lambda + g_{nN - k}^\lambda \bigr)\,.
\end{equation*}
Therefore, since $\chi_k(\lambda)={\mu_k(\lambda)/\mu_0(0)}$,
the NLEP multipliers are given by
\begin{equation}\label{eq:NLEP_multiplier_disk}
  \chi_k(\lambda) = \dfrac{g_k^\lambda + \sum_{n=1}^{\infty}
    \bigl( g_{nN +k}^\lambda + g_{nN - k}^\lambda \bigr)}
  {g_0^0+2 \sum_{n=1}^{\infty} g_{nN}^0}\,,\qquad k=0,\ldots,N-1 \,.
\end{equation}
 
\subsection{Synchronous Instabilities}
\paragraph{}

From \eqref{eq:a_zero_eigenvalue}, and the special form of $\chi_k(\lambda)$
given in \eqref{eq:NLEP_symmetric}, we deduce that
$$
	\mathcal{A}_0(0) = 1 - \frac{mq}{p-1} < 0\,,
$$
where the strict inequality follows from the the usual
  assumption (\ref{eq:gm_expon}) on the Gierer-Meinhardt exponents. As
  a result, synchronous instabilities do not occur through a
  zero-eigenvalue crossing, and can only arise through a Hopf
  bifurcation. To examine whether such a Hopf bifurcation for the
  synchronous mode can occur, we now seek purely imaginary zeros of
  $\mathcal{A}_0(\lambda)$. Classically, in the uncoupled
  case $K=0$, such a threshold occurs along a Hopf bifurcation curve
  $D_v = D_v^{\star}(\tau_s)$ (cf.~\cite{ward_n_2003}). We have an oscillatory
  instability if $\tau_s$ is sufficiently large, and no such instability
  when $\tau_s$ is small (cf.~\cite{ward_n_2003}, \cite{ward_2003}). Bulk-membrane coupling introduces two
additional parameters, $\tau_b$ and $K$, in addition to the
quantities $L$ and $A$ for the well-mixed case, or $R$ and $D_b$ for
  the case of the disk. Thus, it is no longer clear how the existence
of a synchronous instability threshold $D_v = D_v^\star(\tau_s)$ will be modified by the additional parameters. Indeed, the analysis below reveals a
variety of new phenomenon such as the existence of synchronous
instabilities for $\tau_s=0$ and islands of stability for large values
of $\tau_s$. These are two behaviors that do not occur for the
classical uncoupled case $K=0$.

We begin by addressing the question of the existence of
synchronous instability thresholds. The key assumption (supported
below by numerical simulations) underlying this analysis is that
synchronous instabilities persist as either the bulk and/or membrane
diffusivities increase. While this assumption is heuristically
reasonable (large diffusivities make it easier for neighbouring spikes
to communicate) an open problem is to demonstrate it
analytically. With this assumption it suffices to seek parameter
values of $\tau_b$, $\tau_s$, and $K$ for which no Hopf bifurcations
exist when $D_v\rightarrow\infty$ in the well-mixed limit
$D_b\to\infty$.

As a first step, we remark that in \cite{ward_n_2003} it was shown that $\text{Re}\mathcal{F}(i\lambda_I)$ is monotone decreasing when $\lambda_I > 0$ for special choices of the Gierer-Meinhardt exponents (see also \cite{ward_2003}). The monotonicity of this function for general Gierer-Meinhardt exponents is supported by numerical calculations. Thus we expect that $\text{Re}\mathcal{F}(i\lambda_I)$ decreases monotonically from $\text{Re}\mathcal{F}(0) = \tfrac{mq}{p-1}>1$ as $\lambda_I>0$ increases. Furthermore, numerical evidence suggests that $\text{Re}\mathcal{C}_0(i\lambda_I)$ is monotone increasing in
$\lambda_I$. Since $\mathcal{C}_0(0)=1$ there must exist a unique root
$\lambda_I=\lambda_I^{\star}> 0$ to
  $\text{Re}\mathcal{A}_0(i\lambda_I)=0$ bounded above by
$\lambda_I^F$, the unique solution to
$\text{Re}\mathcal{F}(i\lambda_I^F)=1$, which depends solely on the
exponents $(p,q,m,0)$. Therefore in the limit $D_v\rightarrow\infty$
the well-mixed NLEP multiplier, as given in
  (\ref{eq:SL_symmetric_chi_k}), becomes
$$
\chi_0(\lambda)\sim \frac{\mu_{s0}^2(\mu_{b0}^2+\beta) -
  K\beta}{\mu_{s\lambda}^2(\mu_{b\lambda}^2+\beta)-K\beta} \left(
  \frac{\mu_{b\lambda}^2 +\beta}{\mu_{b0}^2+\beta}\right)\,.
$$
Seeking a purely imaginary root of $\mathcal{A}_0(\lambda)=0$ we focus
first on the real part. We calculate
$$
\text{Re}\mathcal{A}_0(i\lambda_I)=
\frac{1+\beta}{1+\beta+K}\biggl(1+K-\frac{K\beta}{1+\beta}\frac{1}{1 +
  \bigl(\tfrac{\tau_b\lambda_I}{1+\beta}\bigr)^2}\bigg) -
\text{Re}\mathcal{F}(i\lambda_I)\,,
$$ 
and note that the root
$\lambda_I=\lambda_I^{\star}(\tau_b,K)$ to
$\text{Re}\mathcal{A}_0(i\lambda_I)=0$ is independent of
$\tau_s$. Next, for the imaginary part we calculate
$$
\text{Im}\mathcal{A}_0(i\lambda_I^{\star}) =
\frac{1+\beta}{1+\beta+K}\biggl(\tau_s +
\frac{K\beta}{1+\beta}\frac{\tfrac{\tau_b}{1+\beta}}
{1+\bigl(\tfrac{\tau_b\lambda_I^{\star}}{1+\beta}\bigr)^2}
\biggr) \lambda_I^{\star} - \text{Im}\mathcal{F}(i\lambda_I^{\star})\,.
$$
Fortunately, at each fixed value of $\tau_s$ the threshold
$K = K(\tau_b)$ can be calculated as the $\tau_s$-level-set of a
function depending only on $K$ and $\tau_b$. Indeed the condition
$\text{Im}\mathcal{A}_0(i\lambda_I^{\star}) = 0$ can equivalently be written
as
\begin{equation}\label{ss:im}
\text{Im}\mathcal{A}_0(i\lambda_I^{\star}) =\frac{1+\beta}{1+\beta+K}\biggl
(\tau_s - \mathcal{M}(\tau_b,K)\biggr)\lambda_I^{\star} = 0\,,
\end{equation}
where we have defined
\begin{equation}\label{eq:synchronous_threshold_general}
  \mathcal{M}(\tau_b,K) \equiv \left(\frac{1+\beta+K}{1+\beta}\right)
  \frac{\text{Im}
    \mathcal{F}(i\lambda_I^{\star})}{\lambda_I^{\star}} -
  \frac{K\beta}{1+\beta}\left(
    \frac{\tfrac{\tau_b}{1+\beta}}
    {1+\bigl(\tfrac{\tau_b\lambda_I^{\star}}{1+\beta}\bigr)^2}\right) \,.
\end{equation}

\begin{figure}[t]
	\centering
	\begin{tabular}{cc}
		\includegraphics[scale=0.65]{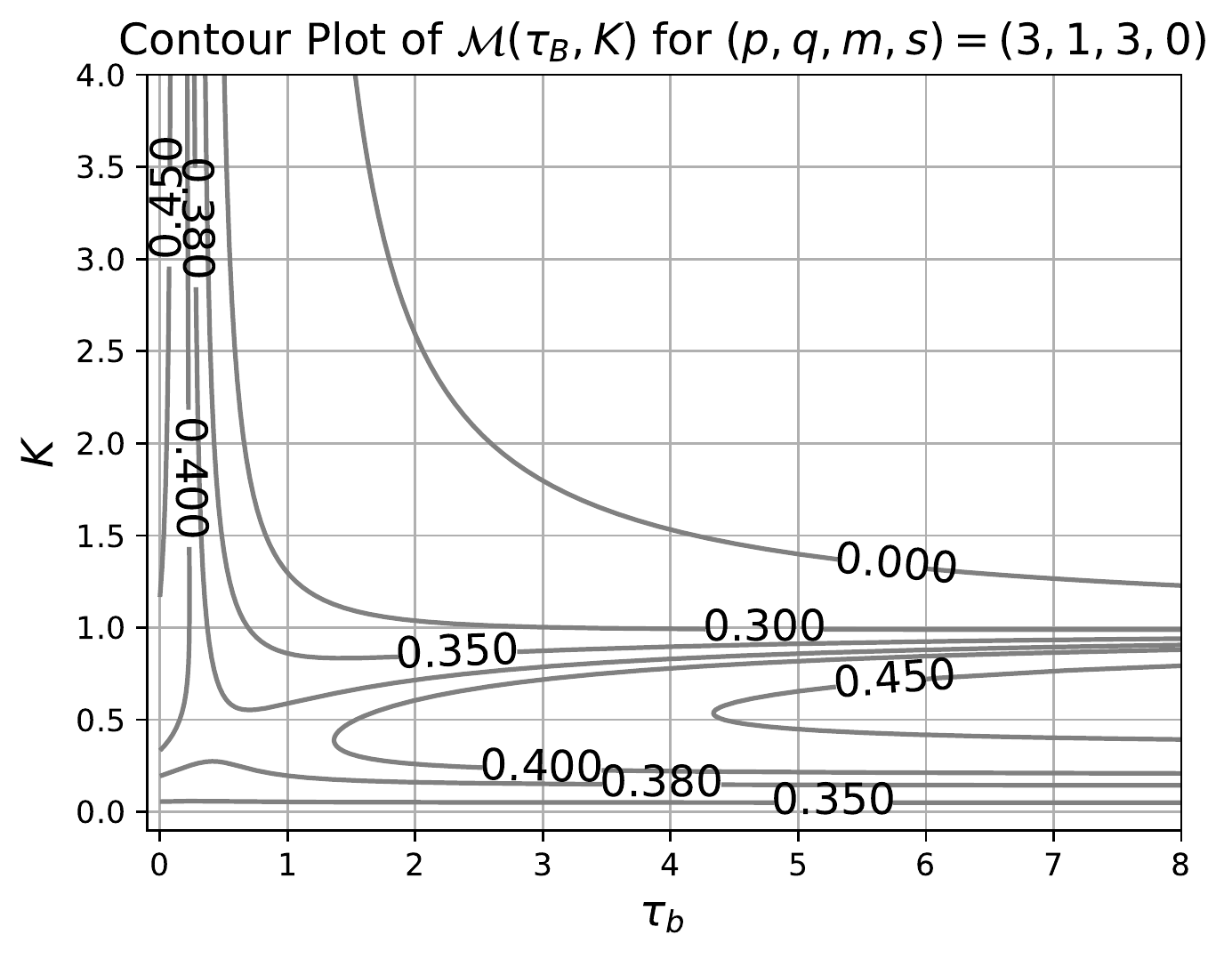} & \includegraphics[scale=0.65]{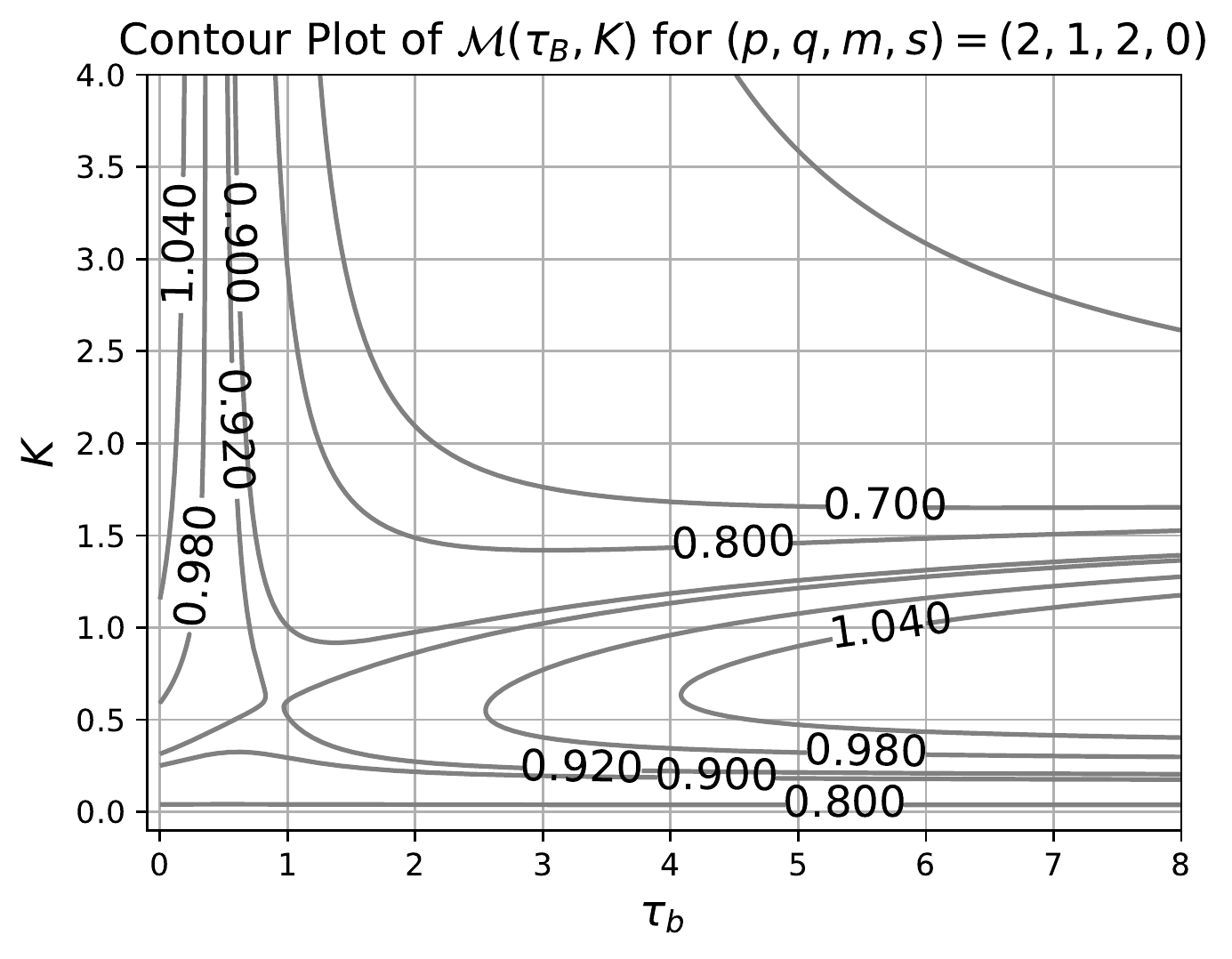}
	\end{tabular}
	\caption{Level sets of $\mathcal{M}(\tau_b,K)$ for
          Gierer-Meinhardt exponents $(p,q,m,s)=(3,1,3,0)$ (left) and
          $(p,q,m,s) = (2,1,2,0)$ (right). In both cases the level set
          value corresponds to a value of $\tau_s = M(\tau_b,K)$. Note
          also the contours tending to a vertical asymptote, and the
          emergence of a horizontal asymptote as $\tau_s$ exceeds some
          threshold. Geometric parameters are $L=2\pi$ and
          $A=\pi$.}\label{fig:M_contours}
\end{figure}

In the $(p,q,m,s)=(3,1,3,0)$ explicitly solvable case we find that
$\text{Im}\mathcal{F}(i\lambda_I^{\star}) =
\tfrac{1}{3}\lambda_I^{\star}\text{Re}\mathcal{F}(i\lambda_I^{\star})$, so that by solving Re$\mathcal{A}_0(i\lambda_I)=0$ for Re$\mathcal{F}(i\lambda_I^\star)$, (\ref{eq:synchronous_threshold_general}) becomes
$$
\mathcal{M}(\tau_b,K) = \frac{1+K}{3} -
\frac{K\beta}{1+\beta}\left( \frac{\tfrac{\tau_b}{1	+\beta}+\tfrac{1}{3}}
{1+\bigl(\tfrac{\tau_b\lambda_I^{\star}}{1+\beta}\bigr)^2}\right) \,.
$$
By substituting this expression into (\ref{ss:im}), we deduce
the existence of two distinct threshold branches obtained by considering
the limits $K\gg 1$ and $K\ll 1$. In this way, we derive
\begin{align*}
  & \tau_s - \mathcal{M}(\tau_b,K) \sim \tau_s - \frac{1}{3} +
    \frac{1}{\beta_0}\biggl(\tau_b - \frac{1}{3}\biggr) +
    {\mathcal O}(K^{-1}) \qquad \mbox{for} \quad K\gg 1\,,\\
  & \tau_s - \mathcal{M}(\tau_b,K) \sim \tau_s - \frac{1}{3} - \frac{1}{3}K +
    {\mathcal O}(K^2) \qquad \mbox{for} \qquad K\ll 1\,,
\end{align*}
where $\beta_0\equiv L/A$. Notice that in ordering both of these
asymptotic expansions we have used that
$0<\lambda_I^{\star}\leq\lambda_I^F$, where the upper bound is independent
of $K$. In the $K\gg 1$ regime we deduce that if
$\tau_b = \tfrac{1}{3} - \beta_0\bigl(\tau_s - \tfrac{1}{3}\bigr)$,
then $\text{Im}\mathcal{A}_0(i\lambda_I^{\star})=0$ forces
$K\rightarrow\infty$, implying the existence of a threshold branch
emerging from $K=\infty$ at these parameter values. Furthermore, since
$\tau_b$ approaches $0$ when $\tau_s$ tends to
$\tfrac{1}{3}\bigl(\tfrac{1}{\beta_0}+1\bigr)$, we deduce
that this branch will disappear for sufficiently large
values of $\tau_s$. In addition, in the $K\ll 1$ regime we
find that a new branch given by $K\approx 3\tau_s - 1$ emerges when
$\tau_s>\tfrac{1}{3}$. The left panel of Figure
\ref{fig:M_contours} shows the numerically-computed contours of
$\mathcal{M}(\tau_b,K)$ for the explicitly solvable case
$(p,q,m,s)=(3,1,3,0)$. The right panel of Figure
\ref{fig:M_contours} shows a qualitatively similar behavior that
occurs for the prototypical Gierer-Meinhardt parameter set
$(p,q,m,s)=(2,1,2,0)$.

The preceding analysis does not directly predict in which regions
synchronous instabilities exist, as it only provides the
boundaries of these regions. We now outline a winding-number
argument, related to that used in \cite{ward_2003}, that provides a
hybrid analytical-numerical algorithm for calculating the
synchronous instability threshold
$D_v = D_v^{\star}(K,\tau_b,\tau_s)$. Furthermore, as we
show below, this algorithm indicates that synchronous instabilities
exist whenever $M(\tau_b,K) < \tau_s$.

\begin{figure}[t]
	\centering
	\setlength\tabcolsep{0.25pt}
	\begin{tabular}{cccc}
		\includegraphics[scale=0.65]{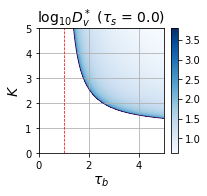} &  \includegraphics[scale=0.65]{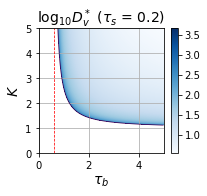} &  \includegraphics[scale=0.65]{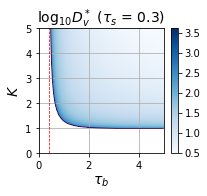} & \includegraphics[scale=0.65]{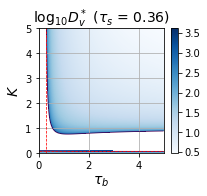} \\
		\includegraphics[scale=0.65]{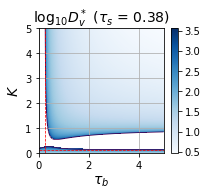} & \includegraphics[scale=0.65]{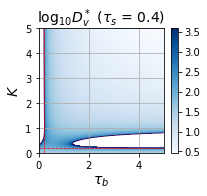} &  \includegraphics[scale=0.65]{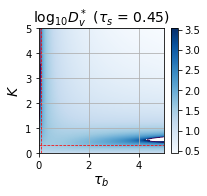} & \includegraphics[scale=0.65]{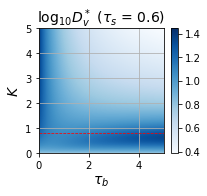} \\
	\end{tabular}
	\caption{Colormap of the synchronous instability threshold
	  $D_v^{\star}$ in the $K$ versus $\tau_b$ parameter
	    plane for the well-mixed explicitly solvable case for
	  various values of $\tau_s$ with $L=2\pi$ and $A = \pi$. The
	  dashed vertical lines indicate the asymptotic predictions
	  for the large $K$ threshold branch, while the dashed
	  horizontal lines indicate the asymptotic predictions for the
	  small $K$ threshold branch. The unshaded regions correspond
	  to those parameter values for which synchronous
	  instabilities are absent.}\label{fig:M_with_Dv_colormap}
\end{figure}

Synchronous instabilities are identified with the zeros to
\eqref{eq:NLEP_algebraic} having a positive real-part when
$\chi(\lambda)$ in \eqref{eq:NLEP_algebraic} is replaced by
$\chi_0(\lambda)$. By using a winding number argument, the search for
such zeros can be reduced from one over the entire right-half plane
$\mbox{Re}(\lambda)>0$ to one along only the positive imaginary
axis. Indeed, if we consider a counterclockwise contour composed of a
segment of the imaginary axis,
$-\rho\leq \text{Im} \lambda \leq \rho$, together with the semi-circle
defined by $|\lambda|=\rho$ and $-\pi/2<\text{arg}\lambda<\pi/2$, then
in the limit $\rho\rightarrow\infty$ the change in argument is
\begin{equation}
\Delta \text{arg}\, \mathcal{A}_0(\lambda) = 2\pi (Z - 1),
\end{equation}
where $Z$ is the number of zeros of $\mathcal{A}_0$ with positive
real-part. Here we have used that $\chi_0(\lambda)\neq 0$
when $\text{Re}(\lambda)\geq 0$, while $\mathcal{F}(\lambda)$ has
exactly one simple (and real) pole in the right-half plane
corresponding to the only positive eigenvalue of the
self-adjoint local operator $\mathcal{L}_0$
(cf.~\cite{wei_1999}). We immediately note that
$\mathcal{F}(\lambda) = {\mathcal O}(\lambda^{-1})$ for
$|\lambda|\gg 1$, $|\arg\lambda|<\pi/2$, whereas
$$
\mathcal{C}_0(\lambda)\sim 2\mu_0(0)\sqrt{\tau_s
  D_v}\lambda^{1/2}\,,\qquad \mathcal{C}_0(\lambda) \sim
\mu_0(0)\frac{N\sqrt{D_v \tau_s}}{\pi R} \lambda^{1/2}\,, \qquad
\mbox{for} \quad |\lambda|\gg 1\,, \,\,\, |\arg\lambda|<\pi/2\,,
$$
for the well-mixed limit and the disk cases,
respectively. Therefore, in both cases we have
$\mathcal{A}_0(\lambda)\sim {\mathcal O}(\lambda^{1/2})$ for
$|\lambda|\gg 1$ with $|\arg\lambda|<\pi/2$, so that the change in
argument over the large semi-circle is $\pi/2$. Furthermore, since
the parameters in $\mathcal{A}_0(\lambda)$ are real-valued,
the change in argument over the segment of the imaginary axis can be
reduced to that over the positive imaginary axis. In this
  way, we deduce that
\begin{equation}
  Z = \frac{5}{4} + \frac{1}{\pi} \Delta \text{arg}\mathcal{A}_0(i\lambda_I)
  \bigr|_{\lambda_I\in(\infty,0]}.
\end{equation}
We readily evaluate the limiting behaviour
$\lim_{\lambda_I\rightarrow\infty}\text{arg}\,
\mathcal{A}_0(i\lambda_I)=\pi/4$. Moreover since $\chi_0(0)=1$ we
evaluate $\mathcal{A}_0(0) = 1 - \tfrac{mq}{p-1} < 0 $ by
  the assumption (\ref{eq:gm_expon}) on the Gierer-Meinhardt
  exponents. Numerical evidence suggests that
$\text{Re}\mathcal{A}_0(i\lambda_I)$ increases monotonically with
$\lambda_I$ and there should therefore be a unique $\lambda_{I}^{\star}$ for
which $\text{Re}\mathcal{A}_0(i\lambda_I^{\star}) = 0$. We conclude that
there are two positive values for the change in argument, and hence
the number of zeros of ${\mathcal A}_0(\lambda)$ in
$\mbox{Re}(\lambda)>0$ is dictated by the sign of
$\text{Im} \mathcal{A}_0(i\lambda_I^{\star})$ as follows:
\begin{equation}\label{ss:crit}
Z =
2\quad\text{if}\quad\text{Im}\mathcal{A}_0(i\lambda_I^{\star})>0\,,\qquad
\text{or}\qquad
Z= 0\quad\text{if}\quad\text{Im}\mathcal{A}_0(i\lambda_I^{\star}) < 0\,.
\end{equation}
Note in particular that, in view of the expression
(\ref{ss:im}) for $\text{Im}\mathcal{A}_0(i\lambda_I)$, this
criterion implies that synchronous instabilities will exist whenever
$M(\tau_b,K)<\tau_s$ in the previous analysis. Within this region,
the criterion (\ref{ss:crit}) suggests a simple numerical algorithm
for iteratively computing the threshold value of
$D_v=D_v^{\star}(K,\tau_b,\tau_s)$. Specifically, with all parameters fixed,
we first solve $\text{Re}\mathcal{A}_0(i\lambda_I)=0$ for
$\lambda_I^0$. Then, we calculate
$\text{Im} \mathcal{A}_0(i\lambda_I^0)$ and increase (resp. decrease)
$D_v$ if $\text{Im} \mathcal{A}_0(i\lambda_I^0)<0$ (resp.
$\text{Im} \mathcal{A}_0(i\lambda_I^0)>0$ until
$\text{Im} \mathcal{A}_0(i\lambda_I^0)=0$. This procedure is repeated
until $|\mathcal{A}_0(i\lambda_I^0)|$ is sufficiently small.

\begin{figure}[t]
	\centering
	\begin{tabular}{cccc}
      \includegraphics[scale=0.7]{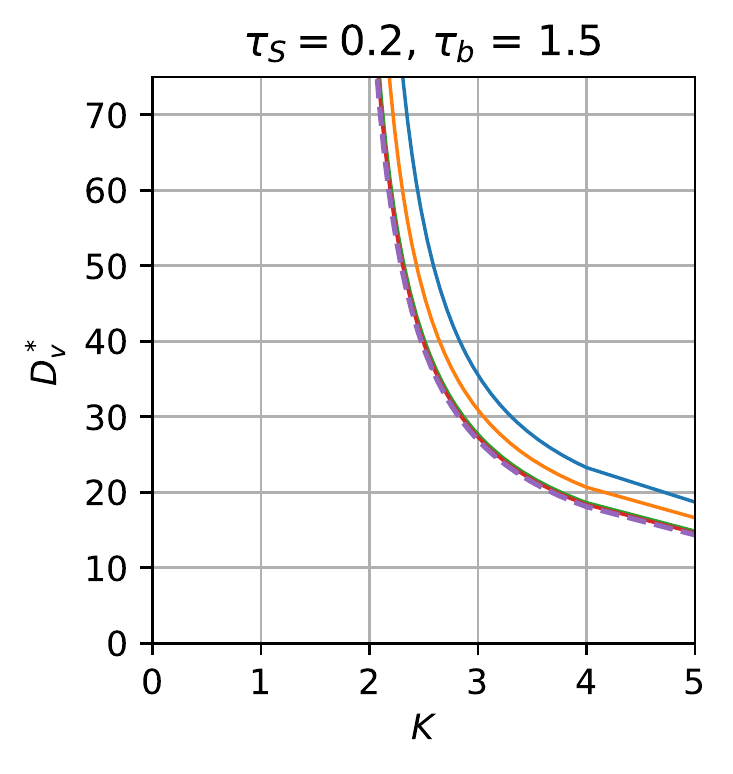} &                                                                        \includegraphics[scale=0.7]{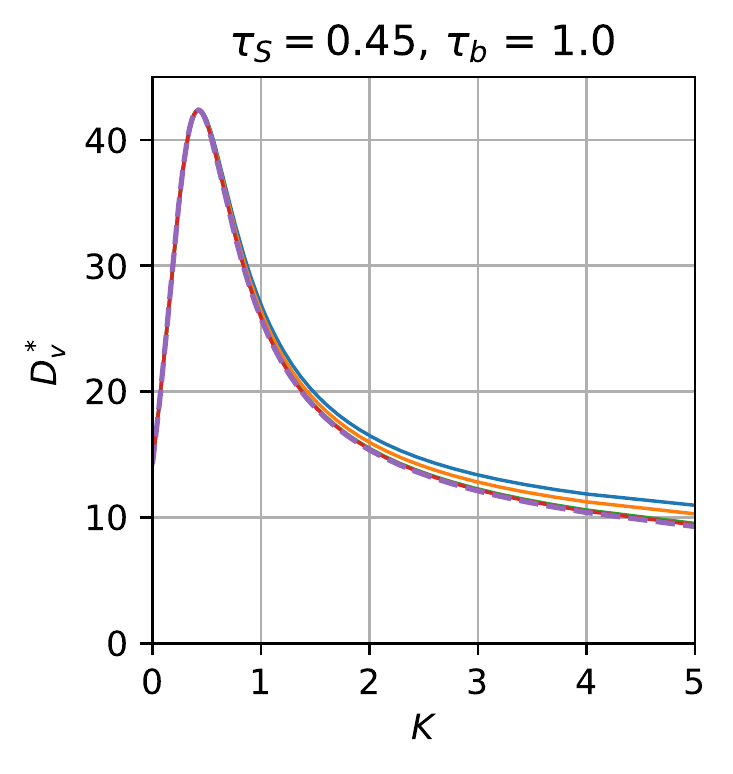} &
      \includegraphics[scale=0.7]{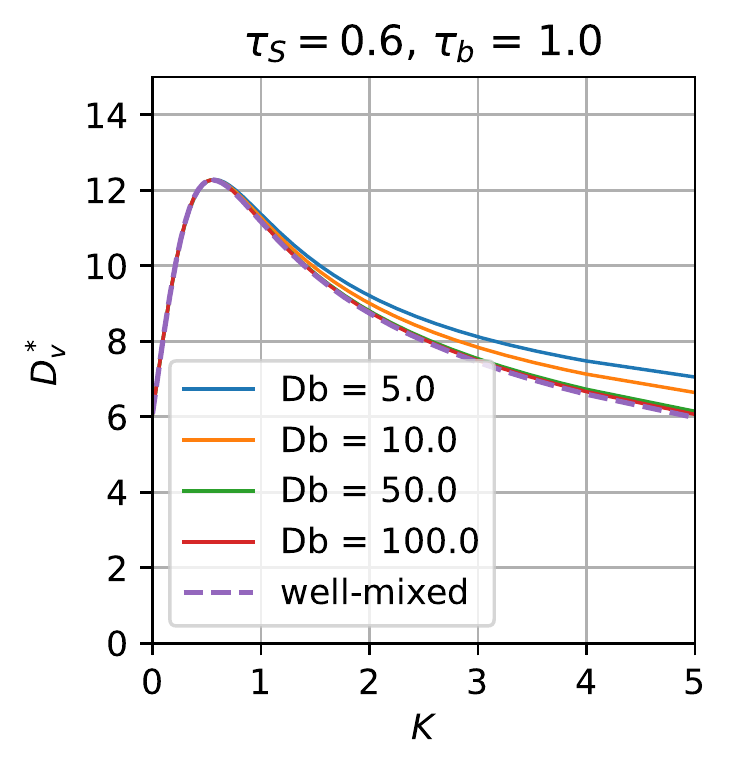}
	\end{tabular}
	\caption{Synchronous instability threshold $D_v^{\star}$
          versus $K$ for three pairs of $(\tau_s,\tau_b)$ for a
          one-spike steady-state ($N=1$) in the unit disk ($R=1$). The
          quality of the well-mixed approximation rapidly improves as
          $D_b$ is increased. The labels for $D_b$ in the
            right panel also apply to the left and middle
            panels.}\label{fig:disk_select_synchronous_3_1_3_0}
\end{figure}

Using the algorithm described above, the results in Figure
\ref{fig:M_with_Dv_colormap} illustrate how the synchronous
instability threshold $D_v^{\star}$ depends on parameters $\tau_s$,
$\tau_b$, and $K$ for the explicitly solvable case in the well-mixed
limit. From these figures we observe that coupling can have
both a stabilizing and a destabilizing effect with respect to
synchronous instabilities. Indeed, on the $K=0$ axis we see, as
expected from the classical theory, that synchronous instabilities
exist beyond some $\tau_s$ value. However, well before this threshold
of $\tau_s$ is even reached it is possible for synchronous
instabilities to exist when both $\tau_b$ and $K$ are sufficiently
large. In contrast, we also see from the panels in
  Fig.~\ref{fig:M_with_Dv_colormap} with $\tau_s=0.36$, $\tau_s=0.38$,
  and $\tau_s=0.4$ that when $\tau_b$ is sufficiently small, there are
  no synchronous instabilities when the coupling $K$ is large enough.
Perhaps the most perplexing feature of this bulk-membrane interaction
is the island of stability that arises around $\tau_s=0.4$ and appears
to persist, propagating to larger values of $\tau_b$ as $\tau_s$
increases (only shown up to $\tau_s=0.6$). Finally in Figure
\ref{fig:disk_select_synchronous_3_1_3_0} we demonstrate how the
synchronous instability threshold behaves for finite
bulk-diffusivity. A key observation from these plots is that that the
instability threshold increases with decreasing value of $D_b$, which
further supports our earlier monotonicity assumption.

\subsection{Asynchronous Instabilities}

\begin{figure}[t]
	\centering
	\includegraphics[scale=0.53]{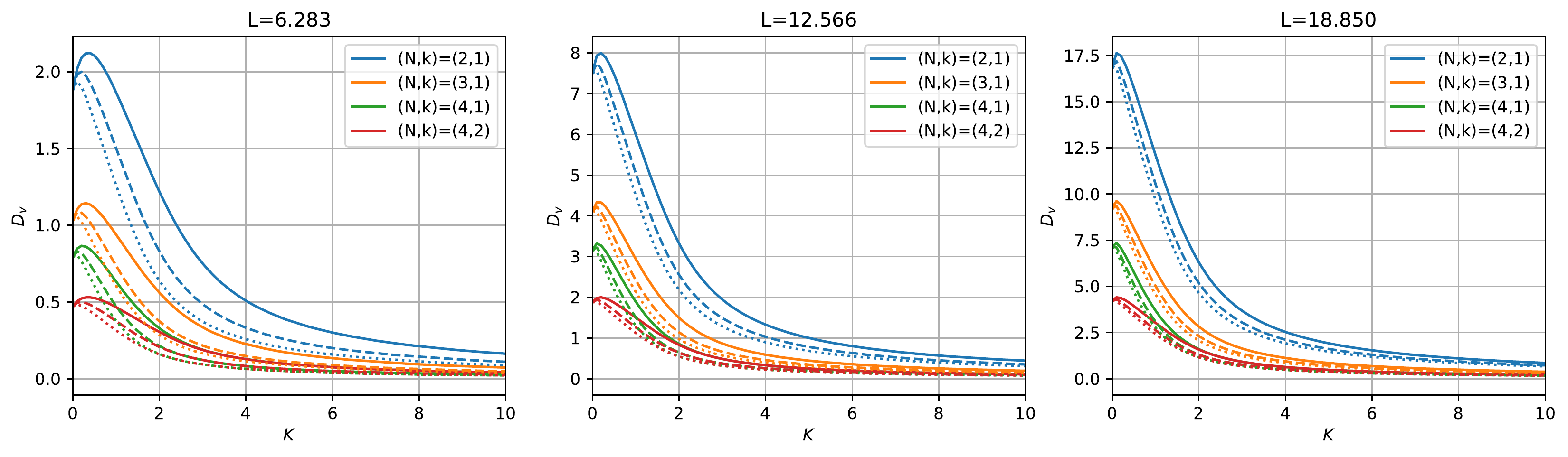}
	\caption{Asynchronous instability thresholds $D_v$ versus the
          coupling $K$ in the well-mixed limit for different values of
          $L$, different $(N,k)$ pairs, and for domain areas $A=3.142$
          (solid), $1.571$ (dashed), and $0.785$ (dotted).}
	\label{fig:wm_asynch_1}
\end{figure}

\begin{figure}[t]
	\centering
\begin{tabular}{cccc}	
 \includegraphics[scale=0.62]{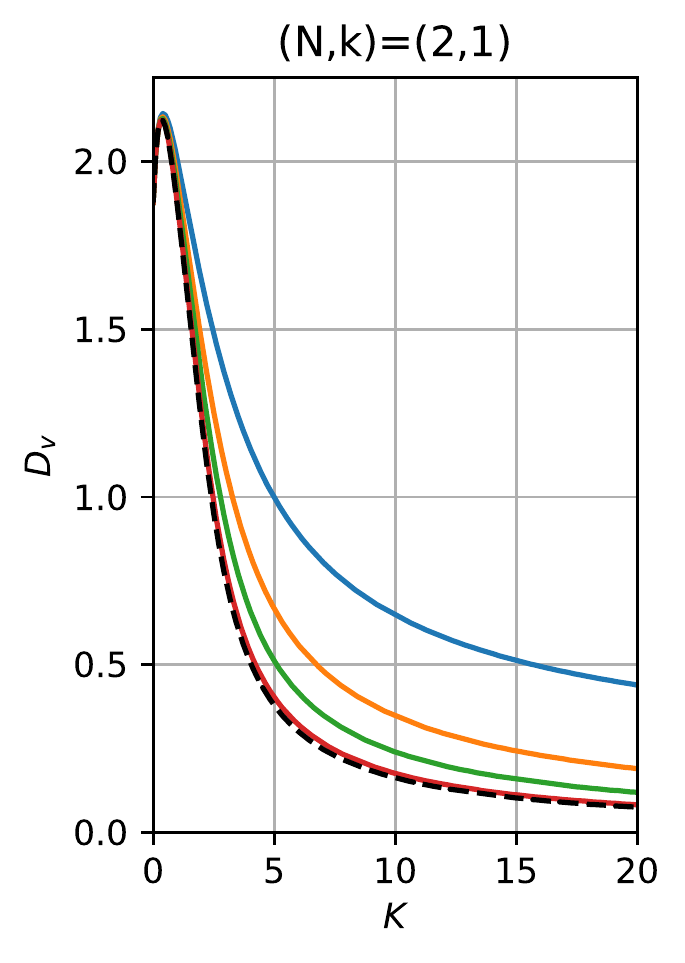}&
\includegraphics[scale=0.62]{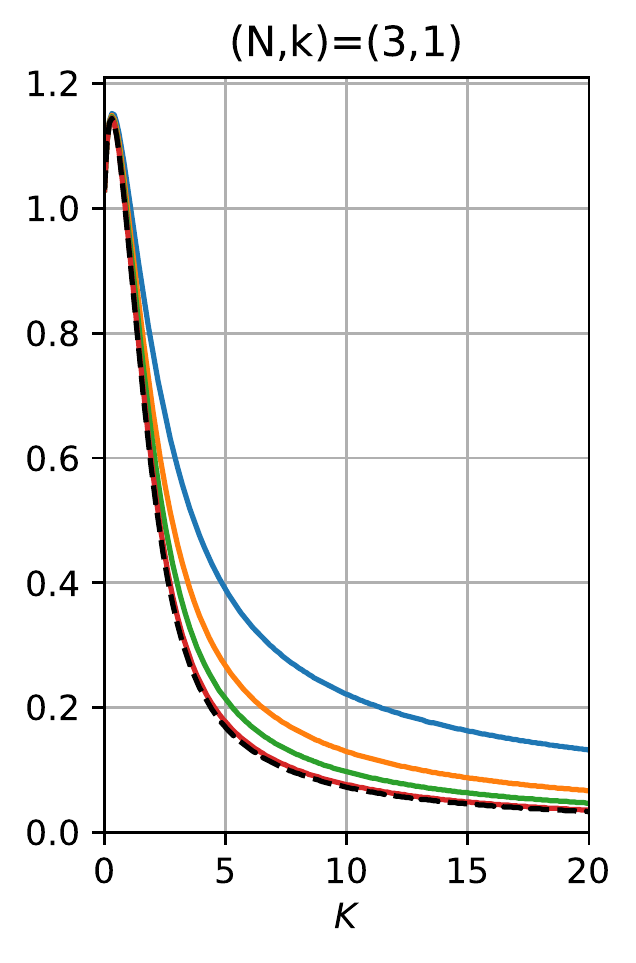}&
\includegraphics[scale=0.62]{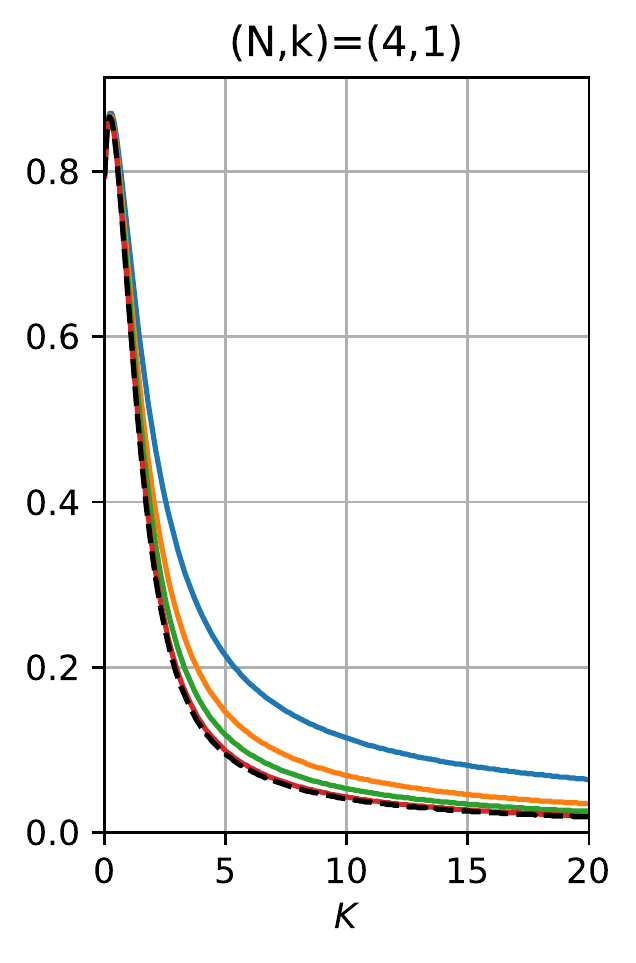}&
\includegraphics[scale=0.62]{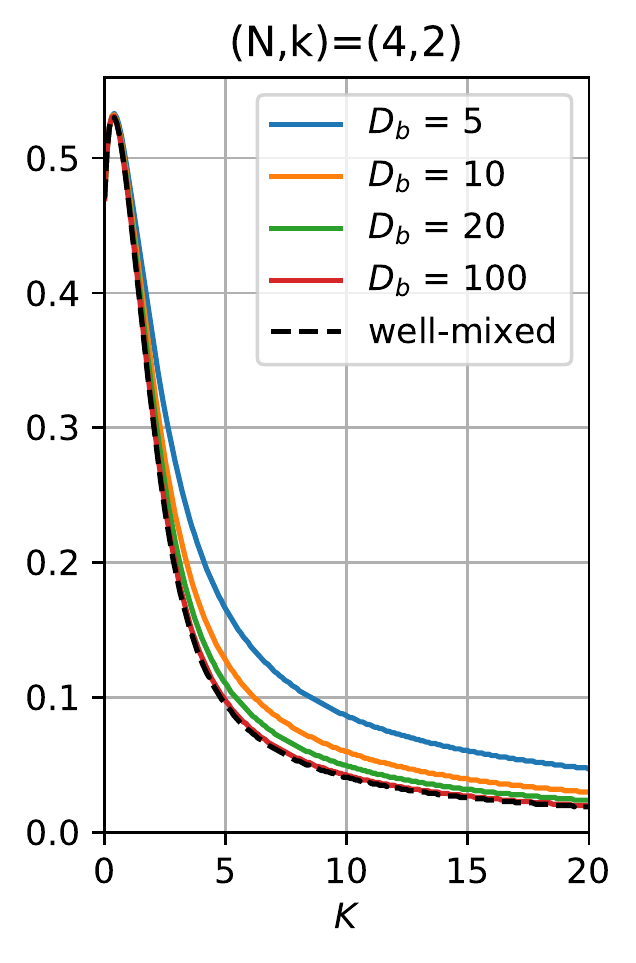}
\end{tabular}
\caption{Asynchronous instability thresholds $D_v$ versus the coupling
  $K$ for the unit disk with Gierer-Meinhardt exponents $(3,1,3,0)$,
  and for different $D_b$. The dashed lines show the corresponding
  thresholds for the well-mixed limit. The legend in the right-most plot applies to each plot.}\label{fig:dk_asynch_1}
\end{figure}

\paragraph{}
Since asynchronous instabilities emerge from a zero-eigenvalue
crossing there are two significant simplifications. Firstly, the
thresholds are determined by the nonlinear algebraic problem
$\mathcal{A}_k(0) = 0$, for each mode $k=1,\ldots,N-1$, as given by
(\ref{eq:NLEP_algebraic}) in which $\chi(\lambda)$ is replaced by
$\chi_k(\lambda)$ as defined in (\ref{eq:NLEP_symmetric}). Secondly,
by setting $\lambda=0$, it follows that all $\tau_s$ and $\tau_b$
dependent terms in $\chi_k(\lambda)$ vanish. Therefore, asynchronous
instability thresholds are independent of these two parameters. The
resulting nonlinear algebraic equations are readily solved with an
appropriate root finding algorithm (e.g.\@ the brentq routine in the
Python library SciPy). Furthermore, in the uncoupled case ($K=0$) the
threshold can be determined explicitly (notice that when $K=0$ the
well-mixed and disk cases coincide). Indeed, defining
$z = \frac{L}{2N\sqrt{D_v}}$ and $y = {\pi k/N}$, the algebraic
problem $\mathcal{A}_k(0)=0$ becomes
$\bigl(\tfrac{mq}{p-1}-1\bigr)\sinh^2(z) = \sin^2(y)$.  From
this relation it readily follows that the competition stability
threshold for $K=0$ is
\begin{equation}
  D_v = \biggl[\frac{2N}{L}\log\biggl(\sqrt{\tfrac{p-1}{mq-p+1}}
  \biggl|\sin\biggl(\frac{\pi k}{N}\biggr)\biggr| +
  \sqrt{\tfrac{p-1}{mq - p + 1}\sin^2\biggl(\frac{\pi k}{N}\biggr) + 1}\biggr)
  \biggr]^{-2}\,.
\end{equation}

Figure \ref{fig:wm_asynch_1} illustrates the dependence of the
asynchronous threshold on the geometric parameters $L$ and $A$ for the
well-mixed limit. In Figure \ref{fig:dk_asynch_1} the effect of finite
bulk diffusivity $D_b$ is explored for the unit disk. This figure also
illustrates that while the asynchronous threshold tends to
zero as $K\rightarrow\infty$ for sufficiently large values of $D_b$
the same is not true for small values of $D_b$. It is
however worth remembering that for large $K$, where the competition
threshold value of $D_v$ appears to approach zero in these figures,
the result is not uniformly valid since the NLEP derivation required
that $D_v \gg \varepsilon^2$.

\subsection{Numerical Support of the Asymptotic Theory}
\paragraph{}

In this subsection we verify some of the predictions of the
steady-state and linear stability theory by performing full
numerical PDE simulations of the coupled bulk-membrane system
\eqref{eq:bsrde2d}. In \S \ref{sec:numerics} we give an outline of
the methods used for computing the full numerical solutions. In \S
\ref{sec:validate_1} and \S \ref{sec:validate_2} we provide
both quantitative and qualitative support for the instability
thresholds predicted by the asymptotic theory.

\subsubsection{Outline of Numerical Methods}\label{sec:numerics}

The spatial discretization of \eqref{eq:bsrde2d} is much simpler for
the well-mixed limit than for the case of the disk with a finite-bulk
diffusivity. Indeed, in the well-mixed limit, the bulk
  inhibitor $V$ is spatially independent to leading order. By
  integrating the bulk PDE \eqref{eq:bsrde2d_V}, and using the
  divergence theorem, we obtain that $V$ satisfies the ODE
\begin{equation}\label{bulk:inf}
\tau_b V_t  = - (\beta - 1) V + \frac{\beta}{L} \int_{0}^{L}v\,d\sigma\,,
\end{equation}
where $\beta \equiv {KL/A}$. For the well-mixed case it
suffices to use a uniform grid in the arc-length coordinate for the
spatial discretization of the membrane problem (\ref{eq:bsrde2d_u})
and (\ref{eq:bsrde2d_v}).  Alternatively, the problem
\eqref{eq:bsrde2d} for finite $D_b$ in the disk requires a full
spatial discretization of the two-dimensional disk. To do so, we use
a finite-element approach where the mesh is chosen in such a way
that the boundary nodes are uniformly distributed. In this way, we
can continue to apply a finite difference discretization for the
membrane problem (\ref{eq:bsrde2d_u}) and (\ref{eq:bsrde2d_v}).
For both the well-mixed case and the disk problem, the
spatially discretized system leads to a large system of ODEs
\begin{equation}\label{eq:discrete_ode}
	\frac{dW}{dt} =  \mathbb{A} W + F(W)\,.
\end{equation}
Here the matrix $\mathbb{A}$ arises from the spatially
discretized diffusion operator, while $F(W)$ denotes the reaction
kinetics and the bulk-membrane coupling terms.

The choice of a time-stepping scheme for reaction diffusion systems is
generally non-trivial. Since the operator $\mathbb{A}$ is stiff, it is
best handled using an implicit time-stepping method. On the other
hand, the kinetics $F(W)$ are typically non-linear so explicit
time-stepping is favourable. Using a purely implicit or explicit
time-stepping algorithm therefore leads to substantial computation
time, either by requiring the use of a non-linear solver to handle the
kinetics in the first case, or by requiring a prohibitively small
time-step to handle the stiff linear operator in the second case. This
difficulty can be circumvented by using so-called mixed methods,
specifically the implicit-explicit methods described in
\cite{ascher_1995}. We will use a second order semi-implicit
  backwards difference scheme (2-SBDF), which employs a second-order
  backwards difference to handle the diffusive term together with an
  explicit time-stepping strategy for the nonlinear term
  (cf.~\cite{ruuth_1995}). This time-stepping strategy is given by
\begin{equation}\label{eq:2-sbdf}
  (3\mathbb{I} - 2\Delta t \mathbb{A}) W^{n+1} = 4 W^n + 4\Delta t F(W^n) -
  W^{n-1} - 2\Delta t F(W^{n-1})\,.
\end{equation}
To initialize this second-order method we bootstrap with a first order
semi-implicit backwards difference scheme (1-SBDF) as follows:
\begin{equation}\label{eq:1-sbdf}
	(\mathbb{I} - \Delta t \mathbb{A}) W^{n+1} = W^{n} + \Delta t F(W^n)\,.
\end{equation}
We will use the numerical method outlined above in the two
proceeding sections.

\subsubsection{Quantitative Numerical Validation: Numerically Computed Synchronous Threshold}\label{sec:validate_1}
\paragraph{}

\begin{figure}[h]
	\centering
	\includegraphics[scale=0.57]{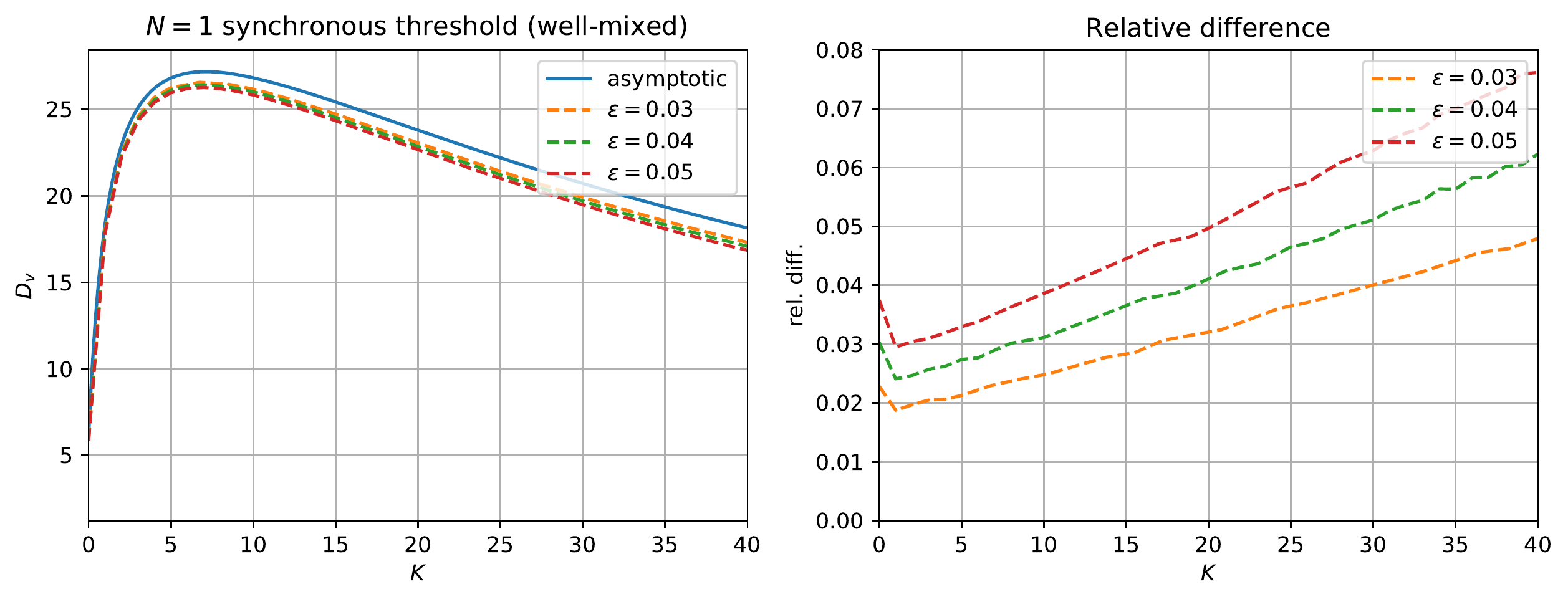}
	\caption{Comparison between numerical and asymptotic
          synchronous instability threshold for $N=1$ with $L=2\pi$,
          $A=\pi$, $\tau_s = 0.6$, and $\tau_b = 0.01$. Notice that, as
          expected, the agreement improves as $\varepsilon$ decreases.}
        \label{fig:numerical_bifurcation_well_mixed}
\end{figure}

We begin by describing a method for numerically calculating the
synchronous instability threshold for a one (or more) spike
pattern. Given an equilibrium solution $(u_0,v_0,V_0)$, for
sufficiently small times the numerical solution will evolve
approximately as the linearization
\begin{equation*}
  u(\sigma,t) = u_0(\sigma)  + e^{\lambda t}\phi(\sigma)\,,\qquad
  v(\sigma,t) = v_0(\sigma) + e^{\lambda t} \psi(\sigma)\,,\qquad
  V(\sigma,t) = V_0(\sigma) + e^{\lambda t}\eta(\sigma)\,.
\end{equation*}
For $\varepsilon>0$ fixed and sufficiently small the
steady-state will be very close to that predicted by the asymptotic
theory. By initializing the numerical solver with one of the
steady-state solutions predicted by the asymptotic theory, and then
tracking its time evolution, we will thus be able to approximate the
value of $\text{Re}(\lambda)$. If we fix a location on the boundary
$\sigma^{\star}$ (e.g.\@ one of the spike locations) and let
$t_1^{\star}<t_2^{\star}<...$ denote the sequence of times at which
$u(\sigma^{\star},t)$ attains a local maximum or minimum in $t$,
then the sequence $u_j^{\star}=u(\sigma^{\star},t_j^{\star})$
($j=1,..,$) will approximate the envelope of
$u(\sigma^{\star},t)$. If this sequence is diverging from its
average then $\text{Re}\lambda\geq 0$, whereas if it is converging
then $\text{Re}\lambda<0$. Furthermore, by writing
\begin{equation*}
  |u_n^{\star} - u_0(\sigma^{\star})| \approx
  e^{t_n \text{Re}(\lambda) } |\phi(\sigma^{\star})|\,,
\end{equation*}
we can solve for $\text{Re}(\lambda)$ by taking two values
$t_n^{\star}>t_m^{\star}$ sufficiently far apart to get
\begin{equation*}
  \text{Re}(\lambda) \approx \frac{\log\bigl| u_n^{\star}
    - u_0(\sigma^{\star})\bigr| -
    \log\bigl| u_m^{\star} - u_0(\sigma^{\star})\bigr|}{t_n^{\star} - t_m^{\star}}\,.
\end{equation*}
This motivates a simple method for estimating the synchronous
instability threshold numerically. Starting with some point in
parameter space (chosen close to the threshold predicted by the
asymptotic theory) we approximate $\text{Re}(\lambda)$ and then
increase or decrease one of the parameters to drive
$\text{Re}(\lambda)$ towards zero. Once $\text{Re}(\lambda)$ is
sufficiently close to zero we designate the resulting point in
parameter space as a numerically-computed synchronous instability
threshold point.

In the well-mixed limit, we fix values of $K$ and vary $D_v$ using
the numerical approach described above until
$\text{Re}(\lambda)$ is sufficiently small. The results in
Figure \ref{fig:numerical_bifurcation_well_mixed} compare the
synchronous instability threshold for $N=1$ in the well-mixed limit
as predicted by the asymptotic theory and by our full numerical
approach for $\varepsilon=0.3,0.4,0.5$. We observe, as expected, that the
asymptotic prediction improves with decreasing values of
$\varepsilon$, but that the agreement is non-uniform in
the coupling parameter $K$.

\subsubsection{Qualitative Numerical Support: A Gallery of Numerical Simulations}\label{sec:validate_2}

\begin{figure}[t]
	\centering
	\begin{tabular}{ccc}	
		\includegraphics[scale=0.715]{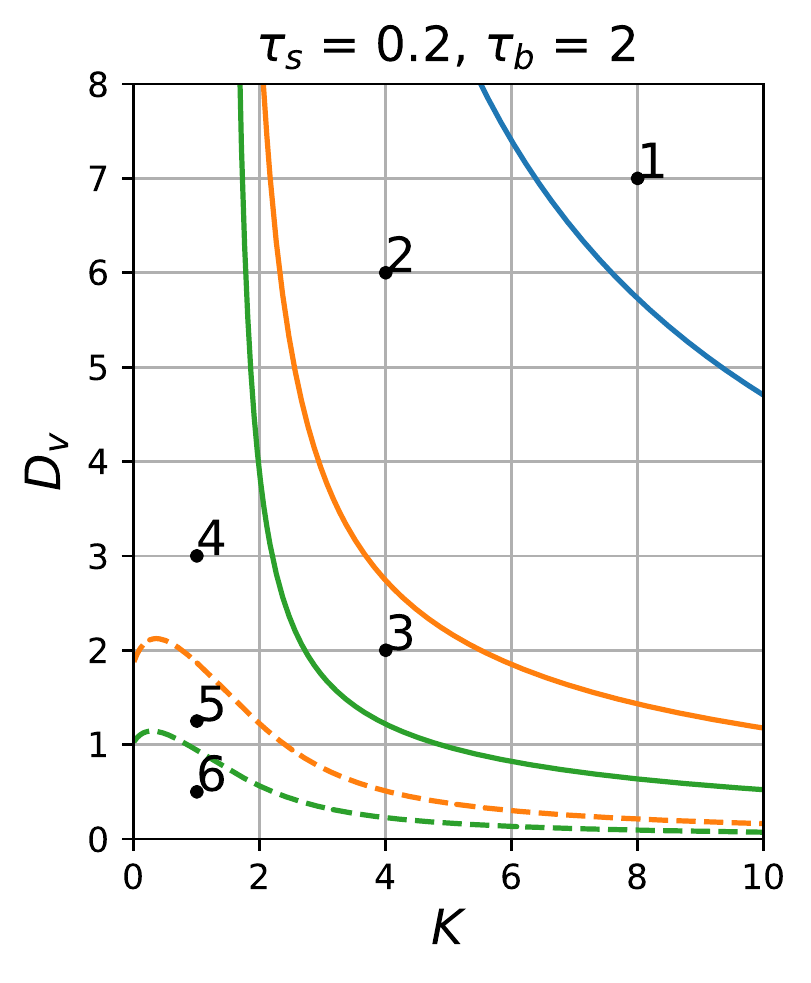}& \includegraphics[scale=0.715]{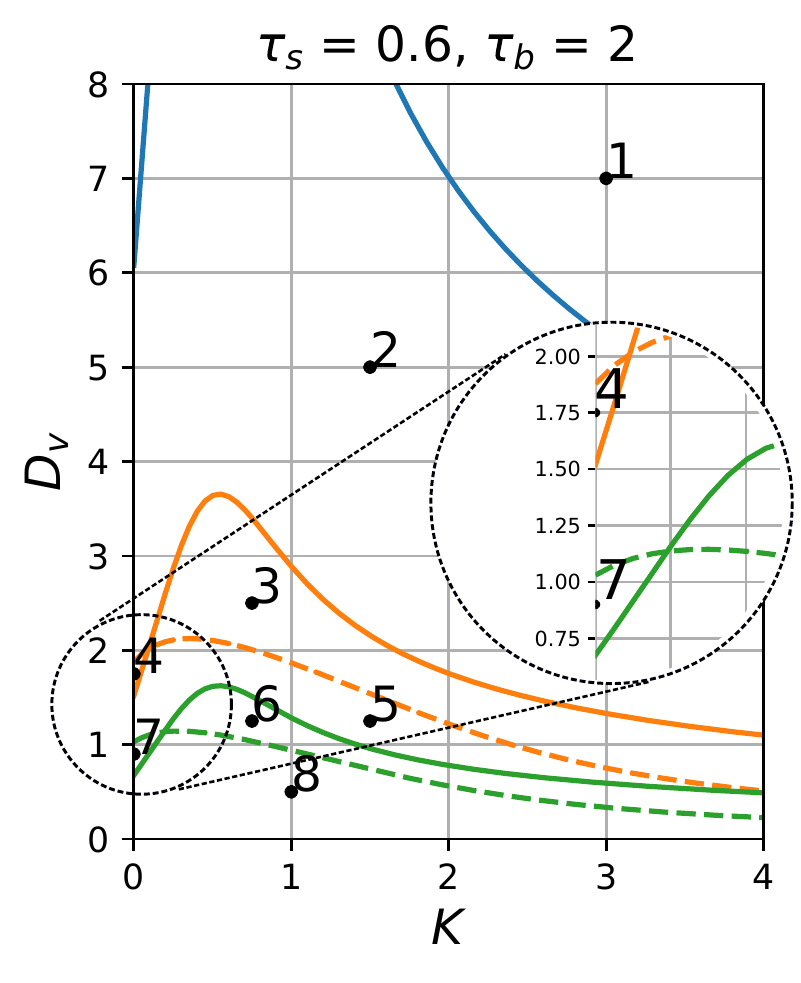} &
		\includegraphics[scale=0.715]{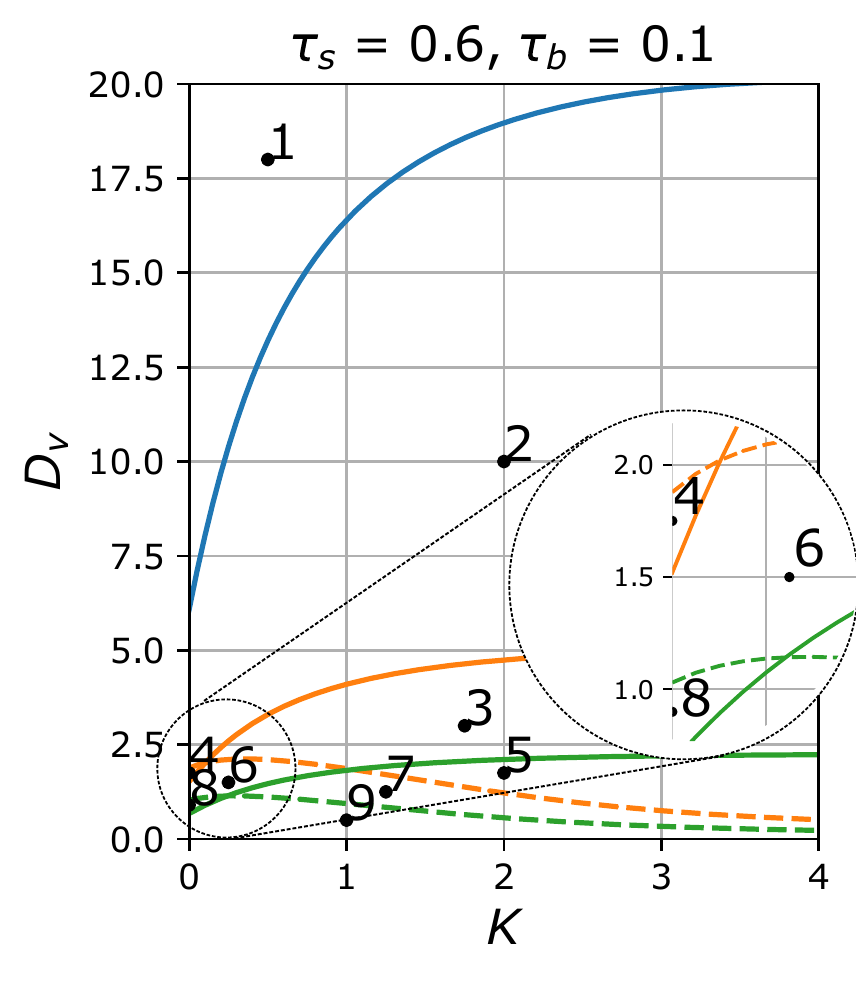}
	\end{tabular}
	\caption{Synchronous (solid) and asynchronous (dashed)
          instability thresholds in the $D_v$ versus $K$
            parameter plane in the well-mixed limit for $N=1$
          ({\color{plot_blue}blue}), $N=2$
          ({\color{plot_orange}orange}), and $N=3$
          ({\color{plot_green}green}). At the top of each of
            the three panels a different pair $(\tau_s,\tau_b)$ is
            specified. See Table \ref{tbl:wm_threshold_legend} for
            $D_v$ and $K$ values at the numbered points in each
            panel. Figures \ref{fig:wm_snapshots_set_1},
            \ref{fig:wm_snapshots_set_2}, and
            \ref{fig:wm_snapshots_set_3} show the corresponding spike
            dynamics from full PDE simulations of (\ref{eq:bsrde2d})
            at the indicated points.}\label{fig:wm_threshold_legends}
\end{figure}

\begin{figure}
	\centering
	\begin{tabular}{cccc}
		\includegraphics[scale=0.6]{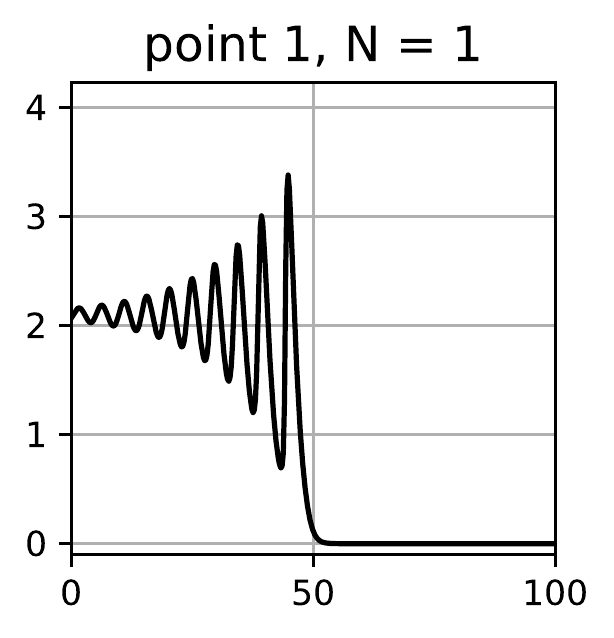} & \includegraphics[scale=0.6]{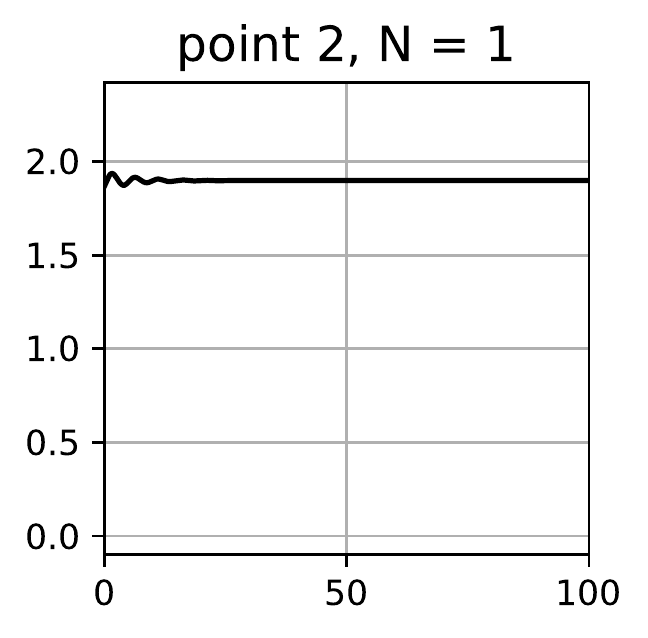} & \includegraphics[scale=0.6]{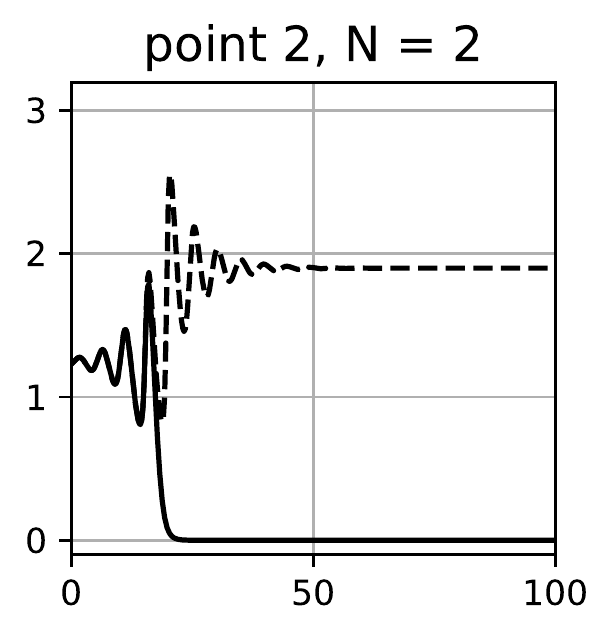} &
		\includegraphics[scale=0.6]{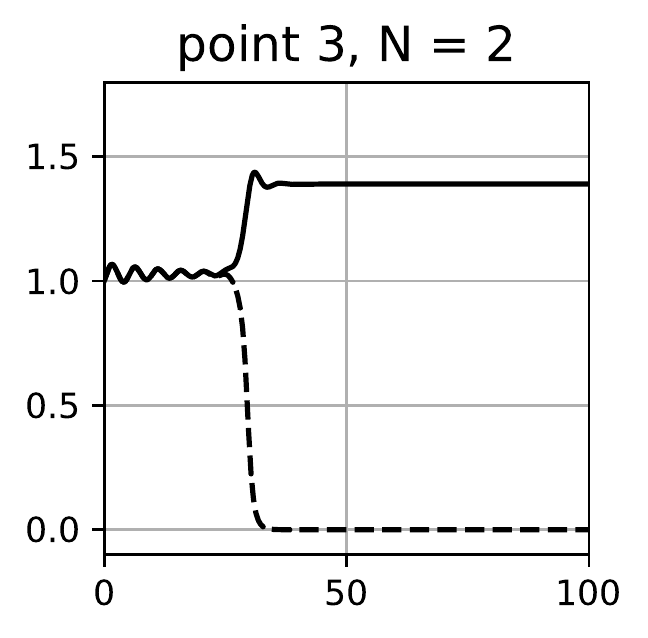} \\ \includegraphics[scale=0.6]{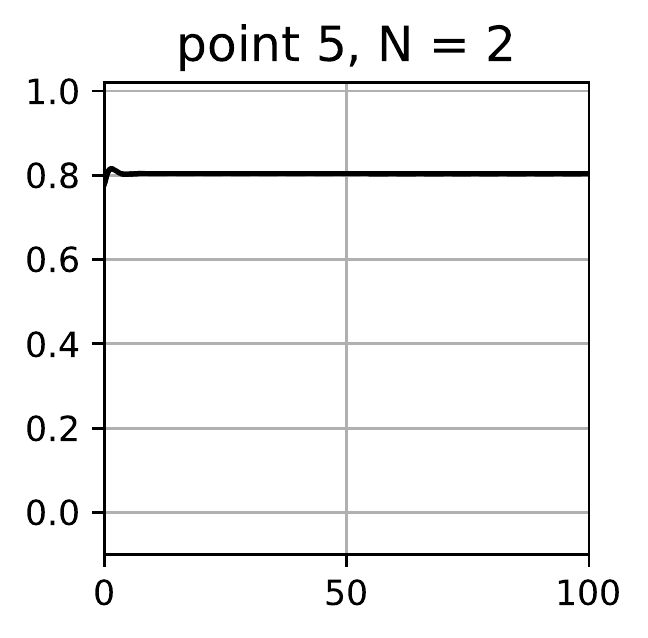} & 
		\includegraphics[scale=0.6]{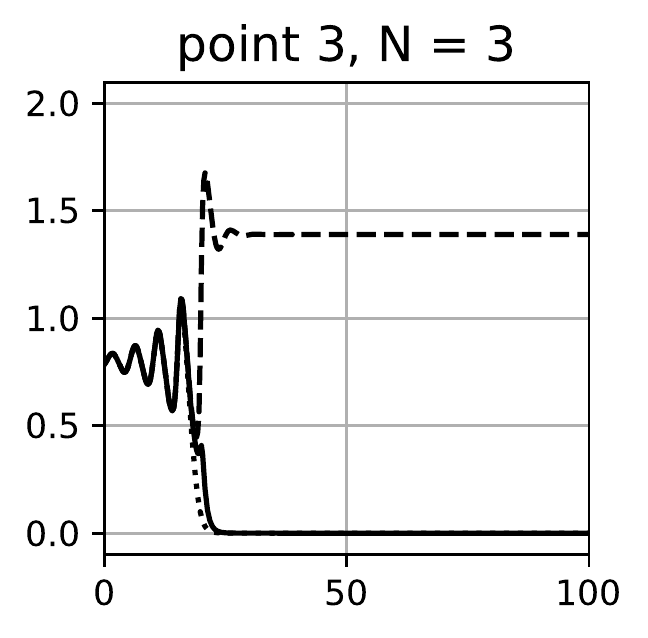} &
		\includegraphics[scale=0.6]{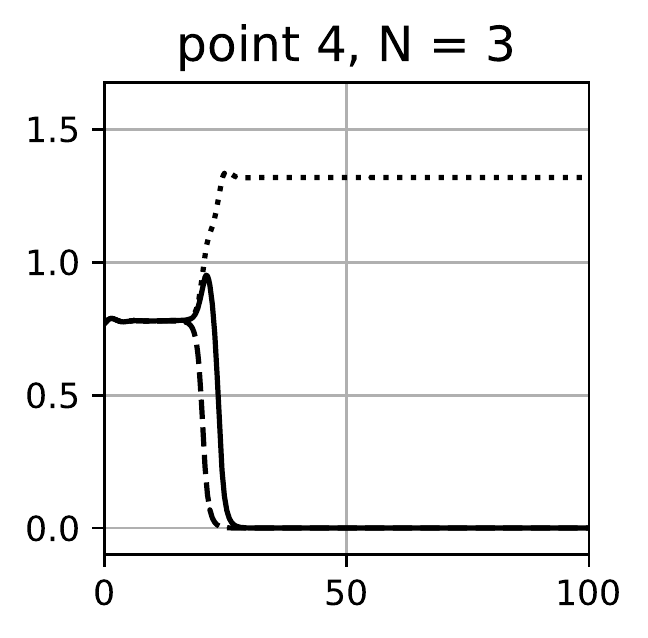} &
		\includegraphics[scale=0.6]{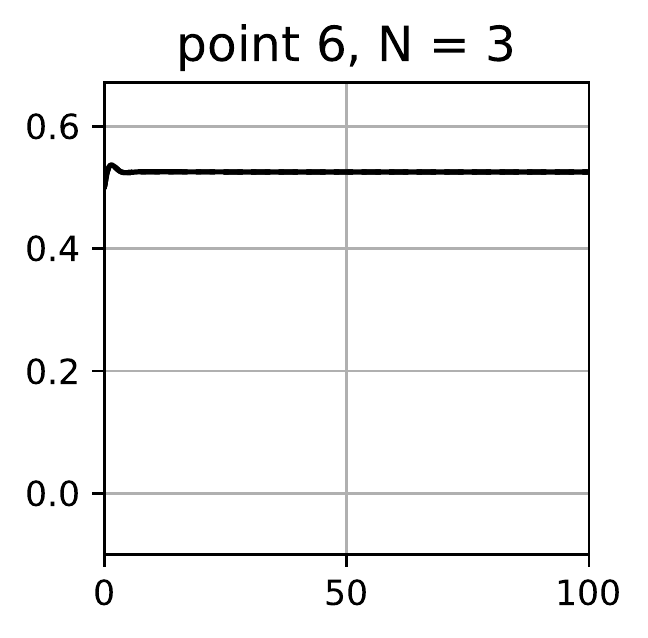}
	\end{tabular}
	\caption{Numerically computed spike heights (vertical axis)
          versus time (horizontal axis) from full PDE simulations of
          (\ref{eq:bsrde2d}) for $\tau_s = 0.2$ and $\tau_b = 2$ at
          the points indicated in the left panel of Figure
          \ref{fig:wm_threshold_legends}. Distinct spike heights are
          distinguished by line types (solid, dashed, and
          dotted).}\label{fig:wm_snapshots_set_1}
\end{figure}

\begin{figure}
	\centering
	\begin{tabular}{ccccc}
		\includegraphics[scale=0.6]{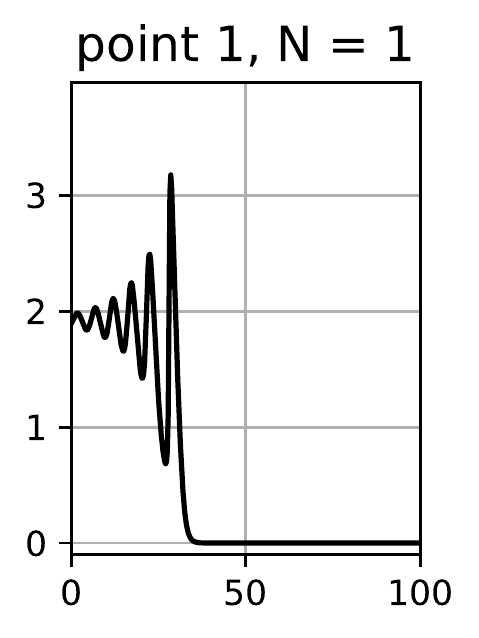} & \includegraphics[scale=0.6]{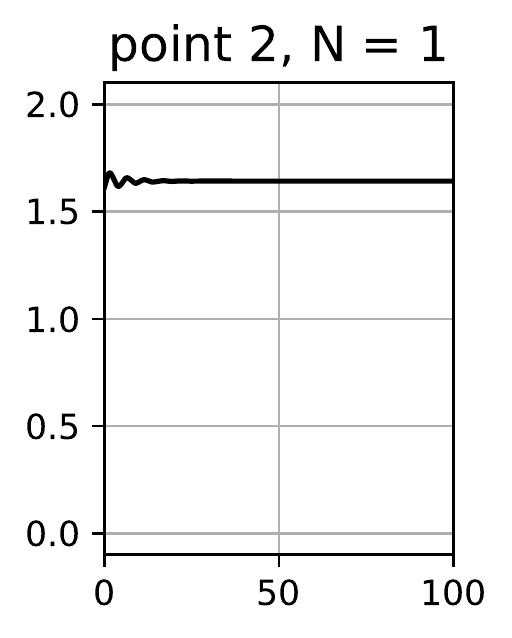} & 
		\includegraphics[scale=0.6]{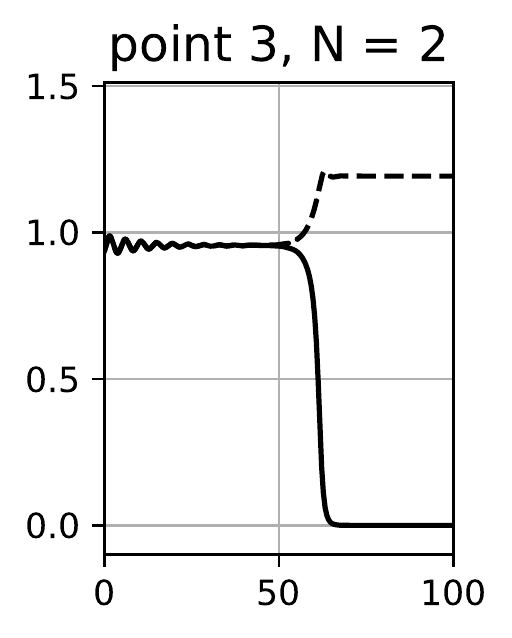} & \includegraphics[scale=0.6]{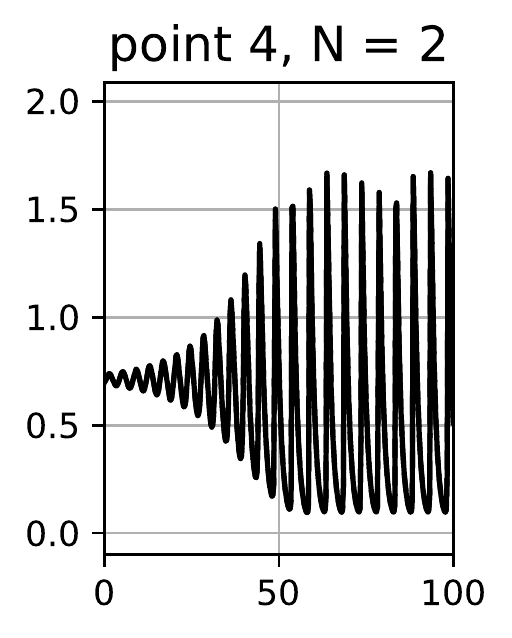} & \includegraphics[scale=0.6]{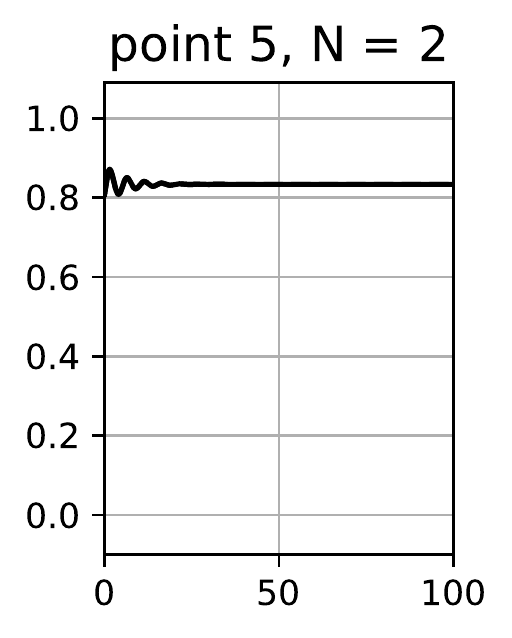} \\ 
		\includegraphics[scale=0.6]{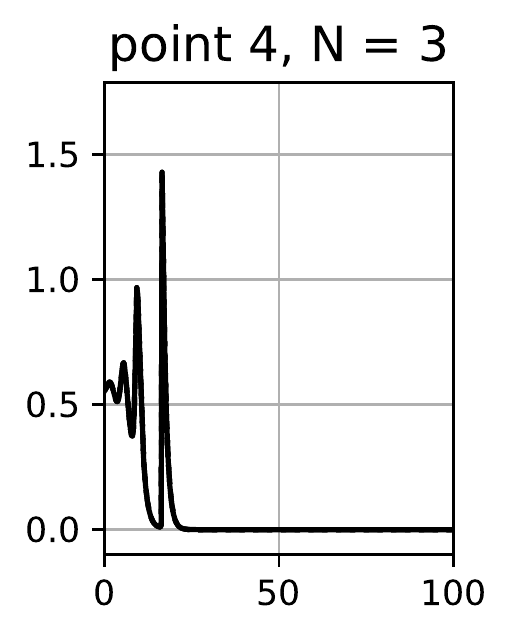} &
		\includegraphics[scale=0.6]{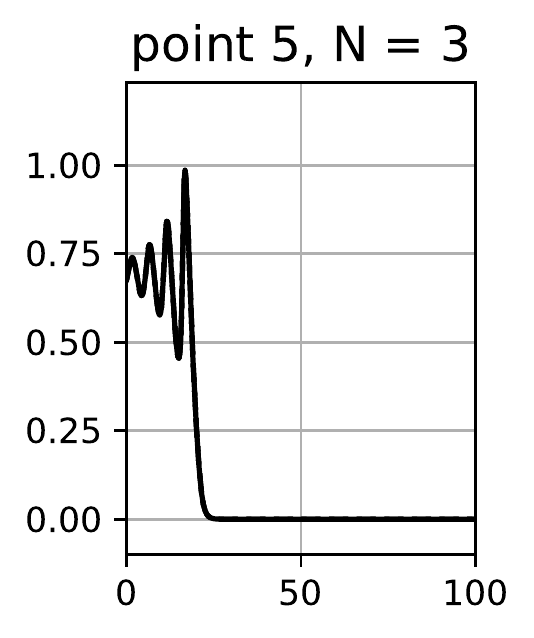} &
		\includegraphics[scale=0.6]{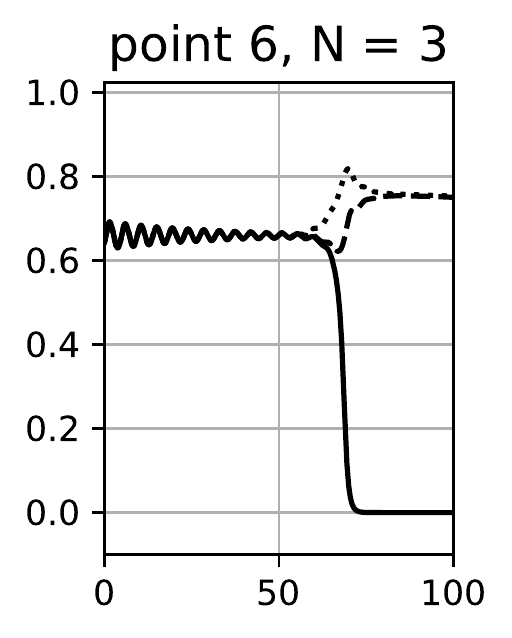} &
		\includegraphics[scale=0.6]{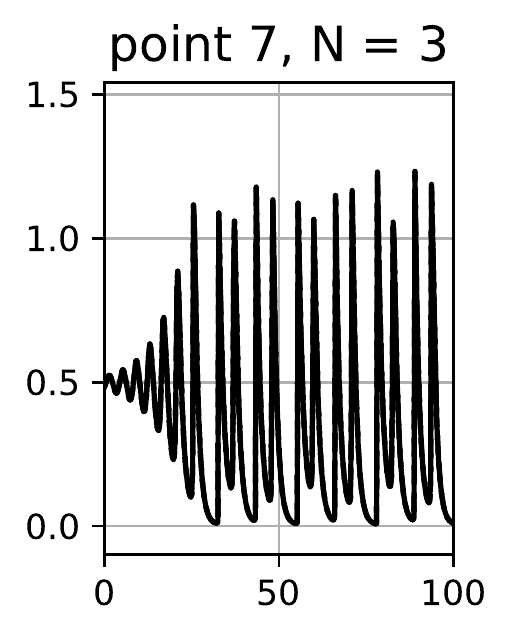} & \includegraphics[scale=0.6]{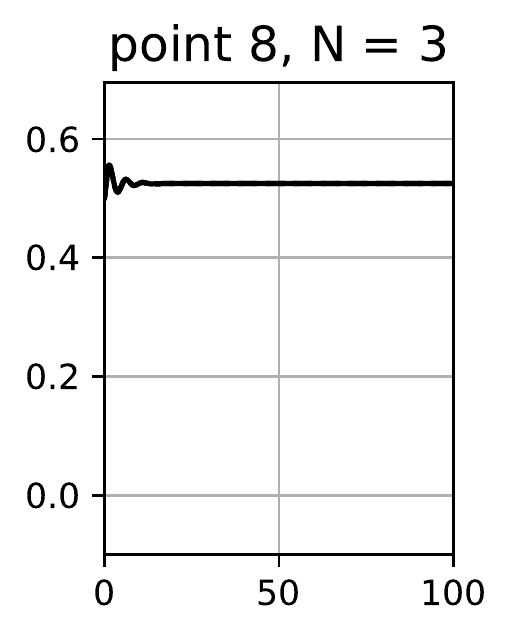}
	\end{tabular}
	\caption{Numerically computed spike heights (vertical axis)
          versus time (horizontal axis) from full PDE simulations of
          (\ref{eq:bsrde2d}) for $\tau_s = 0.6$ and $\tau_b = 2$ at
          the points indicated in the middle panel of Figure
          \ref{fig:wm_threshold_legends}. Distinct spike heights are
          distinguished by line types (solid, dashed, and
          dotted).}\label{fig:wm_snapshots_set_2}
\end{figure}

\begin{figure}
	\centering
	\begin{tabular}{ccccc}
		\includegraphics[scale=0.6]{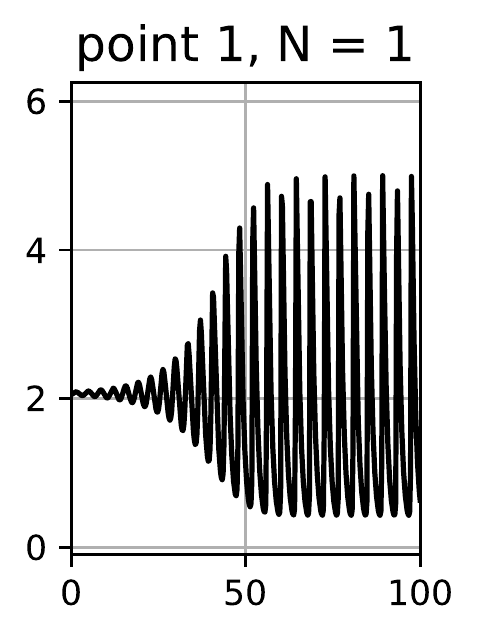} &  \includegraphics[scale=0.6]{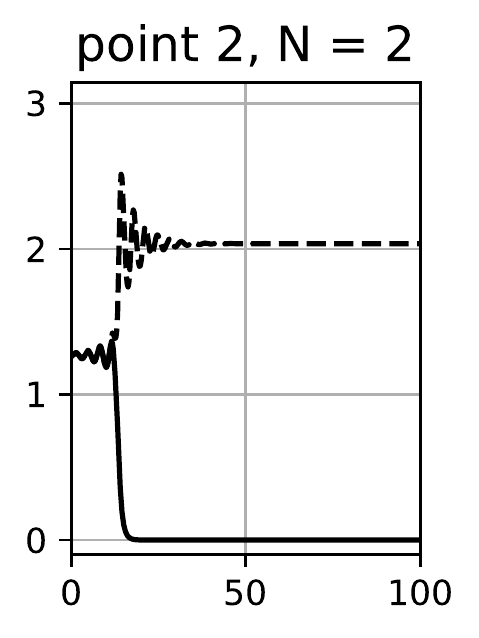} &
		\includegraphics[scale=0.6]{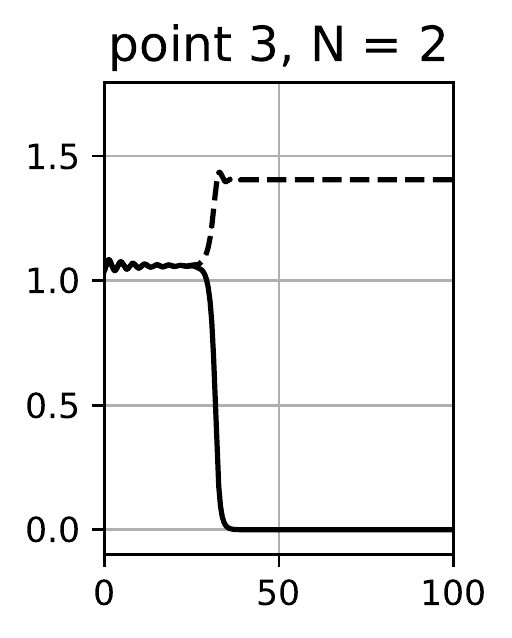} & \includegraphics[scale=0.6]{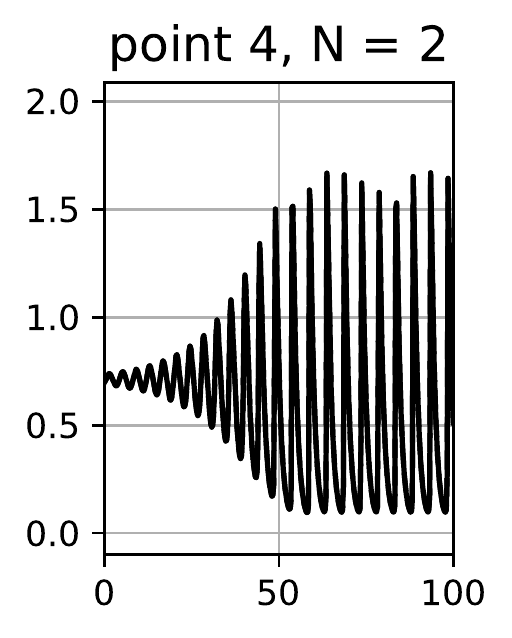} & \includegraphics[scale=0.6]{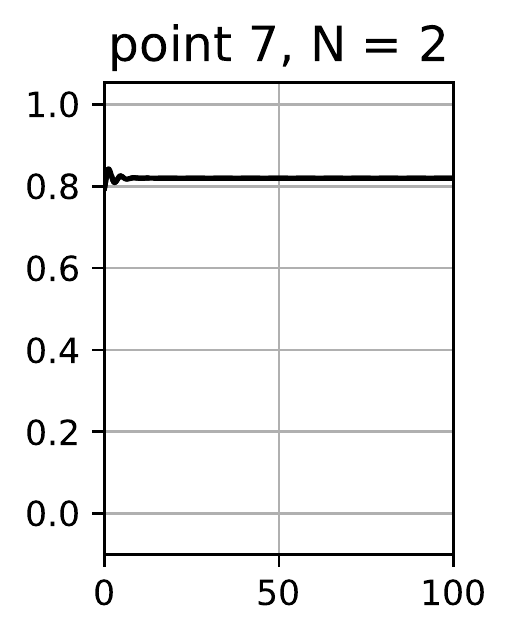} \\ 
		\includegraphics[scale=0.6]{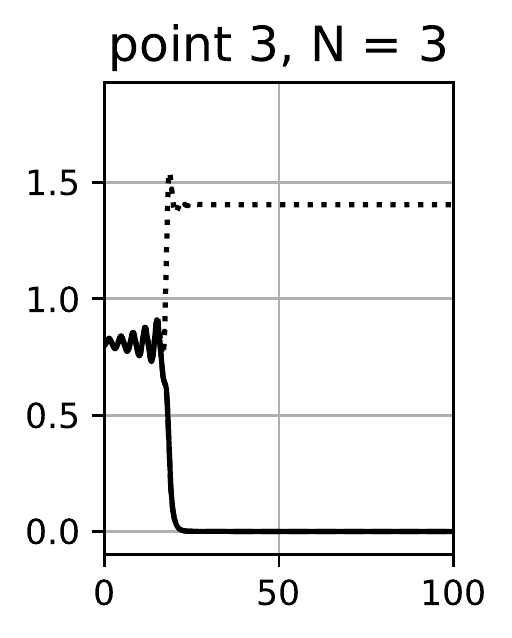} &
		\includegraphics[scale=0.6]{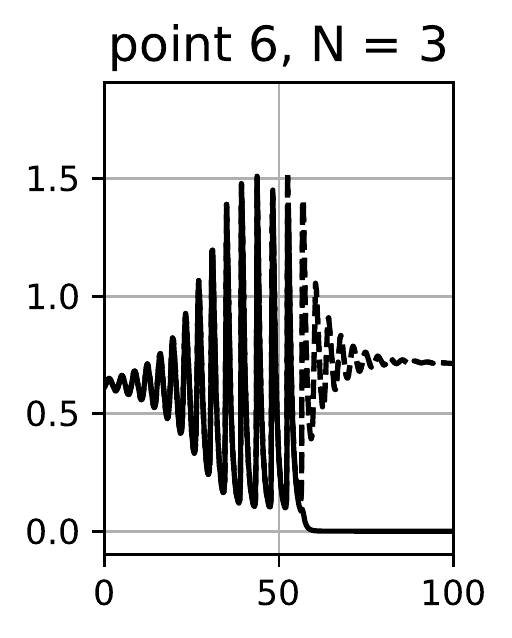} &
		\includegraphics[scale=0.6]{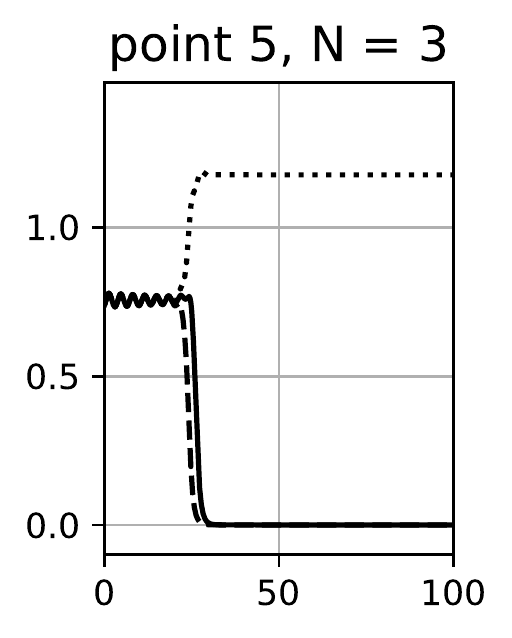} &
		\includegraphics[scale=0.6]{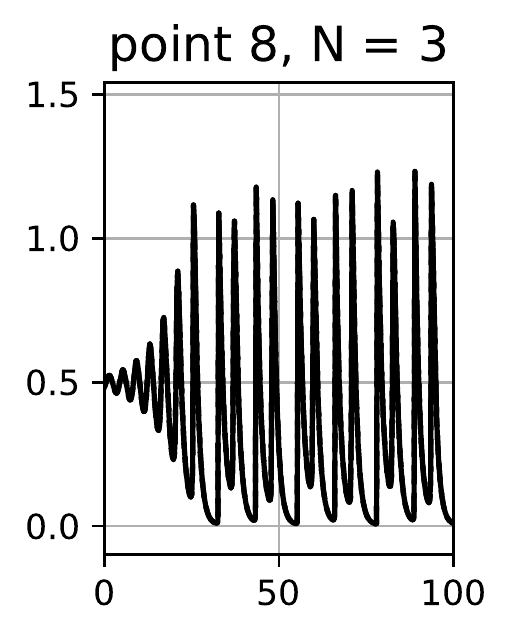} & \includegraphics[scale=0.6]{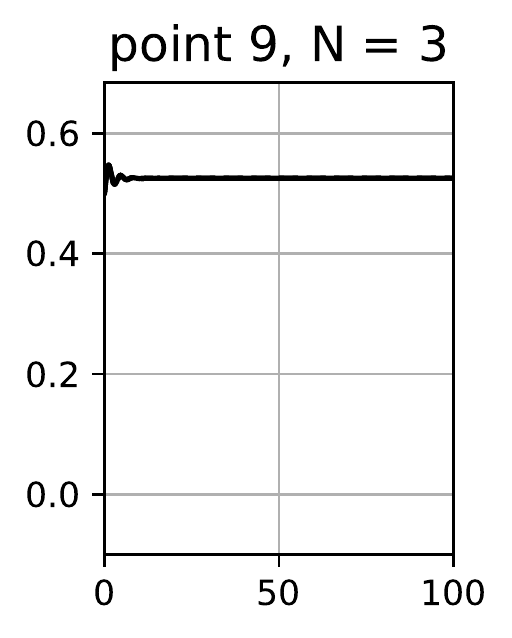}
	\end{tabular}
	\caption{Numerically computed spike heights (vertical axis)
          versus time (horizontal axis) from full PDE simulations of
          (\ref{eq:bsrde2d}) for $\tau_s = 0.6$ and $\tau_b = 0.1$ at
          the points indicated in the right panel of Figure
          \ref{fig:wm_threshold_legends}. Distinct spike heights are
          distinguished by line types (solid, dashed, and
          dotted).}\label{fig:wm_snapshots_set_3}
\end{figure}

We conclude this section by first showcasing the dynamics of multiple
spike patterns for several choices of the parameters $K$, $D_v$,
$\tau_s$, and $\tau_b$ in the well-mixed limit. We will focus
exclusively on the explicitly solvable Gierer-Meinhardt
exponent set $(p,q,m,s)=(3,1,3,0)$ with $\varepsilon=0.05$ and the
geometric parameters $L = 2\pi$ and $A=\pi$. For the numerical
computation we discretized the domain boundary with $1200$ uniformly
distributed points ($\Delta\sigma \approx 0.00524$) and used
trapezoidal integration for the bulk-inhibitor equation
(\ref{bulk:inf}). Furthermore, we used 2-SBDF time-stepping
initialized by 1-SBDF with a time-step size of
$\Delta t = 2.5(\Delta \sigma)^2\approx 6.854\times 10^{-4}$. In
Figure \ref{fig:wm_threshold_legends} we plot the asymptotically
predicted synchronous and asynchronous instability thresholds for
three pairs of time-scale parameters:
$(\tau_s,\tau_b) = (0.2,2), (0.6,2), (0.6,0.1)$. Each plot also
contains several sample points whose $K$ and $D_v$ values are given in
Table \ref{tbl:wm_threshold_legend} below. The corresponding
full PDE numerical simulations, tracking the heights of the spikes
versus time, at these sample points are shown in Figures
\ref{fig:wm_snapshots_set_1}, \ref{fig:wm_snapshots_set_2}, and
\ref{fig:wm_snapshots_set_3}. We observe that the initial
instability onset in these figures is in agreement with that predicted
by the linear stability theory. For example, when
$\tau_s=0.6$ and $\tau_b=2$ an $N=3$ spike pattern at point six should
be stable with respect to an $N=3$ synchronous instability but
unstable with respect to the $N=3$ asynchronous instabilities. Indeed
the initial instability onset depicted in the ``point $6$, $N=3$''
plot of Figure \ref{fig:wm_snapshots_set_2} showcases the
non-oscillatory growth of two spikes and decay of one as expected. In
addition the plots in Figures \ref{fig:wm_snapshots_set_1},
\ref{fig:wm_snapshots_set_2}, and \ref{fig:wm_snapshots_set_3} support
two previously stated conjectures. Firstly, pure Hopf bifurcations for
$N\geq 2$ should be supercritical (see ``Point $4$, $N=2$'', ``Point
$7$, $N=3$'' in Figure \ref{fig:wm_snapshots_set_2}, and ``Point $4$,
$N=2$'', ``Point $8$, $N=3$'' in Figure
\ref{fig:wm_snapshots_set_3}). Secondly, we observe that
asynchronous instabilities lead to the eventual annihilation of some
spikes and the growth of others. As a result, our PDE simulations
suggest that these instabilities are subcritical.

\begin{table}[h!]
	\begin{subtable}{0.24\linewidth}
	\centering
	\begin{tabular}{c|cc}
		\hline
		Point & $K$ & $D_v$ \\ \hline
		1 & 8 & 7 \\
		2 & 4 & 6 \\
		3 & 4 & 2 \\
		4 & 1 & 3 \\
		5 & 1 & 1.25 \\
		6 & 1 & 0.5 \\
		& & \\
		& & \\		
		& & \\ \hline
	\end{tabular}
	\caption{}
	\end{subtable}%
	\begin{subtable}{0.24\linewidth}
	\centering
	\begin{tabular}{c|cc}
		\hline
		Point & $K$ & $D_v$ \\ \hline
		1 & 3 & 7 \\
		2 & 1.5 & 5 \\
		3 & 0.75 & 2.5 \\
		4 & 0 & 1.75 \\
		5 & 1.5 & 1.25 \\
		6 & 0.75 & 1.25 \\
		7 & 0 & 0.9 \\
		8 & 1 & 0.5 \\
		& & \\ \hline
	\end{tabular}
	\caption{}
	\end{subtable}%
	\begin{subtable}{0.24\linewidth}
	\centering
	\begin{tabular}{c|cc}
		\hline
		Point & $K$ & $D_v$ \\ \hline
		1 & 0.5 & 18 \\
		2 & 2 & 10 \\
		3 & 1.75 & 3 \\
		4 & 0 & 1.75 \\
		5 & 2 & 1.75 \\
		6 & 0.25 & 1.5 \\
		7 & 1.25 & 1.25 \\
		8 & 0 & 0.9 \\
		9 & 1 & 0.9 \\ \hline
	\end{tabular}
	\caption{}
	\end{subtable}%
	\begin{subtable}{0.24\linewidth}
	\centering
	\begin{tabular}{c|cc}
		\hline
		Point & $K$ & $D_v$ \\ \hline
		1 & 0.5 & 18 \\
		2 & 2 & 10 \\
		3 & 2 & 3.5 \\
		4 & 1 & 0.5 \\
		5 & 0.025 & 1.8 \\
		 &  &  \\
		 &  &  \\
		 &  &  \\
		 &  &  \\ \hline
	\end{tabular}
	\caption{}
	\end{subtable}
	\caption{$K$ and $D_v$ values at the sampled points
            in the three panels of
            Fig.~\ref{fig:wm_threshold_legends}: (a)
          Left panel: $(\tau_s,\tau_b) = (0.2,2)$, (b)
          Middle panel: $(\tau_s,\tau_b) = (0.6,2)$, and (c)
          Right panel: $(\tau_s,\tau_b) = (0.6,0.1)$. Table (d) shows the $K$ and $D_v$ values at the sampled points for the disk appearing in the left panel of Fig.~\ref{fig:disk_legend_and_snapshots}.}
        \label{tbl:wm_threshold_legend}
\end{table}

\begin{figure}
	\centering
	\begin{subfigure}{0.4\textwidth}
	\includegraphics[scale=0.65]{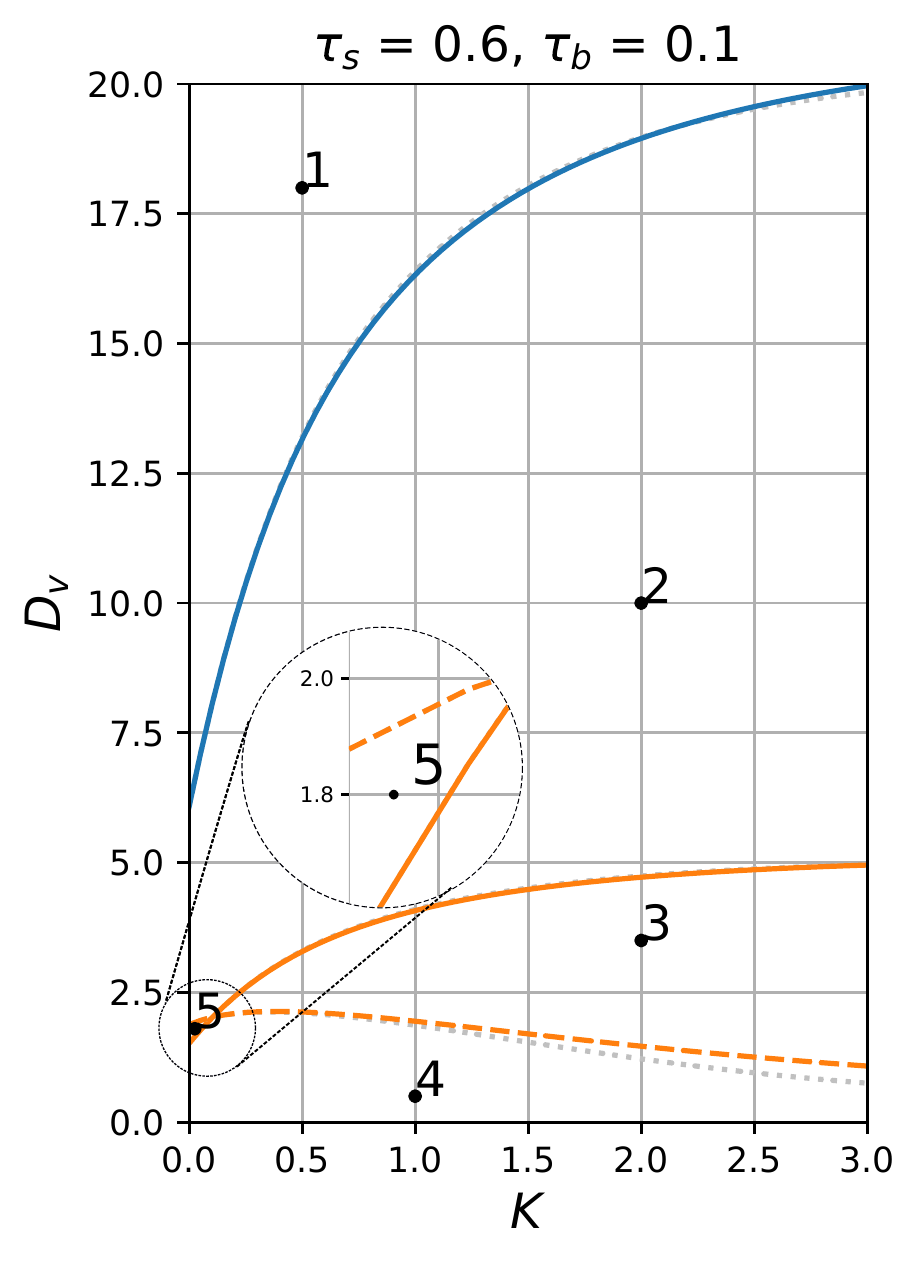}
	\caption{}
	\end{subfigure}%
	\begin{subfigure}{0.6\textwidth}
		\begin{tabular}{ccc}
			\includegraphics[scale=0.6]{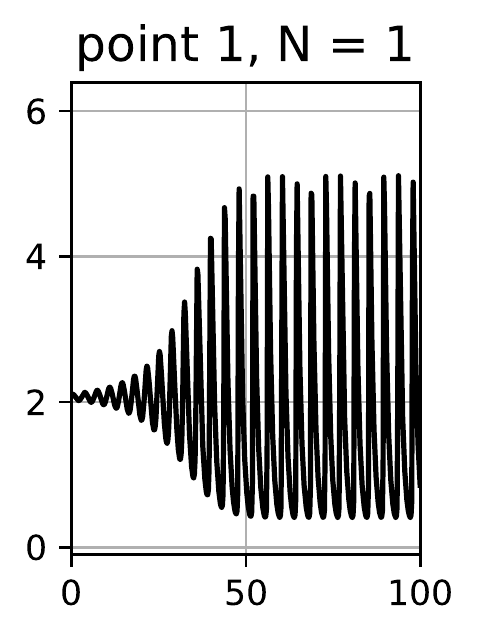} &  \includegraphics[scale=0.6]{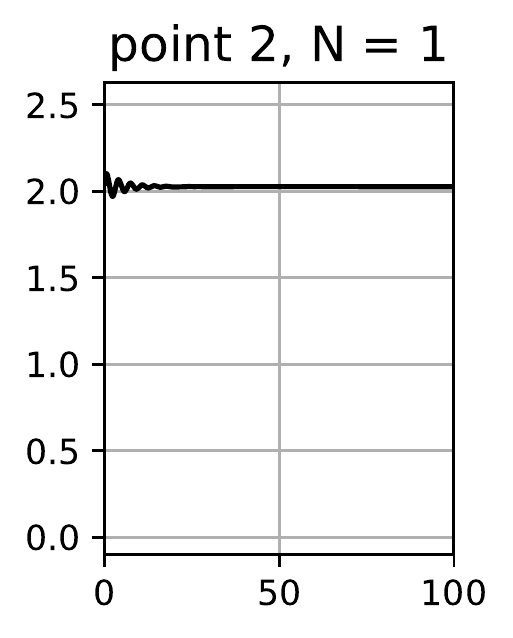} &
			\includegraphics[scale=0.6]{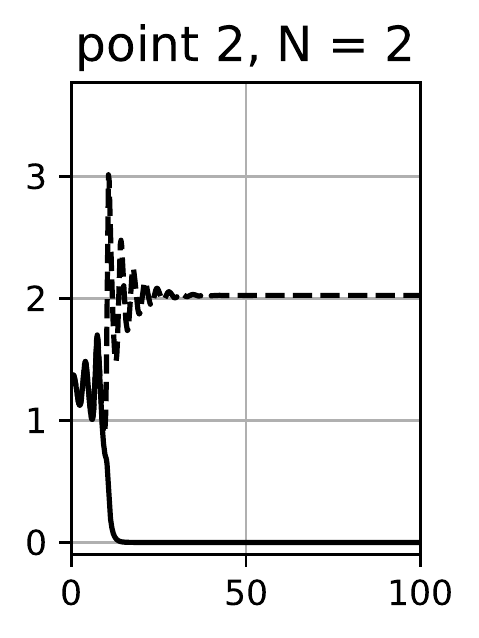} \\ \includegraphics[scale=0.6]{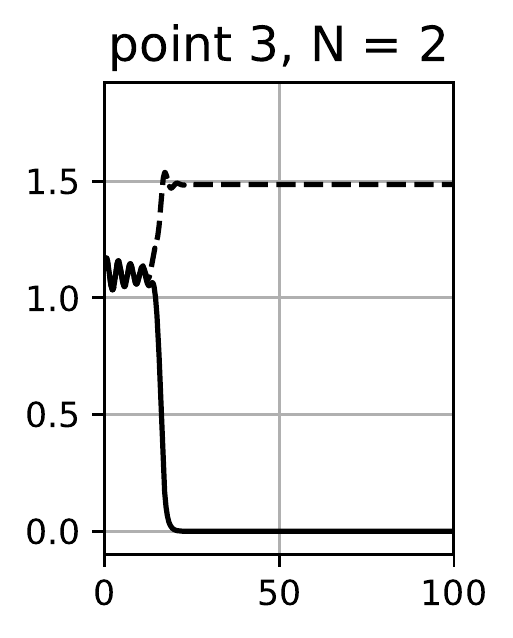} & \includegraphics[scale=0.6]{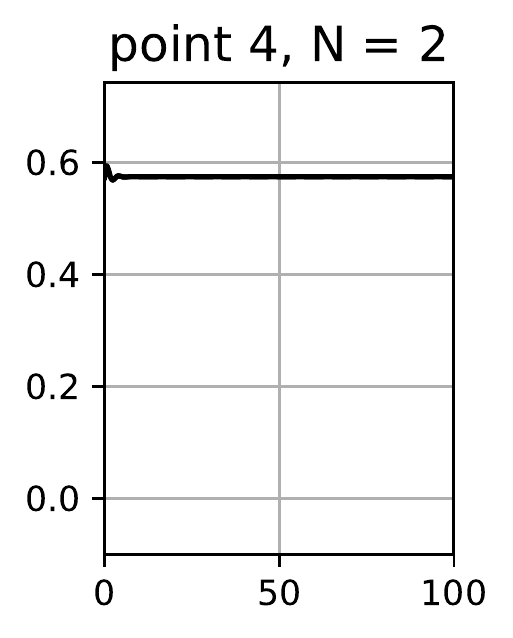} & 
			\includegraphics[scale=0.6]{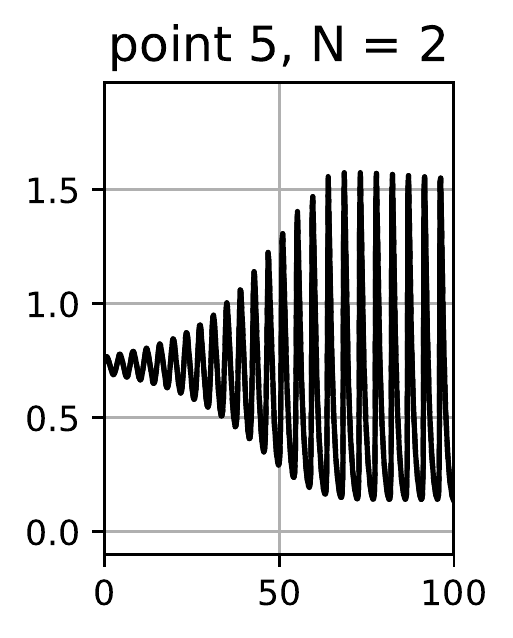}
		\end{tabular}
	\caption{}
	\end{subfigure}
	\caption{Left panel (a): Synchronous (solid) and
            asynchronous (dashed) instability thresholds in the $D_v$
            versus $K$ parameter plane for the unit disk with $D_b=10$
            and $(\tau_s,\tau_b)=(0.6,0.1)$. $N=1$ spike and $N=2$
            spikes correspond to the ({\color{plot_blue}blue}) and
            ({\color{plot_orange}orange}) curves, respectively. The
            faint grey dotted lines are the corresponding well-mixed
            thresholds.  Right panel (b): Numerically computed spike
            heights (vertical axis) versus time (horizontal axis) from
            full PDE simulations of (\ref{eq:bsrde2d}) at the points
            indicated in the left panel for $N=1$ and
            $N=2$ spikes. For videos of the PDE simulations please see the supplementary materials.}\label{fig:disk_legend_and_snapshots}
\end{figure}

We now show that this agreement between predictions of our
linear stability theory and results from full PDE simulations
continues to hold for the case of a finite bulk diffusivity. To
illustrate this agreement, we consider the unit disk with $D_b=10$
for $(\tau_s,\tau_b)=(0.6,0.1)$. For this parameter set, in the left
panel of Figure \ref{fig:disk_legend_and_snapshots} we show the
asymptotically predicted synchronous and asynchronous instability
thresholds in the $D_v$ versus $K$ parameter plane for $N=1$ and
$N=2$.  The faint grey dotted lines in this figure indicate the
corresponding well-mixed thresholds. In the right panel of Figure
\ref{fig:disk_legend_and_snapshots} we plot the spike heights versus
time, as computed numerically from (\ref{eq:bsrde2d}), at the sample
points indicated in the left panel. In each case, the numerically computed solution uses a $2\%$ perturbation away from the asymptotically computed $N$-spike equilibrium. As in the well-mixed case, the
full numerical simulations confirm the predictions of the linear
stability theory. Furthermore, Figures \ref{fig:disk_snapshots_p_2_n_2} and \ref{fig:disk_snapshots_p_5_n_2} depict both the bulk-inhibitor and the two membrane-bound species at certain times for an $N=2$ spike pattern at points $2$ and $5$ in the left panel of Figure \ref{fig:disk_legend_and_snapshots}, respectively. From this figure,
we observe that the bulk-inhibitor field is largely constant except within a
small near region near the spike locations.

\setcounter{equation}{0} 
\setcounter{section}{3}
\section{The Effect of Boundary Perturbations on Asynchronous Instabilities}\label{sec:boundary}
\paragraph{}

\begin{figure}[t]
	\centering
	\begin{subfigure}{0.23\textwidth}
		\centering
		\includegraphics[height=2.25in]{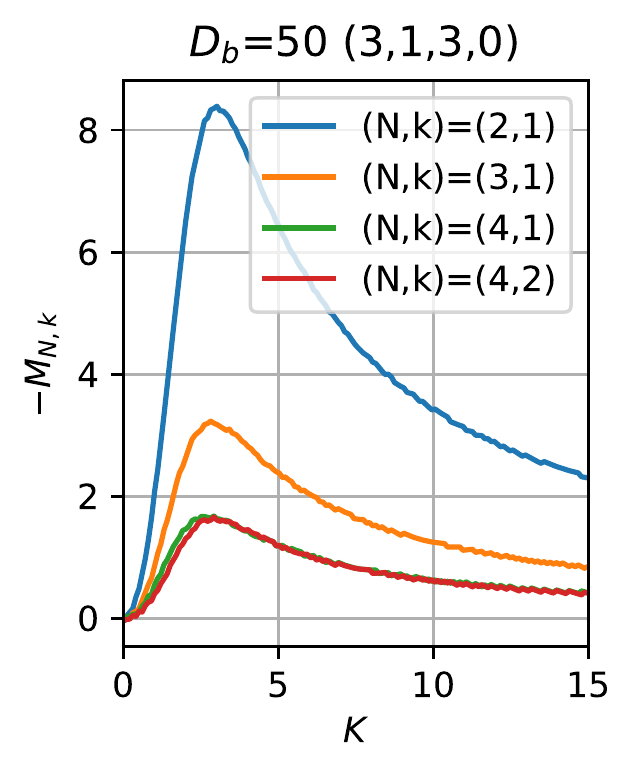}
	\end{subfigure}
	~
	\begin{subfigure}{0.23\textwidth}
		\centering
		\includegraphics[height=2.25in]{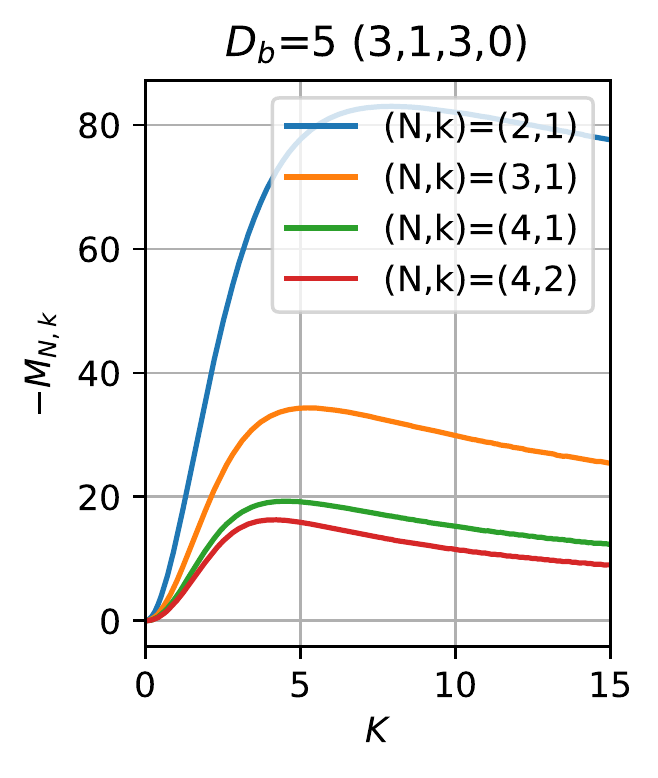}
	\end{subfigure}
	~
	\begin{subfigure}{0.23\textwidth}
		\centering
		\includegraphics[height=2.25in]{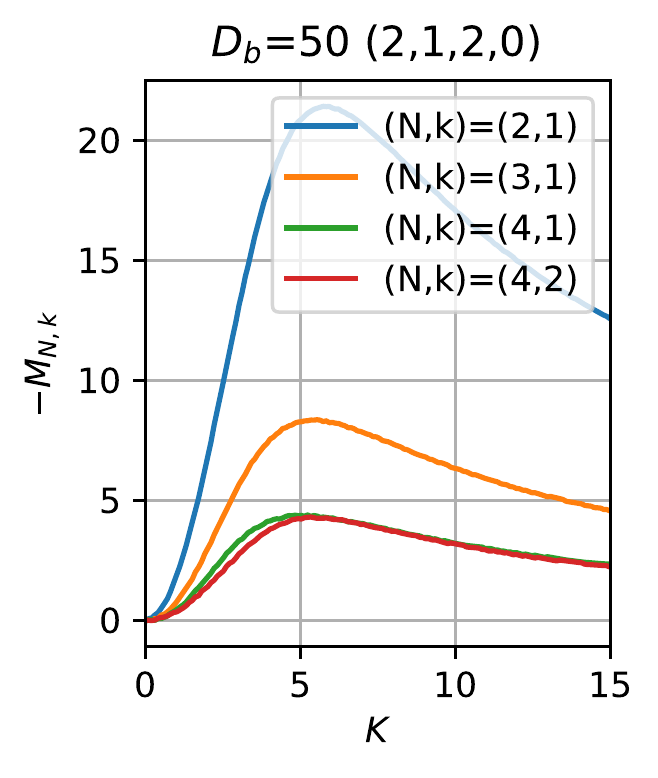}
	\end{subfigure}
	~
	\begin{subfigure}{0.23\textwidth}
		\centering
		\includegraphics[height=2.25in]{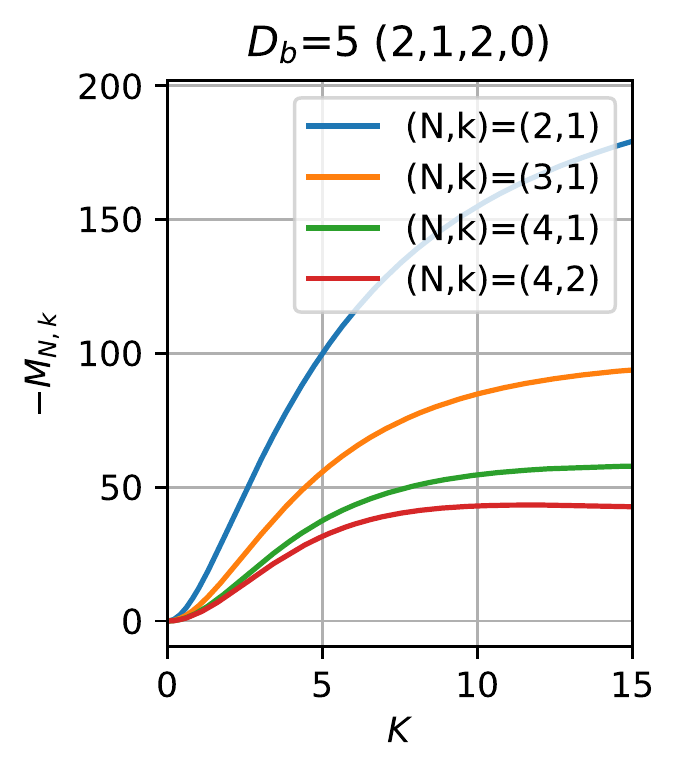}
	\end{subfigure}
	\\
	\begin{subfigure}{0.23\textwidth}
		\centering
		\includegraphics[height=2.25in]{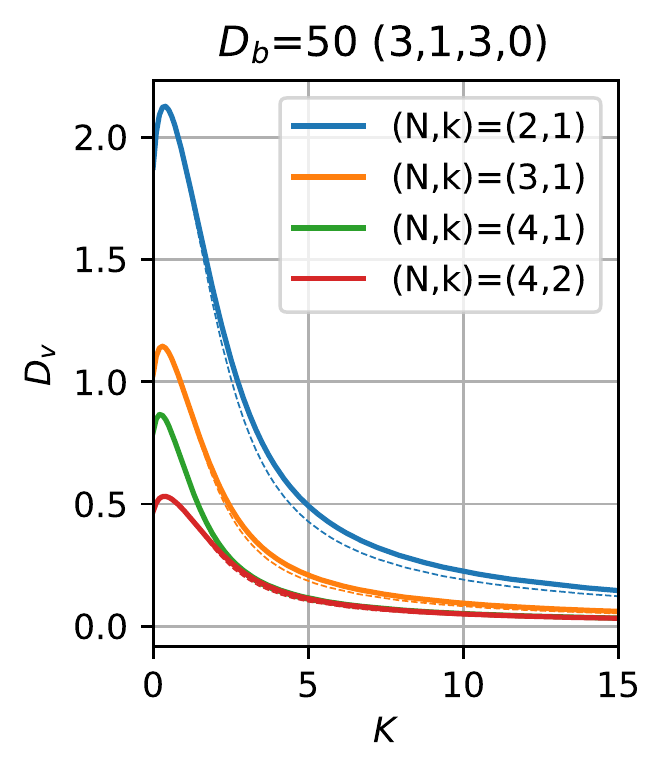}
	\end{subfigure}
	~
	\begin{subfigure}{0.23\textwidth}
		\centering
		\includegraphics[height=2.25in]{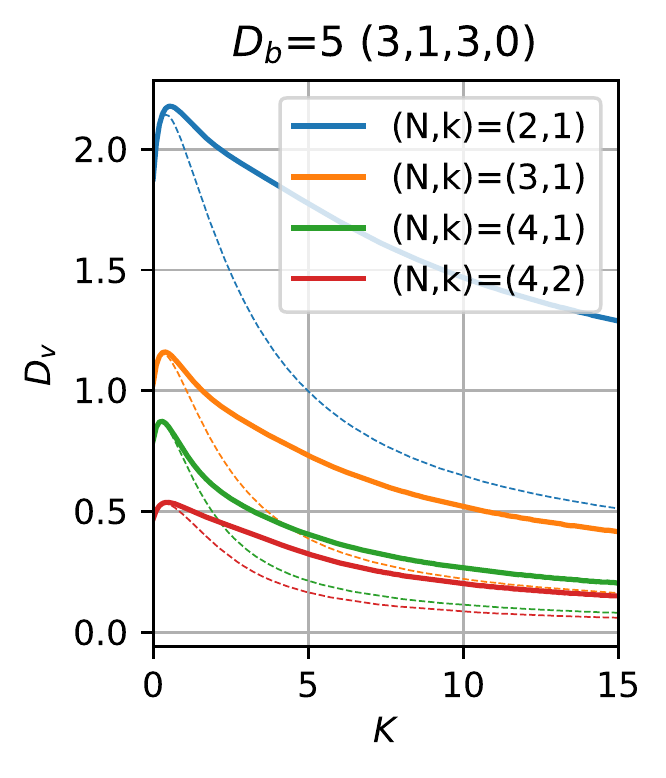}
	\end{subfigure}
	~
	\begin{subfigure}{0.23\textwidth}
		\centering
		\includegraphics[height=2.25in]{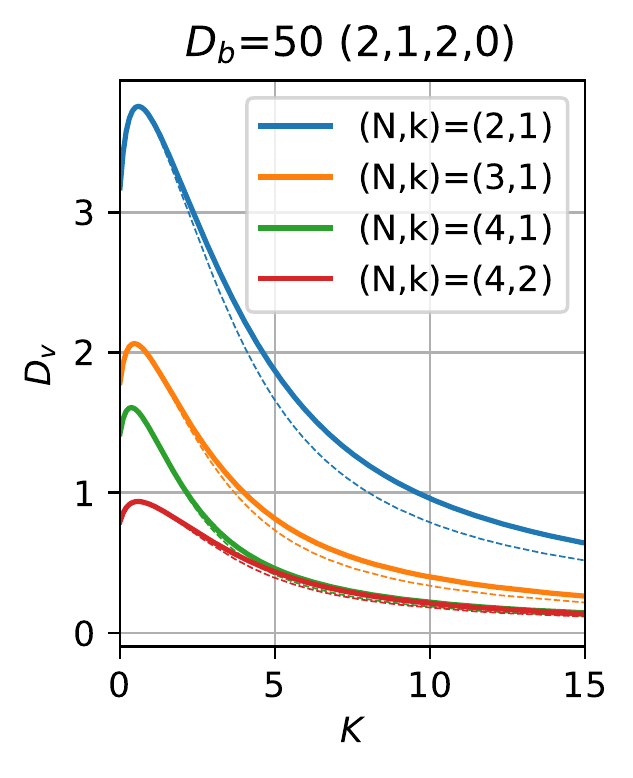}
	\end{subfigure}
	~
	\begin{subfigure}{0.23\textwidth}
		\centering
		\includegraphics[height=2.25in]{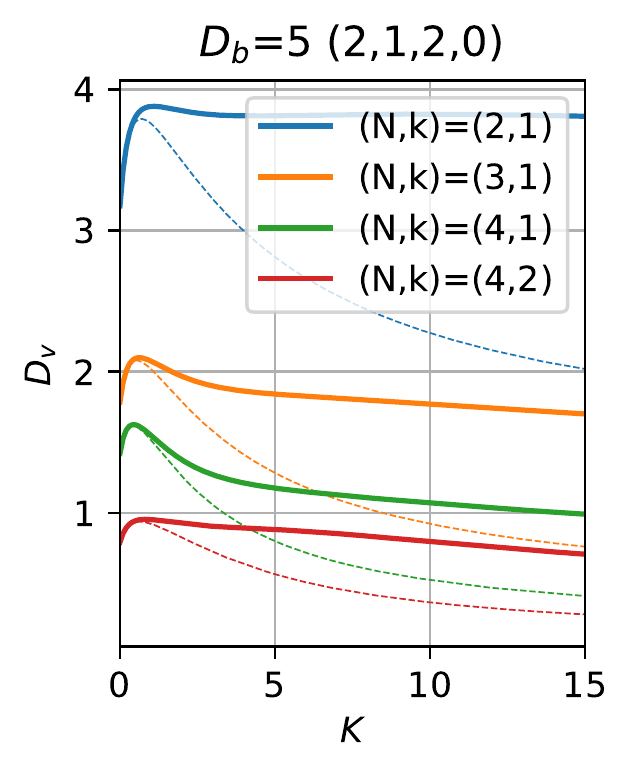}
	\end{subfigure}
	\caption{The effect of boundary perturbations on the asynchronous stability of symmetric $N$-spike patterns for the unit disk. The top row shows the multiplier $M_{N,k}$, defined in \eqref{eq:D_vk1}, as a function of $K$ while the bottom row shows the leading order correction to the asynchronous instability threshold, with the dashed line indicating the unperturbed threshold. Each column correspond to a choice of $D_b = 50$ or $D_b = 5$ with Gierer-Meinhardt exponents of $(p,q,m,s) = (3,1,3,0)$ or $(p,q,m,s) = (2,1,2,0)$. In the second row the boundary perturbation has parameters $\xi = 1$ (indicating an outward bulge at the spike locations), and $\delta = 0.01$.}\label{fig:boundary_perturbations}
\end{figure}

The goal of this section is to calculate the leading order correction
to the asynchronous instability thresholds for a perturbed
disk. Specifically we consider the domain
$$
\Omega_\delta \equiv \{(r,\theta)\,|\,0\leq r< R+ \delta h(\theta)\,,
\quad 0\leq \theta < 2\pi \}\,,
$$
where $h(\theta)$ is a smooth ${\mathcal O}(1)$ function with
a Fourier series
$h(\theta) = \sum_{n=-\infty}^{\infty} h_n e^{i n \theta}$. Although our
final results will be restricted to the specific form
\begin{equation}\label{eq:boundary_perturbation_periodic}
  h(\theta) = 2 R \xi \cos(N\theta) = R \xi e^{i N\theta} + R \xi 
    e^{-iN\theta}\,,
\end{equation}
where $\xi$ is a parameter, there is no additional difficulty in
considering a general Fourier series in the analysis below. However, we remark that in using the general Fourier series given above we must impose appropriate symmetry conditions on $h(\theta)$ so that the symmetric $N$-spike pattern construction, and in particular the resulting NLEP \eqref{eq:NLEP_symmetric}, remain valid.
Our main goal is to determine a two-term asymptotic expansion in powers
of $\delta$ for each asynchronous instability threshold in the form
$$
D_v \sim D_{vk0}^{\star}(D_b,K,R) + D_{vk1}^{\star}(D_b,K,R) \delta +
{\mathcal O}(\delta^2)\,,
$$
such that a zero-eigenvalue crossing is maintained to at least second order,
i.e.~for which $\lambda = {\mathcal O}(\delta^2)$.

Recall that the only component of the asynchronous NLEP
\eqref{eq:NLEP_symmetric} that depends on the problem geometry is the
NLEP multiplier $\chi_k(\lambda)$. To study the effect of boundary
perturbations, it therefore suffices to calculate the leading order
corrections to the corresponding membrane Green's function
satisfying (\ref{eq:g_lam_memb}). Furthermore, we note that since
we are only interested in a first order expansion, whereas
$\lambda={\mathcal O}(\delta^2)$, there is no loss in validity
assuming that $\lambda$ is an independent parameter that we ultimately
set to zero. Upon expanding
$D_v = D_{v0}\bigl(1+\tfrac{D_{v1}}{D_{v0}}\delta\bigr)$, a two-term expansion
for the perturbed membrane Green's function is given by (see Appendix
\ref{app:boundary_perturbation_details})
$$
G_{\partial\Omega}^\lambda(\theta,\theta_0)\sim G_{\partial\Omega
  0}^\lambda(\theta,\theta_0) + G_{\partial\Omega 1}^\lambda
(\theta,\theta_0) \delta + {\mathcal O}(\delta^2),
$$
where $G_{\partial\Omega 0}^\lambda$ is the membrane Green's function
for the unperturbed disk calculated previously in
\eqref{eq:Disk_G_Membrane_Series} and the leading-order correction is
\begin{equation}\label{eq:perturbed_disk_membrane_correction}
  G_{\partial\Omega 1}^\lambda(\theta,\theta_0) = -
  \tfrac{h(\theta_0)}{R} G_{\partial\Omega 0}^\lambda(\theta,\theta_0) +
  \tfrac{1}{2\pi R}\sum_{n=-\infty}^\infty \sum_{k=-\infty}^\infty
  \hat{g}_{n,k}^\lambda h_{n-k}g_{k}^\lambda g_{n}^\lambda e^{in\theta - ik\theta_0}
  - \tfrac{D_{v1}}{2\pi R^3}\sum_{n=-\infty}^{\infty} n^2 (g_{n}^\lambda)^2
  e^{in(\theta-\theta_0)}\,.
\end{equation}
In this expression the coefficients $\hat{g}_{n,k}^\lambda$ are
given by
\begin{equation}
\hat{g}_{n,k}^\lambda =  \tfrac{D_{v0}}{R^3}k(n+k) + K^2
a_{k}^\lambda \bigl(\hat{a}_{n,k}^\lambda + P_{k}^{\prime}(R)\bigr) \,,
\end{equation}
where $g_{k}^\lambda$, $a_{k}^{\lambda}$, and $\hat{a}_{n,k}^\lambda$ are
defined in (\ref{eq:gnlambda}), (\ref{eq:Disk_G_Bulk_Series}), and
(\ref{appb:coeff}), respectively.

Restricting our attention to perturbations of the form
\eqref{eq:boundary_perturbation_periodic}, and considering a symmetric
$N$-spike pattern with spikes centered at
$\theta_j = \tfrac{2\pi (j-1)}{N}$ for $j=1,...,N$, we deduce from
(\ref{eq:perturbed_disk_membrane_correction}) that
\begin{equation}
\begin{split}
  G_{\partial\Omega 1}^\lambda(\theta,\theta_j) = -2\xi G_{\partial\Omega 0}^\lambda
 (\theta,\theta_j) + \frac{\xi}{2\pi}\sum_{n=-\infty}^{\infty}
 \bigl\{ \hat{g}_{n,n+N}^\lambda g_{0,n+N}^\lambda +
 \hat{g}_{n,n-N}^\lambda g_{0,n-N}^\lambda \} g_{0n}^\lambda e^{in(\theta-\theta_j)}\\
 - \tfrac{D_{v1}}{2\pi R^3}\sum_{n=-\infty}^{\infty} n^2 (g_{0n}^\lambda)^2
        e^{in(\theta-\theta_j)}\,.
\end{split}
\end{equation}
Note that by symmetry the consistency and balance equations continue
to hold for a symmetric $N$ spike pattern. Furthermore the perturbed
Green's matrix remains circulant, and therefore its eigenvalues can be
read off as
$$
\mu_k(\lambda) = \sum_{j=0}^{N-1}G_{\partial\Omega}^\lambda\bigl(
\tfrac{2\pi}{N}j, 0 \bigr) e^{\frac{2\pi i j k}{N}} \sim
\mu_{k0}(\lambda) + \delta\biggl\{ -2\xi \mu_{k0}(\lambda) + \xi
\mu_{k11}(\lambda) + D_{v1}\mu_{k12}(\lambda) \biggr\} \equiv
\mu_{k0}(\lambda) + \delta \mu_{k1}(\lambda),
$$
where
\begin{subequations}
\begin{align}
 \mu_{k0}(\lambda)  & = \tfrac{N}{2\pi R}\sum_{n=-\infty}^\infty g_{nN-k}^\lambda\,,\\
  \mu_{k11}(\lambda) & = \tfrac{N}{2\pi}\sum_{n=-\infty}^\infty
         \bigl\{ \hat{g}_{nN-k,(n+1)N-k}^\lambda g_{(n+1)N-k}^\lambda +
 \hat{g}_{nN-k,(n-1)N-k}^\lambda g_{(n-1)N-k}^\lambda  \bigr\}g_{nN-k}^\lambda\,, \\
  \mu_{k12}(\lambda) & = -\tfrac{N}{2\pi R^3}
                       \sum_{n=-\infty}^{\infty} (nN-k)^2 (g_{nN-k}^\lambda)^2\,.
\end{align}
\end{subequations}
      
Finally, upon setting $\lambda=0$ in the zero-eigenvalue crossing
condition $\mathcal{A}_k(0)=\left[\chi_k(0)\right]^{-1}-{mq/(p-1)}$
for the asynchronous modes $k=1,\ldots,N-1$ (see
(\ref{eq:a_zero_eigenvalue})), and noting
$\chi_k(0)={\mu_k(0)/\mu_0(0)}$ from (\ref{eq:NLEP_symmetric}), we
obtain that
\begin{equation}\label{bc:delta}
   \frac{\mu_{00}(0) + \delta \left[-2\xi\mu_{00}(0) + \xi \mu_{011}(0) +
    D_{v1}\mu_{012}(0)\right]}{\mu_{k0}(\lambda) +
    \delta \left[-2\xi\mu_{k0}(\lambda) + \xi \mu_{k11}(\lambda) +
    D_{v1}\mu_{k12}(\lambda)\right]} - \frac{mq}{p-1} = 0 \,, \qquad
  \mbox{for} \quad k=1,\ldots,N-1 \,.
\end{equation}
The leading-order problem is satisfied by the previously determined
threshold $D_{v0}=D_{vk0}^{\star}(K,D_b,R)$. On the other hand, by
expanding (\ref{bc:delta}) in powers of $\delta$, we obtain from equating
${\mathcal O}(\delta)$ terms in this expansion that
$$
\xi \bigl( \mu_{011}(0) - \tfrac{mq}{p-1}\mu_{k11}(0) \bigr) + D_{v1}
\bigl( \mu_{012}(0) - \tfrac{mq}{p-1}\mu_{k12}(0) \bigr) = 0 \,.
$$
Upon solving for $D_{v1}=D_{vk1}^{\star}(K,D_b,R)$ in this expression,
we conclude that
\begin{equation}\label{eq:D_vk1}
  D_{vk1}^{\star} = - M_{N,k} \xi\,,\qquad \mbox{where} \quad
  M_{N,k} \equiv \frac{\mu_{011}(0) -
    \tfrac{mq}{p-1}\mu_{k11}(0)}{\mu_{012}(0) - \tfrac{mq}{p-1}\mu_{k12}(0)}\,.
\end{equation}
Therefore, the sign and magnitude of the multiplier $M_{N,k}$
determines how the asynchronous instability threshold changes when the
boundary is perturbed by a single Fourier mode of the form
\eqref{eq:boundary_perturbation_periodic}.

Figure \ref{fig:boundary_perturbations} illustrates the effect of boundary perturbations of the form \eqref{eq:boundary_perturbation_periodic} by plotting the multiplier $-M_{N,k}$ in the top row, and the leading order corrected asynchronous threshold $D_v \sim D_{vk0}^\star + D_{vk1}^\star \delta$ in the bottom row. Note that the (positive) maximums of $h(\theta)$ correspond with the quasi-equilibrium spike locations $\theta_j$ for each $j=1,...,N$. From \eqref{eq:D_vk1} we therefore conclude that positive values of $-M_{N,k}$ indicate an increase in stability when spike locations bulge out ($\xi > 0$), and a decrease in stability otherwise. The results of Figure \ref{fig:boundary_perturbations} thus indicate that an outward bulge at the location of each spike in a symmetric $N$-spike pattern leads to an improvement in stability of the pattern with respect to asynchronous instabilities. In addition, the magnitude of $-M_{N,k}$ shows that this stabilizing effect is most pronounced at some finite value of $K$ corresponding to a maximum of $-M_{N,k}$. Furthermore, comparing the $D_b=50$ and $D_b=5$ plots we see that decreasing the bulk diffusivity further accentuates the effect of boundary perturbations as is clear from the relative magnitude of $-M_{N,k}$ in these two cases. These numerical observations lead us to propose the following numerically supported proposition.
\begin{theorem} Consider a symmetric $N$-spike pattern for the Gierer-Meinhardt system \eqref{eq:bsrde2d} on the unit disk. Then a domain perturbation of the form \eqref{eq:boundary_perturbation_periodic}, which creates an outward bulge at each spike location, will increase the asynchronous instability threshold of the symmetric $N$-spike pattern.
\end{theorem}

\setcounter{equation}{0}
\setcounter{section}{4}
\section{Discussion}\label{sec:discussion}
\paragraph{}
We have introduced a coupled bulk-membrane PDE model in
which a scalar linear 2-D bulk diffusion process is coupled through
a linear Robin boundary condition to a two-component 1-D RD system
with Gierer-Meinhardt (nonlinear) reaction kinetics defined on the
domain boundary. For this coupled bulk-membrane PDE model, in the
singularly perturbed limit of a long-range inhibition and
short-range activation for the membrane-bound species, we have
studied the existence and linear stability of localized steady-state
multi-spike patterns defined on the membrane. Our primary goal was
to study how the bulk diffusion process and the bulk-membrane
coupling modifies the well-known linear stability properties of
steady-state spike patterns for the 1-D Gierer-Meinhardt model in
the absence of coupling.

By using a singular perturbation analysis on our coupled model
(\ref{eq:bsrde2d}) we first derived a nonlinear algebraic system
(\ref{eq:equi_alg_system}) characterizing the locations and heights
of steady-state multi-spike patterns on the membrane. Then we
derived a new class of NLEPs (nonlocal eigenvalue problems)
characterizing the linear stability on ${\mathcal O}(1)$ time-scales
of these steady-state patterns. In this NLEP, the multiplier of the
nonlocal term is determined in terms of the model parameters
together with a new coupled nonlocal Green's function problem. More
specifically, a novel feature of our steady-state and linear
stability analysis is the appearance of a nonlocal 1-D membrane
Green's function $G_{\partial\Omega}^\lambda(\sigma,\zeta)$ (see
(\ref{eq:g_lam_memb})), satisfying
$$
D_v\partial_\sigma^2 G_{\partial\Omega}^\lambda(\sigma,\zeta) -
(1+ K + \tau_s \lambda) G_{\partial\Omega}^\lambda(\sigma,\zeta) +
K^2\int_0^L G_\Omega^\lambda(\sigma,\tilde{\sigma})G_{\partial\Omega}^\lambda
(\tilde{\sigma},\zeta)\, d\tilde{\sigma} = - \delta(\sigma-\zeta)\,,\qquad
0<\sigma,\zeta< L\,,
$$
which is coupled to a 2-D bulk Green's function $G_\Omega^\lambda$ satisfying
(see (\ref{eq:g_lam_surf}))
$$
D_b\Delta G_\Omega^\lambda - (1+\tau_b\lambda) G_\Omega^\lambda =
0\,,\quad \mbox{in} \,\,\,\, \Omega\,;\qquad D_b\partial_n
G_\Omega^\lambda + K G_\Omega^\lambda =
\delta_{\partial\Omega}(x-x_0)\,,\quad \mbox{on} \,\,\,\,
\partial\Omega\,.
$$
Recall (\ref{eq:bsrde2d}) for the description of all the model parameters
including, the time constants $\tau_s$ and $\tau_b$, the diffusivities
$D_v$ and $D_b$, and the coupling constant $K$.

To proceed with a more explicit linear stability theory we
restricted our analysis to symmetric multi-spike patterns, which are
characterized by equidistantly (in arc-length) separated spikes of
equal height, for two analytically tractable cases. The
first case is when $\Omega$ is a disk of radius $R$, while the
second case is when the bulk is well mixed (i.e.~$D_b\gg 1$). For
these two specific cases, we obtained analytical expressions for
the relevant Green's function, and consequently the NLEP multipliers,
in the form of infinite series for the disk and explicit formulae for
the well-mixed limit. Parameter thresholds for two distinct forms of
linear instabilities, corresponding to either synchronous or
asynchronous perturbations of the heights of the steady-state
spikes, were then computed from the NLEP. Our results indicate a
non-monotonic dependence on the bulk-membrane coupling strength $K$
for both modes of instability, together with an intricate
relationship between the time-scale and coupling parameters for the
synchronous instabilities. Specifically, for the asynchronous
instability modes the coupling has the effect of improving stability
for smaller values of $K$ by raising the instability threshold for
$D_v$, but reducing the range of stability for larger
values of $K$. This effect is amplified in the synchronous case where
for certain choices of $\tau_s$ a small region in the $K$
versus $\tau_b$ parameter space can be found for which no
instabilities exist (see Figure
\ref{fig:M_with_Dv_colormap}). Finally, by using a Finite Element /
Finite Difference mixed IMEX scheme, we confirmed our linear stability
thresholds with full numerical PDE simulations.

We conclude the discussion by highlighting some open problems and
directions for future research. Firstly, for our coupled model,
additional work is required to calculate and study the linear
stability of asymmetric spike patterns. Secondly, we have neglected
the role of small $O(\varepsilon^2)$ eigenvalues corresponding to weak
drift instabilities, which can be studied either through a more
detailed asymptotic analysis or by deriving and analyzing a
corresponding slow spike-dynamics ODE system. Thirdly, the numerical
evidence provided by our PDE simulations suggests that, when $N\geq 2$ in the absence of competition instabilities, the Hopf bifurcation is
supercritical, and leads to the emergence of a small amplitude
time-periodic solution near the bifurcation point. The numerical
evidence also suggests that competition instabilities are
subcritical, and result in the annihilation of one or more spikes in
a multi-spike pattern. It would be worthwhile to analytically
establish these conjectured branching behaviors from a weakly
nonlinear analysis that is valid either near a Hopf bifurcation
point or near a zero-eigenvalue crossing.

Finally, there are several directions for extending our
model and applying a similar methodology. One direction would be
to analyze similar problems in higher space dimensions, such as a
3-D linear bulk diffusion process coupled to a nonlinear RD system
on a 2-D surface. A further direction would be to consider a
two-component bulk diffusion process, with nonlinear bulk
kinetics.  For this more complicated model it would be interesting
to study the interplay between 1-D membrane-bound and 2-D
bulk-bound localized patterns. Additionally it would be
instructive to asymptotically construct and analyze the localized
patterns observed in the numerical study of Madzvamuse et.\@
al. \cite{madzvamuse_2015,madzvamuse_2016} as well those of
R{\"a}tz et.\@ al.\@ \cite{ratz_2012,ratz_2013,ratz_2014}.

\addcontentsline{toc}{section}{References}
\bibliographystyle{abbrv}
\bibliography{paper-2d_bibliography_new}

\begin{thebibliography}{10}

\bibitem{ascher_1995}
U.~M. Ascher, S.~J. Ruuth, and B.~T.~R. Wetton.
\newblock Implicit-explicit methods for time-dependent partial differential
  equations.
\newblock {\em SIAM J. Numer. Anal.}, 32(3):797--823, 1995.

\bibitem{madz2018}
D.~Cusseddu, L.~Edelstein-Keshet, J.~A. Mackenzie, S.~Portet, and
  A.~Madzvamuse.
\newblock A coupled bulk-surface model for cell polarisation.
\newblock {\em J. Theor. Bio.}, 2018.

\bibitem{doelgm}
A.~Doelman, R.~A. Gardner, and T.~Kaper.
\newblock Large stable pulse solutions in reaction-diffusion equations.
\newblock {\em Indiana U. Math. Journ.}, 50(1):443--507, 2001.

\bibitem{vpd}
A.~Doelman, R.~A. Gardner, and T.~Kaper.
\newblock Stability of spatially periodic pulse patterns in a class of
  singularly perturbed reaction-diffusion equations.
\newblock {\em Indiana U. Math. Journ.}, 54(5):1219--1301, 2005.

\bibitem{elliott}
C.~M. Elliott, T.~Ranner, and C.~Venkataraman.
\newblock Coupled bulk-surface free boundary problems arising from a
  mathematical model of receptor-ligand dynamics.
\newblock {\em SIAM J. Math. Anal.}, 49(1):360--397, 2017.

\bibitem{gierer_1972}
A.~Gierer and H.~Meinhardt.
\newblock A theory of biological pattern formation.
\newblock {\em Kybernetik}, 12(1):30--39, Dec 1972.

\bibitem{gomez2007}
A.~Gomez-Marin, J.~Garcia-Ojalvo, and J.~M. Sancho.
\newblock Self-sustained spatiotemporal oscillations induced by membrane-bulk
  coupling.
\newblock {\em Phys. Rev. Lett.}, 98:168303, Apr 2007.

\bibitem{iron_2001}
D.~Iron, M.~J. Ward, and J.~Wei.
\newblock The stability of spike solutions to the one-dimensional
  {G}ierer-{M}einhardt model.
\newblock {\em Phys. D}, 150(1-2):25--62, 2001.

\bibitem{levine_2005}
H.~Levine and W.-J. Rappel.
\newblock Membrane-bound {T}uring patterns.
\newblock {\em Phys. Rev. E (3)}, 72(6):061912, 5, 2005.

\bibitem{macdonald2013}
C.~B. Macdonald, B.~Merriman, and S.~J. Ruuth.
\newblock Simple computation of reaction-diffusion processes on point clouds.
\newblock {\em Proc. Natl. Acad. Sci. USA}, 110(23):9209--9214, 2013.

\bibitem{madzvamuse_2016}
A.~Madzvamuse and A.~H. Chung.
\newblock The bulk-surface finite element method for reaction diffusion systems
  on stationary volumes.
\newblock {\em Finite Elements in Analysis and Design}, 108:9--21, 2016.

\bibitem{madzvamuse_2015}
A.~Madzvamuse, A.~H.~W. Chung, and C.~Venkataraman.
\newblock Stability analysis and simulations of coupled bulk-surface
  reaction-diffusion systems.
\newblock {\em Proc. A.}, 471(2175):20140546, 18, 2015.

\bibitem{nec}
Y.~Nec and M.~J. Ward.
\newblock An explicitly solvable nonlocal eigenvalue problem and the stability
  of a spike for a class of reaction-diffusion system.
\newblock {\em Math. Mod. Nat. Phen.}, 8(2):55--87, 2013.

\bibitem{nishiura}
Y.~Nishiura.
\newblock {\em Far-from Equilibrium dynamics: {T}ranslations of mathematical
  monographs}, volume 209.
\newblock AMS Publications, {P}rovidence, {R}hode {I}sland, 2002.

\bibitem{ratz_2012}
A.~R\"atz and M.~R\"oger.
\newblock Turing instabilities in a mathematical model for signaling networks.
\newblock {\em J. Math. Biol.}, 65(6-7):1215--1244, 2012.

\bibitem{ratz_2013}
A.~R\"atz and M.~R\"oger.
\newblock Erratum to: {T}uring instabilities in a mathematical model for
  signaling networks [mr2993944].
\newblock {\em J. Math. Biol.}, 66(1-2):421--422, 2013.

\bibitem{ratz_2014}
A.~R\"atz and M.~R\"oger.
\newblock Symmetry breaking in a bulk-surface reaction-diffusion model for
  signalling networks.
\newblock {\em Nonlinearity}, 27(8):1805--1827, 2014.

\bibitem{ruuth_1995}
S.~J. Ruuth.
\newblock Implicit-explicit methods for reaction-diffusion problems in pattern
  formation.
\newblock {\em J. Math. Biol.}, 34(2):148--176, 1995.

\bibitem{turing_1952}
A.~M. Turing.
\newblock The chemical basis of morphogenesis.
\newblock {\em Philos. Trans. Roy. Soc. London Ser. B}, 237(641):37--72, 1952.

\bibitem{ward_n_2003}
M.~J. Ward and J.~Wei.
\newblock Hopf bifurcation and oscillatory instabilities of spike solutions for
  the one-dimensional {G}ierer-{M}einhardt model.
\newblock {\em J. Nonlinear Science}, 13(2):209--264, 2003.

\bibitem{ward_2003}
M.~J. Ward and J.~Wei.
\newblock Hopf bifurcation of spike solutions for the shadow
  {G}ierer-{M}einhardt model.
\newblock {\em European J. Appl. Math.}, 14(6):677--711, 2003.

\bibitem{wei_1999}
J.~Wei.
\newblock On single interior spike solutions of the {G}ierer-{M}einhardt
  system: uniqueness and spectrum estimates.
\newblock {\em European J. Appl. Math.}, 10(4):353--378, 1999.

\bibitem{wei-book}
J.~Wei and M.~Winter.
\newblock {\em Mathematial aspects of pattern formation in biological systems},
  volume 189.
\newblock Applied Mathematical Sciences Series, Springer, 2014.

\end{thebibliography}

\appendix

\renewcommand{\theequation}{\Alph{section}.\arabic{equation}}
\setcounter{equation}{0}
\section{Green's Functions in the Well-Mixed Limit and for the Disk}\label{app:Greens_Functions}

In this appendix we collect all the relevant Green's functions and
indicate some of their key properties. We focus specifically on the
uncoupled ($K=0$) Green's function, the well-mixed Green's function
($D_b\rightarrow\infty$), and the disk Green's function
($\Omega = B_R(0)$). For the first two cases explicit formulae can be
derived, while for the final case we must rely on a Fourier
  series expansion representation.

\subsection{Uncoupled Membrane Green's Function}

When the bulk and membrane are uncoupled there is no direct dependence
on the bulk Green's function. Indeed the only relevant geometric
dependent parameter becomes the perimeter of the domain
$L=|\partial\Omega|$. Thus, $\Omega$ may be an arbitrary bounded and
simply connected subset of $\mathbb{R}^2$. We define the uncoupled
Green's function $\Gamma^\lambda$ as the solution to
\begin{equation}\label{eq:Uncoupled_Membrane_PDE}
  D_v\partial_\sigma^2\Gamma - \mu^2 \Gamma = -\delta(\sigma-\zeta)\,,\qquad
  0<\sigma<L\,,\qquad \Gamma\quad \mbox{is}\,\, L\text{-periodic} \,.
\end{equation}
The solution to (\ref{eq:Uncoupled_Membrane_PDE}) is readily
calculated as
\begin{equation}\label{eq:Uncoupled_G_Membrane}
  \Gamma(\sigma,\zeta) = \frac{1}{2\sqrt{D_v}\mu }\coth\biggl(\frac{\mu L}
  {2\sqrt{D_v}}\biggr)\cosh\biggl(\frac{\mu}{\sqrt{D_v}}|\sigma-\zeta|\biggr)
  - \frac{1}{2\sqrt{D_v}\mu}\sinh\biggl(\frac{\mu}{\sqrt{D_v}}
  |\sigma-\zeta|\biggr).
\end{equation}

\subsection{Bulk and Membrane Green's functions in the Well-Mixed Limit}

We now derive the leading order expression for the membrane Green's
function, defined by \eqref{eq:g_lam_memb}, when $D_b\rightarrow\infty$. To
leading order $G_{\Omega}^\lambda$, defined by \eqref{eq:g_lam_surf},
is constant and from the divergence theorem we find
\begin{equation}\label{app:g0}
G_{\Omega}^\lambda(\sigma,\tilde{\sigma}) \sim G_{\Omega 0}^\lambda =
\frac{1}{KL + \mu_{b\lambda}^2A} = \frac{\beta/K}{\mu_{b\lambda}^2 +
  \beta}\frac{1}{L}\,,\qquad \mbox{where} \quad \beta\equiv K\frac{L}{A}\,.
\end{equation}
Here $L \equiv |\partial\Omega|$ and $A \equiv |\Omega|$. The leading order
problem for the membrane Green's function in \eqref{eq:g_lam_memb}
is then
\begin{equation}\label{app:gmemb}
  D_v \partial_\sigma^2 G_{\partial\Omega}^\lambda - \mu_{s\lambda}^2
  G_{\partial\Omega}^\lambda + K^2 G_{\Omega 0}^\lambda\int_0^L G_{\partial\Omega }^\lambda
  (\tilde{\sigma};\zeta)\, d\tilde{\sigma} = -\delta(\sigma-\zeta)\,.
\end{equation}
Upon integrating this equation and using the periodic boundary
conditions we get
\begin{equation*}
  \int_0^L G_{\partial\Omega}^\lambda(\tilde{\sigma};\zeta)\, d\tilde{\sigma} =
  \frac{1}{\mu_{s\lambda}^2 - K^2 L G_{\Omega 0}^\lambda} =
  \left(\frac{1}{\mu_{s\lambda}^2(\mu_{b\lambda}^2+\beta) - K\beta}\right)
  \frac{1} {A G_{\Omega 0}^\lambda}\,,
\end{equation*}
where $G_{\Omega 0}^\lambda$ is defined in
(\ref{app:g0}). Therefore, from (\ref{app:gmemb}),
we find that $G_{\partial\Omega}^\lambda$ satisfies
\begin{equation*}
  D_v \partial_\sigma^2 G_{\partial\Omega}^\lambda - \mu_{s\lambda}^2
  G_{\partial\Omega}^\lambda  = -\delta(\sigma-\zeta) -
  \frac{K^2/A}{\mu_{s\lambda}^2(\mu_{b\lambda}^2+\beta) - K\beta}\,.
\end{equation*}
This problem is readily solved in terms of the uncoupled Green's function of
\eqref{eq:Uncoupled_G_Membrane} by defining
$$
\Gamma^\lambda(\sigma,\zeta) := \Gamma(\sigma,\zeta)\bigr|_{\mu = \mu_{s\lambda}}
\,,
$$
and then using the decomposition
\begin{equation}\label{eq:SL_G_Membrane_expansion}
  G_{\partial\Omega}^\lambda(\sigma,\zeta) = \Gamma^\lambda(\sigma,\zeta) +
  \frac{K^2}{\mu_{s\lambda}^2 A} \frac{1}{\mu_{s\lambda}^2(\mu_{b\lambda}^2+\beta)-
    K\beta} = \Gamma^\lambda(\sigma,\zeta) +
  \frac{\gamma_\lambda}{\mu_{s\lambda}^2}\,,\qquad \gamma_\lambda \equiv
  \frac{K^2/A}{\mu_{s\lambda}^2(\mu_{b\lambda}^2+\beta)-K\beta} \,.
\end{equation}

\subsection{Bulk and Membrane Green's functions in the Disk}

Here we consider the bulk Green's function defined by
(\ref{eq:g_lam_surf}). By using separation of variables (in polar
coordinates), and applying the boundary condition in
(\ref{eq:g_lam_surf}), we can write this Green's function as a
Fourier series
\begin{equation}\label{eq:Disk_G_Bulk_Series}
  G_{\Omega}^\lambda(r,\sigma,\tilde{\sigma}) = \frac{1}{2\pi R}
  \sum_{n=-\infty}^\infty a_{n}^\lambda P_n(r) e^{\tfrac{in}{R}(\sigma-\tilde{\sigma})}\,,
  \quad P_n(r) \equiv \frac{ I_{|n|}(\omega_{b\lambda} r) }
  { I_{|n|}(\omega_{b\lambda} R)}\,, \quad a_{n}^\lambda \equiv
  \frac{1}{D_b P_{n}^{\prime}(R) + K} \,, \quad
    \omega_{b\lambda} \equiv \frac{\mu_{b\lambda}}{\sqrt{D_b}} \,.
\end{equation}
We remark that the singularity lies on the
boundary and for this reason the radial dependence is given only in
terms of the modified Bessel functions of the first kind
$I_n(z)$. Similarly, we can represent the membrane Green's
function in \eqref{eq:g_lam_memb} for the disk in terms of the
Fourier series
\begin{equation}\label{eq:Disk_G_Membrane_Series}
  G_{\partial \Omega}^\lambda(\sigma,\sigma_0) = \frac{1}{2\pi R}
  \sum_{n=-\infty}^{\infty} g_{n}^\lambda e^{\tfrac{in}{R}(\sigma-\sigma_0)}\,,\qquad
  g_{n}^{\lambda} \equiv \frac{1}{D_{v}\tfrac{n^2}{R^2} + \mu_{s\lambda}^2 - K^2
    a_{n}^\lambda }\,.
\end{equation}

\subsection{A Useful Summation Formula for the Disk Green's Functions}

We make note here of a useful summation formula for numerically
evaluating the Green's function eigenvalues for the disk. By
integrating the function
$(\zeta^2 + z^2)^{-1}\cot\bigl(\tfrac{\pi}{N}(\zeta - k)\bigr)$ over
the contour enclosing $[-R,R]\times[-R,R]$, and then taking
  the limit $R\rightarrow\infty$, we obtain
\begin{equation}
  S(z;N,k) := \sum_{n=-\infty}^{\infty}\frac{1}{(nN+k)^2 + z^2} =
  \frac{\pi}{2Nz}\biggl[\coth\biggl(\frac{\pi}{N}(z+ik)\biggr) +
  \coth\biggl(\frac{\pi}{N}(z-ik)\biggr)\biggr]\,.
\end{equation}

\renewcommand{\theequation}{\Alph{section}.\arabic{equation}}
\setcounter{equation}{0}
\section{Derivation of Membrane Green's Function for the Perturbed Disk}\label{app:boundary_perturbation_details}

In this appendix we provide the details for calculating the
leading-order correction to the perturbed disk Green's function given
in \eqref{eq:perturbed_disk_membrane_correction}. Recall that the bulk
Green's function solves
\begin{equation}\label{appb:g}
D_b\Delta G_{\Omega}^\lambda - \mu_{b\lambda}^2 G_{\Omega}^\lambda =
0,\quad\text{in}\quad\Omega_\delta\,, \qquad D_b\partial_n
G_{\Omega}^\lambda + K G_{\Omega}^\lambda =
\delta_{\partial\Omega_\delta}(x-\tilde{x}),\quad
\text{on}\quad\partial\Omega_\delta\,.
\end{equation}
On the boundary $r = R + \delta h(\theta)$ of the perturbed disk we
calculate in terms of polar coordinates that
$$
\hat{n}(\theta) = \bigl[1 + \bigl(\tfrac{\delta h^{\prime}(\theta)}{R + \delta
  h(\theta)}\bigr)^2\bigr]^{-\frac{1}{2}}\bigl(\hat{e}_r -
\tfrac{\delta h^{\prime}(\theta)}{R+ \delta h(\theta)}\hat{e}_\theta
\bigr)\,,\quad \nabla = \hat{e}_r \partial_r +
\frac{1}{r}\hat{e}_\theta\partial_\theta\,,\quad
\delta_{\partial\Omega_\delta}(x-\tilde{x}) = \bigl[1 +
\bigl(\tfrac{\delta h^{\prime}(\theta)}{R + \delta
  h(\theta)}\bigr)^2\bigr]^{-\frac{1}{2}}\frac{\delta(\theta-\tilde{\theta})}{R
  + \delta h(\theta)}\,,
$$
which yields the following asymptotic behaviour as
$\delta \rightarrow 0$:
$$
\hat{n}(\theta) \sim \hat{e}_r -
\delta\frac{h^{\prime}(\theta)}{R}\hat{e}_\theta + {\mathcal O}(\delta^2)\,,\qquad
\delta_{\partial\Omega_\delta}(x-\tilde{x}) \sim
\frac{1}{R}\delta(\theta-\tilde{\theta}) -
\delta\frac{h(\theta)}{R^2}\delta(\theta-\tilde{\theta}) +
{\mathcal O}(\delta^2)\,.
$$
Next, for $\delta\ll 1$,  we seek a solution of the form
$$
G_{\Omega}^\lambda(r,\theta,\tilde{\theta} \sim G_{\Omega
  0}^\lambda(r,\theta,\tilde{\theta}) + G_{\Omega
  1}^\lambda(r,\theta,\tilde{\theta}) \delta + {\mathcal O}(\delta^2)\,.
$$
Upon substituting these expansions into (\ref{appb:g}), and
collecting powers of $\delta$, we obtain the following
zeroth-order and first-order problems:
\begin{subequations}
  \begin{align}
  & D_b \Delta G_{\Omega 0}^\lambda - \mu_{b\lambda}^2 G_{\Omega 0}^\lambda = 0\,,\quad
 \text{in}\quad\Omega_0\,,\qquad \mathcal{B}_0 G_{\Omega 0}^\lambda =
\frac{\delta(\theta-\tilde{\theta})}{R}\,,&\text{on}\quad \partial\Omega_0\,,&\\
    & D_b \Delta G_{\Omega 1}^\lambda - \mu_{b\lambda}^2 G_{\Omega 1}^\lambda  = 0\,,\quad
      \text{in}\quad\Omega_0\,,\qquad \mathcal{B}_0 G_{\Omega 1}^\lambda = -
      \frac{h(\theta)}{R}\frac{\delta(\theta-\tilde{\theta})}{R} -
      \mathcal{B}_1 G_{\Omega 0}^\lambda,&\text{on}\quad \partial\Omega_0\,,&
	\end{align}
\end{subequations}
where the boundary operators $\mathcal{B}_0$ and
$\mathcal{B}_1$ are defined by
$$
\mathcal{B}_0 \equiv D_b\partial_r + K\,\qquad \mathcal{B}_1 \equiv
D_b\biggl(h(\theta)\partial_r^2 -
\frac{h^{\prime}(\theta)}{R^2}\partial_\theta\biggr) + K h(\theta)\partial_r.
$$
The zeroth-order solution is the unperturbed disk bulk Green's function
given in \eqref{eq:Disk_G_Bulk_Series}. For the problem for
the leading order correction, we use linearity to decompose its solution
in the form
\begin{equation}\label{appb:g1}
G_{\Omega 1}^\lambda(r,\theta,\tilde{\theta}) =
-\frac{h(\tilde{\theta})}{R} G_{\Omega
  0}^\lambda(r,\theta,\tilde{\theta}) + \tilde{G}_{\Omega
  1}^\lambda(r,\theta,\tilde{\theta})\,,\qquad \tilde{G}_{\Omega
  1}^\lambda(r,\theta,\tilde{\theta}) = \frac{1}{2\pi
  R}\sum_{n=-\infty}^{\infty} \tilde{a}_{1n}^\lambda (\tilde{\theta})
P_{n}(r) e^{in\theta}\,,
\end{equation}
for some coefficients $\tilde{a}_{1n}^\lambda$ to be
found. To determine an expression for these coefficients, we first
multiply the boundary condition
$\mathcal{B}_0 \tilde{G}_{\Omega 1}^\lambda = - \mathcal{B}_1
G_{\Omega 0}^\lambda$ by $e^{-in\theta}$, and then integrate from $0$ to
$2\pi$. This gives
\begin{equation}\label{appb:ii}
\frac{1}{R}\bigl( D_b P_n^{\prime}(R) + K)
\tilde{a}_{1n}^\lambda(\tilde{\theta}) = - \int_0^{2\pi}
e^{-in\theta}\mathcal{B}_1 G_{\Omega 0}^\lambda \, d\theta \,.
\end{equation}
Then, by using the differential equation satisfied by
$G_{\Omega 0}^\lambda$ we calculate the right-hand side of this
expression as
\begin{equation}\label{app:g_int}
\begin{split}
  \int_0^{2\pi} e^{-in\theta}\mathcal{B}_1 G_{\Omega
    0}^\lambda(R,\theta,\tilde{\theta}) \, d\theta &= D_b
  \int_0^{2\pi}h(\theta)G_{\Omega 0
    rr}^\lambda(R,\theta,\tilde{\theta}) e^{-in\theta}\, d\theta -
  \tfrac{D_b}{R^2}\int_0^{2\pi}h^{\prime}(\theta)
  G_{\Omega 0\theta}^\lambda(R,\theta,\tilde{\theta})e^{-in\theta}\, d\theta \\
   & \qquad +K \int_0^{2\pi} h(\theta)G_{\Omega 0
    r}^\lambda(R,\theta,\tilde{\theta}) e^{-in\theta}\, d\theta\,.
\end{split}
\end{equation}
Next, we assume that the boundary perturbation $h(\theta)$ is
sufficiently smooth so that each of the following hold:
\begin{equation}\label{app:hexp}
h(\theta) = \sum_{n=-\infty}^{\infty} h_n e^{in\theta}\,,\quad
h^{\prime}(\theta) = i \sum_{n=-\infty}^{\infty} n h_n e^{in\theta}\,,\qquad
h^{\prime\prime}(\theta) = - \sum_{n=-\infty}^{\infty} n^2 h_n e^{in\theta}\,.
\end{equation}
This allows us to calculate the individual terms on the right-hand
side of (\ref{app:g_int}) as
\begin{align*}
  &\int_0^{2\pi}h(\theta)G_{\Omega 0 rr}^\lambda(R,\theta,\tilde{\theta})
    e^{-in\theta}d\theta = \frac{1}{R}\sum_{k=-\infty}^{\infty} P_k^{\prime\prime}(R)
    a_{k}^\lambda h_{n-k}e^{-ik\tilde{\theta}}\,,\\
  &\int_0^{2\pi}h^{\prime}(\theta) G_{\Omega 0\theta}^\lambda(R,\theta,\tilde{\theta})
    e^{-in\theta}d\theta = -\frac{1}{R}\sum_{k=-\infty}^{\infty}k(n-k)a_{k}^\lambda
    h_{n-k}e^{-ik\tilde{\theta}}\,,\\
  &\int_0^{2\pi} h(\theta)G_{\Omega 0 r}^\lambda(R,\theta,\tilde{\theta})
    e^{-in\theta}d\theta = \frac{1}{R}\sum_{k=-\infty}^{\infty} P_k^{\prime}(R)
    a_{k}^\lambda h_{n-k} e^{-ik\tilde{\theta}}\,,
\end{align*}
where $a_{k}^\lambda$ are the Fourier coefficients of the
leading-order Green's function, as defined in
(\ref{eq:Disk_G_Bulk_Series}). By substituting these relations into
(\ref{app:g_int}), and then using (\ref{appb:ii}), we determine the
coefficients as
\begin{equation}\label{appb:coeff}
\tilde{a}_{1n}^\lambda(\tilde{\theta}) = \sum_{k=-\infty}^{\infty}
\hat{a}_{n,k}^\lambda a_{k}^\lambda h_{n-k} e^{-ik\tilde{\theta}} \,, \qquad
\mbox{where} \qquad \hat{a}_{n,k}^\lambda \equiv
-\frac{ D_{b} P_k^{\prime\prime}(R) + K P_k^{\prime}(R) + \tfrac{D_b}{R^2}k(n-k)}
{D_b P_n^{\prime}(R) + K}\,.
\end{equation}
In (\ref{appb:coeff}), to calculate various derivatives of
$P_n(R)$, as defined in (\ref{eq:Disk_G_Bulk_Series}), we make
repeated use of the identity
$$
I_{n}^{\prime}(z) = \frac{n}{z} I_{n}(z) + I_{n+1}(z) \,,
$$
to readily derive that
$$
P_{n}^{\prime}(R) = \frac{|n|}{R} +
\omega_{b\lambda}\frac{I_{|n+1|}(\omega_{b\lambda}R)}{I_{|n|}(\omega_{b\lambda}R)}\,,
\qquad
P_{n}^{\prime\prime}(R) = \frac{|n|(|n|-1)}{R^2} +
\frac{2|n|+1}{R}\omega_{b\lambda}
\frac{I_{|n+1|}(\omega_{b\lambda}R)}{I_{|n|}(\omega_{b\lambda}R)} + \omega_{b\lambda}^2
\frac{I_{|n+2|}(\omega_{b\lambda}R)}{I_{|n|}(\omega_{b\lambda}R)}\,.
$$
This completes the derivation of the leading-order correction for the
bulk Green's function, defined in (\ref{appb:g1}).

Next, we derive a two-term approximation for the membrane
Green's function problem on the perturbed disk. This Green's function
satisfies
\begin{equation}
  D_v\partial_\sigma^2 G_{\partial\Omega}^\lambda(\sigma,\sigma_0) -
  \mu_{s\lambda}^2 G_{\partial\Omega }^\lambda(\sigma,\sigma_0) +
  K^2\int_0^{|\partial\Omega_\delta|} G_{\Omega}^\lambda(\sigma,\tilde{\sigma})
  G_{\partial\Omega}^\lambda(\tilde{\sigma},\sigma_0)\,d\tilde{\sigma} = -
  \delta(\sigma-\sigma_0),\qquad 0\leq\sigma<|\partial\Omega_\delta|.
\end{equation}
Repeated use of the chain rule to the arc-length formula
$$
\sigma(\theta) = \int_0^\theta \bigl( R + \delta h(\vartheta)
\bigr)\sqrt{ 1 + \biggl(\frac{\delta h^{\prime}(\vartheta)}{R + \delta
    h(\vartheta)}\biggr)^2 }\, d\vartheta \,,
$$
gives
$$
\partial_\sigma^2 = \frac{1}{(R+\delta h(\theta))^2 + (\delta
  h^{\prime}(\theta))^2 } \partial_\theta^2 -
\delta h^{\prime}(\theta)\frac{R +\delta
  h(\theta) + \delta h^{\prime\prime}(\theta)}{[(R+\delta h(\theta))^2 + (\delta
  h^{\prime}(\theta))^2]^2}\partial_\theta\,.
$$
Multiplying the membrane equation through by
$(R+\delta h(\theta))^2 + (\delta h^{\prime}(\theta))^2$, writing
$D_v = D_{v0}\bigl(1+\tfrac{D_{v1}}{D_{v0}}\delta\bigr)$, and then
dividing through by $R^2\bigl(1+\tfrac{D_{v1}}{D_{v0}}\delta\bigr)$, we
obtain the perturbed problem {\small
\begin{align*}
  & \tfrac{D_{v0}}{R^2} \partial_\theta^2 G_{\partial\Omega}^\lambda(\theta,\theta_0)
    - \tfrac{D_{v0}}{R^2}\delta h^{\prime}(\theta) \tfrac{R + \delta[h(\theta) +
  h^{\prime\prime}(\theta)]}{(R+\delta h(\theta))^2 +
    (\delta h^{\prime}(\theta))^2}\partial_\theta
    G_{\partial\Omega}^\lambda(\theta,\theta_0)
    - \tfrac{\mu_{s\lambda}^2}{R^2}\tfrac{(R+\delta h(\theta))^2 +
    (\delta h^{\prime}(\theta))^2}{1 + \tfrac{D_{v1}}{D_{v0}}\delta }
    G_{\partial\Omega}^\lambda(\theta,\theta_0) \\
  & + \tfrac{K^2}{R^2}\tfrac{(R+\delta h(\theta))^2 +
    (\delta h^{\prime}(\theta))^2 }{1+\tfrac{D_{v1}}{D_{v0}}\delta}\int_0^{2\pi}
    \bigl( G_{\Omega 0}^\lambda(R,\theta,\tilde{\theta}) + \delta
    G_{\Omega 1}^\lambda(R,\theta,\tilde{\theta}) + \delta h(\theta)
    G_{\Omega 0 r}^\lambda(R,\theta,\tilde{\theta}) \bigr)
    G_{\partial\Omega}^\lambda(\tilde{\theta},\theta_0)
    \sqrt{(R+\delta h(\tilde{\theta}))^2+
    (\delta h^{\prime}(\tilde{\theta}))^2}\, d\tilde{\theta} \\
  & \qquad = - \tfrac{1}{R^2}\tfrac{\sqrt{(R+\delta h(\theta))^2 +
    (\delta h^{\prime}(\theta))^2}}{1+\tfrac{D_{v1}}{D_{v0}}\delta} \delta
    (\theta - \theta_0)\,.
\end{align*}
}
To determine a two-term asymptotic solution to this problem, we
expand the membrane Green's function as
$$
G_{\partial\Omega}^\lambda(\theta,\theta_0)\sim G_{\partial\Omega
  0}^\lambda(\theta,\theta_0) + \delta G_{\partial\Omega 1}^\lambda
(\theta,\theta_0) + {\mathcal O}(\delta^2)\,.
$$
Upon substituting this expansion into the perturbed problem,
and collecting powers of $\delta$, we obtain the following
zeroth-order and first-order problems:
\begin{subequations}
\begin{equation}
  \mathcal{M}_{0} G_{\partial\Omega 0}^\lambda(\theta,\theta_0) =
  -\tfrac{1}{R}\delta(\theta-\theta_0)\,,\qquad \mathcal{M}_0
  G_{\partial\Omega 1}^\lambda(\theta,\theta_0) =
  -\bigl(\tfrac{h(\theta)}{R}-\tfrac{D_{v1}}{D_{v0}}\bigr)
  \tfrac{1}{R}\delta(\theta-\theta_0) -
  \mathcal{M}_1 G_{\partial\Omega 0}^\lambda(\theta,\theta_0)\,.
\end{equation}
Here we have defined the unperturbed membrane operator
$\mathcal{M}_0$ by
\begin{equation}
  \mathcal{M}_0 \psi(\theta,\theta_0) \equiv
  \tfrac{D_{v0}}{R^2}\partial_\theta^2\psi(\theta,\theta_0) -
  \mu_{s\lambda}^2 \psi(\theta,\theta_0) + K^2\int_0^{2\pi}
  G_{\Omega 0}^\lambda(R,\theta,\tilde{\theta})\psi(\tilde{\theta},\theta_0)
  R\, d\tilde{\theta}\,,
\end{equation}
and its leading-order correction $\mathcal{M}_1$ by
\begin{align}\label{eq:m1}
\begin{split}
  \mathcal{M}_1 \psi(\theta,\theta_0) \equiv & -
  \tfrac{D_{v0}}{R^3}h^{\prime}(\theta)\partial_\theta \psi(\theta,\theta_0) -
  \mu_{s\lambda}^2\bigl(\tfrac{2h(\theta)}{R} -
  \tfrac{D_{v1}}{D_{v0}}\bigr)\psi(\theta,\theta_0) \\
  & + K^2\bigl(\tfrac{2h(\theta)}{R} - \tfrac{D_{v1}}{D_{v0}}\bigr)\int_0^{2\pi}
  G_{\Omega 0}^\lambda(R,\theta,\tilde{\theta})\psi(\tilde{\theta},\theta_0)
  R \, d\tilde{\theta} +
  K^2\int_0^{2\pi} G_{\Omega 1}^\lambda(R,\theta,\tilde{\theta})
  \psi(\tilde{\theta},\theta_0)R\, d\tilde{\theta} \\
  & + K^2 h(\theta)\int_0^{2\pi} G_{\Omega 0 r}^\lambda(R,\theta,\tilde{\theta})
  \psi(\tilde{\theta},\theta_0)R \, d\tilde{\theta} + K^2\int_0^{2\pi}
  G_{\Omega 0}^\lambda(R,\theta,\tilde{\theta})\psi(\tilde{\theta},\theta_0)
  h(\tilde{\theta}) \, d\tilde{\theta} \,.
\end{split}
\end{align}
\end{subequations}

The zeroth-order solution is that of the unperturbed disk and is given
by \eqref{eq:Disk_G_Membrane_Series}. By linearity, we then seek
the solution for the leading order correction in the form
\begin{equation}\label{appb:memb_g1_1}
G_{\partial\Omega 1}^\lambda(\theta,\theta_0) =
\bigl(\tfrac{h(\theta_0)}{R} - \tfrac{D_{v1}}{D_{v0}} \bigr)
G_{\partial\Omega 0}^\lambda(\theta,\theta_0) +
\tilde{G}_{\partial\Omega 1}^\lambda(\theta,\theta_0)\,,
\end{equation}
where $\tilde{G}_{\partial\Omega 1}^\lambda(\theta,\phi)$ now satisfies
$$
\mathcal{M}_0 \tilde{G}_{\partial\Omega 1}^\lambda(\theta,\theta_0) =
- \mathcal{M}_1 G_{\partial\Omega 0}^\lambda(\theta,\theta_0)\,.
$$
We will represent the solution
$\tilde{G}_{\partial\Omega 1}^\lambda$ in terms of a Fourier series
as
\begin{equation}\label{appb:memb_g1_2}
\tilde{G}_{\partial\Omega 1}^\lambda(\theta,\theta_0) = \frac{1}{2\pi
  R}\sum_{n=-\infty}^{\infty} \tilde{g}_{1n}^\lambda(\theta_0)
e^{in\theta}\,,
\end{equation}
for some coefficients $\tilde{g}_{1n}^\lambda(\theta_0)$ to be
found. Similar to the calculation provided above for the perturbed
bulk Green's function, we obtain that
\begin{equation}\label{appb:gint}
\tilde{g}_{1n}^\lambda(\theta_0) = R g_{0n}^\lambda
\int_0^{2\pi}e^{-in\theta}\mathcal{M}_1 G_{\partial\Omega
  0}^\lambda(\theta,\theta_0)\, d\theta\,.
\end{equation}
By using (\ref{eq:m1}) we calculate the right-hand side of
this expression as
\begin{align}\label{appb:g_memb_int}
  \begin{split}
  \int_0^{2\pi} e^{-in\theta} \mathcal{M}_1 G_{\partial\Omega 0}^\lambda(\theta,\theta_0)
  d\theta & = -\tfrac{D_{v0}}{R^3} J_{1n}(\theta_0) -
  \tfrac{2\mu_{s\lambda}^2}{R} J_{2n}(\theta_0) + \tfrac{\mu_{s\lambda}^2 D_{v1}}
            {D_{v0}} J_{3n}(\theta_0) \\
  & \qquad + 2K^2 J_{4n}(\theta_0) - \tfrac{K^2 R D_{v1}}{D_{v0}} J_{5n}(\theta_0) +
  K^2 R J_{6n}(\theta_0) + K^2 R J_{7n}(\theta_0)\,,
  \end{split}
\end{align}
where the various integrals $J_{1n},\ldots,J_{7n}$ are defined by
\begin{align*}
&
 J_{1n}(\theta_0) = \int_0^{2\pi} h^{\prime}(\theta) G_{\partial\Omega 0 \theta}^\lambda
                 (\theta,\theta_0) e^{-in\theta}d\theta\,,\qquad
 J_{2n}(\theta_0) = \int_0^{2\pi} h(\theta) G_{\partial\Omega 0}(\theta,\theta_0)
                 e^{-in\theta}d\theta \,, \\
&
 J_{3n}(\theta_0) = \int_0^{2\pi} G_{\partial\Omega 0}^\lambda(\theta,\theta_0)
                 e^{-in\theta}d\theta\,, \qquad
 J_{4n}(\theta_0) = \int_0^{2\pi}\int_0^{2\pi} h(\theta) G_{\Omega 0}^\lambda
     (R,\theta,\tilde{\theta})G_{\partial\Omega 0}^\lambda(\tilde{\theta},\theta_0)
     e^{-in\theta}d\tilde{\theta}d\theta\,,\\
&
 J_{5n}(\theta_0) = \int_0^{2\pi}\int_0^{2\pi} G_{\Omega 0}^\lambda
     (R,\theta,\tilde{\theta})G_{\partial\Omega 0}^\lambda(\tilde{\theta},\theta_0)
     e^{-in\theta}d\tilde{\theta}d\theta\,, \qquad
 J_{6n}(\theta_0) = \int_0^{2\pi}\int_0^{2\pi} \tilde{G}_{\Omega 1}^\lambda
     (R,\theta,\tilde{\theta})G_{\partial\Omega 0}^\lambda(\tilde{\theta},\theta_0)
     e^{-in\theta}d\tilde{\theta}d\theta\,, \\
  &
    J_{7n}(\theta_0) = \int_0^{2\pi}\int_0^{2\pi} h(\theta) G_{\Omega 0 r}^\lambda
     (R,\theta,\tilde{\theta})G_{\partial\Omega 0}^\lambda(\tilde{\theta},\theta_0)
     e^{-in\theta}d\tilde{\theta}d\theta\,.
\end{align*}
By using the Fourier series representations for the
leading-order bulk and membrane Green's functions given in
(\ref{eq:Disk_G_Bulk_Series}) and (\ref{eq:Disk_G_Membrane_Series}),
respectively, together with (\ref{app:hexp}) for $h(\theta)$, we
calculate explicitly that
\begin{align*}
&
 J_{1n}(\theta_0) = -\frac{1}{R} \sum_{k=-\infty}^{\infty} k(n-k) h_{n-k}g_{k}^\lambda
                 e^{-ik\theta_0}\,,\quad
 J_{2n}(\theta_0) = \frac{1}{R} \sum_{k=-\infty}^{\infty} h_{n-k}g_{k}^\lambda
                 e^{-ik\theta_0}\,,\quad
 J_{3n}(\theta_0) = \frac{1}{R} g_{n}^\lambda e^{-in\theta_0}\,,\\
& J_{4n}(\theta_0) = \frac{1}{R^2}\sum_{k=-\infty}^{\infty} h_{n-k}a_{k}^\lambda
    g_{k}^\lambda e^{-ik\theta_0}\,,\quad
 J_{5n}(\theta_0) = \frac{1}{R^2} a_{n}^\lambda g_{n}^\lambda e^{-in\theta_0}\,, \\
  & J_{6n}(\theta_0) = \frac{1}{R^2}\sum_{k=-\infty}^{\infty} h_{n-k}
    \hat{a}_{n,k}^\lambda a_{k}^\lambda g_{k}^\lambda e^{-ik\theta_0}\,,\quad
    J_{7n}(\theta_0) = \frac{1}{R^2}\sum_{k=-\infty}^{\infty} h_{n-k} P_k^{\prime}(R)
    a_{k}^\lambda g_{k}^\lambda e^{-ik\theta_0}\,.
\end{align*}
Upon substituting these expressions into
(\ref{appb:g_memb_int}), and then recalling (\ref{appb:gint}), we
conclude that
\begin{align*}
  \tilde{g}_{1n}^\lambda(\theta_0) = &g_{n}^\lambda
  \sum_{k=-\infty}^{\infty}\bigl\{\tfrac{D_{v0}}{R^3}k(n-k) -
   \tfrac{2\mu_{s\lambda}^2}{R} + \tfrac{2K^2}{R} a_{k}^\lambda +
   K^2 \hat{a}_{n,k}^\lambda a_{k}^\lambda + K^2 P_{k}^{\prime}(R) a_{k}^\lambda\bigr\}
  h_{n-k}g_{k}^\lambda e^{-ik\theta_0} \\
  & \qquad +  \tfrac{D_{v1}}{D_{v0}} g_{n}^\lambda \bigl(\mu_{s\lambda}^2 -
    2\pi K^2 R a_{n}^\lambda \bigr) g_{n}^\lambda e^{-in\theta_0}\,,
\end{align*}
where the coefficients $a_{k}^\lambda$ are defined in
(\ref{eq:Disk_G_Bulk_Series}). We can use the definition of the coefficients
$g_{n}^\lambda$, as given in (\ref{eq:Disk_G_Membrane_Series}), to write
$\mu_{s\lambda}^2 - K^2 a_{n}^\lambda = \tfrac{1}{g_{n}^\lambda} -
\tfrac{D_{v0}}{R^2}n^2$. In this way, we get
\begin{align*}
  \tilde{g}_{1n}^\lambda(\theta_0) = \sum_{k=-\infty}^{\infty}\hat{g}_{n,k}^\lambda
  h_{n-k}g_{k}^\lambda e^{-ik\theta_0} g_{n}^\lambda + \bigl(\tfrac{D_{v1}}{D_{v0}} -
  \tfrac{2h(\theta_0)}{R} \bigr)g_{n}^\lambda e^{-in\theta_0} -
  \tfrac{D_{v1}}{R^2} n^2 (g_{n}^\lambda)^2 e^{-in\theta_0}\,,
\end{align*}
where
$$
\hat{g}_{n,k}^\lambda = \tfrac{D_{v0}}{R^3}k(n+k) + K^2 a_{k}^\lambda
\bigl(\hat{a}_{n,k}^\lambda + P_{k}^{\prime}(R)\bigr).
$$	
Finally, from (\ref{appb:memb_g1_1}) and (\ref{appb:memb_g1_2}), we
conclude that the first order correction for the membrane Green's
function is
$$
G_{\partial\Omega 1}^\lambda(\theta,\theta_0) =
-\tfrac{h(\theta_0)}{R} G_{\partial\Omega 0}^\lambda(\theta,\theta_0)
+ \tfrac{1}{2\pi R}\sum_{n=-\infty}^\infty \sum_{k=-\infty}^\infty
\hat{g}_{n,k}^\lambda h_{n-k}g_{k}^\lambda g_{n}^\lambda e^{in\theta
  - ik\theta_0} - \tfrac{D_{v1}}{2\pi R^3}\sum_{n=-\infty}^{\infty}
n^2 (g_{n}^\lambda)^2 e^{in(\theta-\theta_0)}\,.
$$

\end{document}